\documentclass[runningheads]{llncs}

\usepackage{graphicx}
\usepackage{subfigure}
\usepackage{comment}
\usepackage{multirow}
\usepackage{xcolor} 
\usepackage{arydshln} 

\usepackage{graphicx}
\usepackage{float}
\usepackage{bbding}
\usepackage{enumerate}
\usepackage{color,soul}
\usepackage{graphicx}
\usepackage{float}
\usepackage{bbding}
\usepackage{amsmath}
\usepackage{xcolor}
\usepackage{colortbl}
\usepackage{fancybox}
\usepackage{amssymb}
\usepackage{pdfpages}
\usepackage{tikz}
\usepackage{array}
\usepackage{multirow}
\usepackage{capt-of}
\usepackage{color,soul}
\usepackage{longtable}
\usepackage{forest}
\usepackage{multirow}
\usepackage{tabularx}
\usepackage{eqnarray}
\usepackage{color,soul}
\usepackage{ltablex} 
\usepackage{ltablex}
\usepackage{supertabular}
\usepackage{arydshln}
\usepackage[linesnumbered,ruled]{algorithm2e}
\usepackage{url}

\begin{document}
\title{{\em Bharatanatyam} Dance Transcription using Multimedia Ontology and Machine Learning}

\author{Tanwi Mallick \and Patha Pratim Das \and
Arun Kumar Majumdar}
\authorrunning{T. Mallick et al.}
%
\institute{Indian Institute of Technology Kharagpur
\email{tanwimallick@gmail.com, ppd@cse.iitkgp.ac.in, akmj@cse.iitkgp.ac.in}}

\maketitle

\begin{abstract}
Indian Classical Dance is an over 5000 years' old multi-modal language for expressing emotions. Preservation of dance through multimedia technology is a challenging task. In this paper, we develop a system to generate parse-able representation of a dance performance. The system will help to preserve intangible heritage, annotate performances for better tutoring, and synthesize dance performances. We first attempt to capture the concepts of the basic steps of an Indian Classical Dance form, named {\em Bharatanatyam Adavu}s, in an ontological model. Next, we build an event-based low-level model that relates the ontology of {\em Adavu}s to the ontology of multi-modal data streams (RGB-D of Kinect in this case) for a computationally realizable framework. Finally, the ontology is used for transcription into Labanotation. We also present a transcription tool for encoding the performances of {\em Bharatanatyam Adavu}s to Labanotation and test it on our recorded data set. Our primary aim is to document the complex movements of dance in terms of Labanotation using the ontology. 
\end{abstract}

%
%

\keywords{Multimedia Ontology, Dance Transcription, Machine Learning, \\ {\em Bharatanatyam Adavu}, Laban XML, Labanotation}




\section{Introduction}
Dance is a form of art that may a tell story, set a mood, or express emotions. Indian Classical Dance (ICD) is an ancient heritage of India, which is more than 5000 years old. With the passage of time, the dance has been performed, restructured, reformulated, and re-expressed by several artists. New choreography has been composed using the basic forms. Hence, the dance forms have been associated with rich set of rules, formations, postures, gestures, stories, and other artifacts. But, till date it has been passed on to the students by the teacher, from one generation to the next, through the traditional method of {\em Guru-Shishya Parampara}, which is the typically acknowledged Indian style of education where the teacher ({\em Guru}) personally trains her / his disciple ({\em Shishya}) to keep up a continuity ({\em Parampara}) of education, culture, learning, or skills. Hence, there is a need to preserve the intangible heritage of the dance artifacts.

Recently many significant systems have been developed to preserve cultural heritage through digital multimedia technology. Preservation of the tangible heritage resources like monuments, handicrafts, and sculpture can be done through digitization, and 2D and 3D modeling techniques. Preservation of intangible resources like language, art and culture, music and dance, is more complex and requires a knowledge intensive approach.  Therefore, only little work has been carried out for preservation of the dance heritage.

These dance forms embody a collection of knowledge which can be preserved either through creating digital transcription of the performances or by annotating the video recordings of performances. Analysis of dance can help convert the audio-visual information of dance into a graphical notation. Dance transcription, still a rarity, can be handy to preserve the heritage of a country like India which boasts of diverse types of classical dance forms. Transcription can also help exchanging dance ideas between performers. Another way of preserving the intangible heritage of dance is dance media annotation or to attach conceptual metadata to the collection of digital artifacts. The collection of digital artifacts with conceptual metadata can help in semantic access to the heritage collection. 

Mallik et al.~\cite{mallik2011nrityakosha} present an ontology based approach for designing a cultural heritage repository. A Multimedia Web Ontology Language (MOWL) is proposed to encode the domain knowledge of a choreography. The suggested architectural framework includes a method to construct the ontology with a labeled set of training data and the use of the ontology to automatically annotate new instances of digital heritage artifacts. The annotations enable creation of a semantic navigation environment in the repository. The efficacy of the approach is demonstrated by constructing an ontology for the domain of ICD in an automated fashion, and with a browsing application for semantic access to the heritage collection of ICD videos.

Use of notation is another way of recording dance for future use. A system of notation is required for recording the details of postures and movements in domain of dance. Labanotation~\cite{guest2014labanotation} is a widely used notation system for recording human movements in terms of graphical primitives and symbols. Karpen~\cite{Karpen90} first attempted to manually encode the movements of {\em Bharatanatyam} on paper using Labanotation. It has been demonstrated by examples that the body movement, space, time, and dynamics of the ICD, in particular {\em Bharatanatyam}, can be described through Labanotation. Hence it is argued that the Labanotation, coupled with video filming, is a good way to record ICD. According to the author, hand gestures can also be easily implemented in Labanotation together with palm facing and specification of the quality of movement. However, no attempt was made in this paper to automate the process and for the next about three decades no work was done in transcribing ICD in Labanotation. Some research on dance preservation using notation has been carried out in other dance forms like Thai dance, Contemporary dance etc. Raheb et al.~\cite{el2011labanotation} use {\em Web Ontology Language} (OWL) to encode the knowledge of dance. The semantics of the Labanotation system is used to build elements of the ontology. Tongpaeng et al.~\cite{tongpaeng2017thai} propose a system to archive the knowledge of Thai dance using Labanotation and then use the score of the notation  to represent the dance in 3D Animation. Till date automatic generation of Labanotation from the recorded dance video has not been attempted.

This work has been inspired by the idea of musical notations. Similar dance transcription systems may be useful in several way. The system can generate parse-able representation of a dance performance, help to preserve intangible heritage, help to annotate performances for better tutoring, and can be used as a front-end for dance synthesis. We first attempt to capture the concepts in {\em Bharatanatyam Adavu}s in an ontological model. At the top level, a {\em Bharatanatyam Adavu} can be expressed as a dance (sequence of visual postures) accompanied by music. Further, we identify the concepts of audio and video structures of {\em Bharatanatyam Adavu}. We next build an event-based low-level model that relates the ontology of {\em Adavu}s to the ontology of multi-modal data streams (RGB-D of Kinect in this case) for a computationally realizable framework. An event denotes the occurrence of an activity (called {\em Causal Activity}) in the audio or the video stream of an {\em Adavu}. The events of audio, video and their synchronization, thus, are related to corresponding concepts of the ontological model. We use this ontology and event characterization for transcription into Labanotation using Laban ontology. We also present a transcription tool for encoding the performances of {\em Bharatanatyam Adavu}s to Labanotation and test it on our recorded data set. Our aim is to examine the ways in which Labanotation can be used for documenting the dance movements.

\section{Indian Classical Dance: {\em Bharatanatyam} and its {\em Adavu}s}
\label{lab_bn}
We introduce the domain of {\em Bharatanatyam Adavu} in the context of our knowledge capture and heritage preservation scheme of the article for the ease of understanding the through the entire paper.

{\em Bharatanatyam} is one of the eight\footnote{\scriptsize ICD has eight distinct styles as recognized by the Ministry of Culture, Government of India: namely, {\em Bharatanatyam}, {\em Kathak}, {\em Odissi}, {\em Kathakali}, {\em Kuchipudi}, {\em Manipuri}, {\em Mohiniyattam}, {\em Sattriya}.} Indian Classical Dance forms. Like most dance forms, {\em Bharatanatyam Adavu} too is deeply intertwined with music. It is usually accompanied by instrumental ({\em Tatta Kazhi}\footnote{\scriptsize A wooden stick is beaten on a wooden block to produce instrumental sound.}, {\em Mridangam}, Flute, Violin, {\em Veena}, etc.) and / or vocal music (Carnatic style -- with or without lyrics) called {\em Sollukattu}. {\em Adavu}s are the basic units of {\em Bharatanatyam} that are combined to create a dance performance. An {\em Adavu} involves various postures and gestures of the body including torso, head, neck, hands, fingers, arms, legs and feet, and eyes.
While performing {\em Adavu}s, the dancer stamps, rubs, touches, slides on the ground in different ways in synchronization with the {\em Sollukattu}. There are 15 basic {\em Adavu}s in {\em Bharatanatyam} -- most having one or more {\em Variants}. In total, we deal with 58 {\em Adavu} variants. There exists a many-to-one mapping from the {\em Adavu}s to the {\em Sollukattu}s. 

\subsection{{\em Sollukattu}s and {\em Bol}s -- the Music of {\em Adavu}s} \label{sec:music}
{\em Bharatanatyam} is deeply intertwined with music. It is usually accompanied by Instrumental ({\em Tatta Kazhi}, {\em Mridangam}, Flute, Violin, {\em Veena}, etc.) and / or Vocal music (Carnatic style -- with or without lyrics). The music is strung together in sequences to create different rhythmic patterns, called {\em Taalam}\footnote{{\em Taalam} is the Indian system for organizing and playing metrical music.}, to accompany dance performances. A repeated cycle of {\em Taalam} consists of a number of equally spaced beats, which are grouped into combinations of patterns. Time interval between any two beats is always equal. 
The specific way they mark the beats (by tapping their laps with their fingers, palm, and back of the hand; or by a specific instrument) are determined by these patterns of the beats or the {\em Taalam}. 


\renewcommand{\baselinestretch}{1} 
\begin{table}
\caption{List of {\em Sollukattu}s with {\em bol} compositions\label{tbl:bn_Bols}}
\begin{scriptsize}
\begin{center}
\begin{tabular}{|l|l|p{9cm}|} \hline
\multicolumn{1}{|c}{\bf {\em Sollukattu}}	& \multicolumn{1}{|c}{\bf \#}	& \multicolumn{1}{|c|}{\bf Description of {\em Bol}s}	\\
\multicolumn{1}{|c}{ }	& \multicolumn{1}{|c}{\bf Beats}	& \multicolumn{1}{|c|}{}	\\ \hline \hline
{\em Joining A}	&	8	&	tat dhit ta [B] tat dhit ta	[B] \\ \hline
{\em Joining B}	&	6	&	[dhit dhit] tei [dhit dhit] tei [dhit dhit] tei [dhit dhit] tei 	\\ \hline
{\em Joining C}	&	8	&	tei tei [dhit dhit] tei tei tei [dhit dhit] tei	\\ \hline
{\em KUMS}	&	6	&	[tan gadu] [tat tat] [dhin na] [tan gadu] [tat tat] [dhin na]	\\ \hline 
{\em Mettu}	&	8	&	tei hat tei hi tei hat tei hi	\\ \hline
{\em Nattal A}	&	8	&	tat tei tam [B] dhit tei tam [B]	\\ \hline
{\em Nattal B}	&	8	&	[tat tei] tam [dhit tei] tam [tat tei] tam [dhit dhit] tei 	\\ \hline
{\em Tattal}	&	8	&	tat tei ta ha dhit tei ta ha	\\ \hline
{\em Natta}	&	8	&	[tei yum] [tat tat] [tei yum] ta [tei yum] [tat tat] [tei yum] ta	\\ \hline
{\em Paikkal}	&	8	&	[dhit tei da] [ta tei] [dhit tei da] [ta tei] \newline [dhit tei da] [ta tei] [dhit tei da] [ta tei]	\\ \hline
{\em Pakka}	&	8	&	ta tei tei tat dhit tei tei tat	\\ \hline
{\em Sarika}	&	8	&	tei a tei e tei a tei e	\\ \hline
{\em Tatta A}	&	8	&	[tei ya] tei [tei ya] tei	[tei ya] tei [tei ya] tei \\ \hline
{\em Tatta B}	&	6	&	tei tei tam tei tei tam	\\ \hline
{\em Tatta C}	&	8	&	[tei ya] [tei ya] [tei ya] tei [tei ya] [tei ya] [tei ya] tei	\\ \hline
{\em Tatta D}	&	8	&	tei tei [tei tei] tam tei tei [tei tei] tam \\ \hline
{\em Tatta E}	&	8	&	tei tei tam [B] tei tei tam [B]	\\ \hline
{\em Tatta F}	&	8	&	tei tei tat tat tei tei tam [B]	\\ \hline
{\em Tatta G}	&	6	&	tei tei tei tei [dhit dhit] tei	\\ \hline
{\em TTD}	&	8	&	[tei tei] [dhat ta] [dhit tei] [dhat ta]  [tei tei] [dhat ta] [dhit tei] [dhat ta]	\\ \hline
{\em Tirmana A}	&	12	&	ta [tat ta] jham [ta ri] ta [B] jham [ta ri] jag [ta ri] tei [B]	\\ \hline
{\em Tirmana B}	&	12	&	[tat ding] [gin na] tom [tak ka] [tat ding] [gin na] \newline tom [tak ka] [dhi ku] [tat ding] [gin na] tom	\\ \hline
{\em Tirmana C}	&	12	&	[ki ta ta ka] [dha ri ki ta] tom tak [ki ta ta ka] [dha ri ki ta] \newline tom [tak ka] [dhi ku] [ki ta ta ka] [dha ri ki ta] tom	\\ \hline
\multicolumn{3}{l}{ } \\
\multicolumn{3}{l}{$\bullet$ Multiple {\em bol}s at the same beat are enclosed within []} \\
\multicolumn{3}{l}{$\bullet$ [B] stands for a beat without any {\em bol}s -- typically called stick-beat} \\
\multicolumn{3}{p{11.5cm}}{$\bullet$ KUMS, Mettu, Nattal, Tattal, and TTD stand for Kartati--Utsanga--Mandi--Sarikkal, Kuditta Mettu, Kuditta Nattal, Kuditta Tattal, and Tei Tei Dhatta respectively}
\end{tabular}
\end{center}
\end{scriptsize}
\end{table}
\renewcommand{\baselinestretch}{1.3} 

{\em Taalam}s necessarily synchronize the movements of various parts of the body with the music through a structured harmonization of four elements, namely -- (a) Rhythmic beats of {\em Taalam}, (b) {\em Mridangam} beats from percussion, (c)  Musical notes or {\em Swara}s\footnote{{\em Swara}, in Sanskrit, connotes a note in the successive steps of the octave.}, and (d)  Steps of the {\em Adavu}s.
It may be noted that a number of different {\em Taalam}s are used in {\em Bharatanatyam}. The {\em Taalam}s\footnote{{\em Adavu}s can be performed in all 7 {\em taalam}s as well; but the rest are less popular.} commonly used in {\em Adavu} are -- {\em Adi taalam} (8 beats' pattern) and {\em Roopakam taalam} (6 beats' pattern).
Finally, a {\em Taalam} is devoid of a physical unit of time and is acceptable as long as it is rhythmic in some temporal unit. With a base time unit, however, {\em Bharatanatyam} deals with three speeds,  called {\em Kaalam} or {\em Tempo}. The {\em Taalam}s are played mainly in 3 different tempos -- {\em Vilambitha Laya} or slow speed,
 {\em Madhya Laya} (double of {\em Vilambitha Laya}) or medium speed, and {\em Drutha Laya} (quadruple of {\em Vilambitha Laya}) or fast speed.


A phrase of rhythmic syllables ({\em Sollukattu}), is linked to specific units of dance movement in an {\em Adavu}. A {\em Sollukattu}\footnote{`{\em sollukattu}' = {\em sollum} (syllables) + {\em kattu} (speaking). A {\em Sollukattu} means a phrase of rhythmic syllables linked to specific units of dance movement ({\em Adavu}). 
} is a specific rhythmic musical pattern created by combination of instrumental and vocal sounds. Traditionally, a {\em Tatta Kazhi} (wooden stick) is beaten on a {\em Tatta Palahai} (wooden block) for the instrumental sound and an accomplice of the dancer speaks out a distinct vocalization of rhythm, like {\em tat}, {\em tei}, {\em ta} etc., called {\em Bol}s\footnote{{\em Bol}s (or {\em bolna} = {\em to speak}), are mnemonic syllables for beats in the {\em taalam}.}.
In a {\em Sollukattu}, both the instrument and the voice follow in sync to create a pattern of beats. Every beat is usually marked by a synchronous beating (instrumental) sound, though some beats may be silent. In some cases, there may be beating (instrumental) sound at positions that are not beats (according to the periodicity). The list of {\em Sollukattu}s are given in Table \ref{tbl:bn_Bols}. As 
{\em Adavu}s are performed along with the rhythmic syllables of a {\em Sollukattu} that continues to repeat in cycles. Rhythm performs the role of a timer (with beats as temporal markers). Between the interval of beats, the dancer changes her posture.

\subsection{{\em Adavu}s -- the Postures and Movements} 
\label{sec:adavus}
{\em Adavu}s are the basic unit of Bharatanatyam that are combined to form a dance sequence in Bharatanatyam. {\em Adavu}s form the foundation stone on which the entire {\em Nritta} rests. It involves various postures, gestures of the body, hand, arms, feet, and eyes\footnote{Current work does not consider hand and eye movements for limitations of sensors.}. While performing {\em Adavu}s the dancer stamps, rubs, touches, slides on the ground in different ways in synchronization with the {\em Sollukattu} (bol) or the syllables used. 
The {\em Adavu}s are classified according to the rhythmic syllables on which they are based and  the style of footwork employed. According to {\em Kalakshetra} school of training there are 15 {\em Adavu}s. Most {\em Adavu}s have two or more {\em Variants}. Variants of an {\em Adavu} bear similarity of intent and style, but differ in details. A total 58 {\em Adavu}s and 23 {\em Sollukattu}s are used in {\em Kalakshetra}\footnote{There are four major styles of {\em Bharatanatyam} -- {\em Thanjavur}, {\em Pandanallur}, {\em Vazhuvoor}, and {\em Mellatur}. {\em Kalakshetra}, promulgated by the {\em Kalakshetra Foundation} founded by Rukmini Devi, is the modern style of {\em Bharatanatyam} and is reconstructed from {\em Pandanallur} style.}. The details are listed in Table \ref{tbl:adavus_sollukattu}. Every (variant of an) {\em Adavu} uses a fixed  {\em Sollukattu} while a given {\em Sollukattu} may be used in multiple {\em Adavu}s. Each posture of {\em Adavu}s is a combination of leg support ({\em Mandalam}), legs position ({\em Pada Bheda}), arms position ({\em Bahu Bheda}), head position ({\em Shiro Bheda}), hand position ({\em Hasta Mudra}s), neck position ({\em Griba Bheda}), eyes position ({\em Drishti Bheda}).

\renewcommand{\baselinestretch}{1} 
\begin{table}[!ht]
\caption{List of {\em Adavu}s with accompanying {\em Sollukattu}}\label{tbl:adavus_sollukattu}
\centering
\begin{scriptsize}
\begin{tabular}{|r|l|l|l|l|} 
\hline
\multicolumn{1}{|c|}{\bf \#}	& 
\multicolumn{2}{c|}{\bf \em Adavu} & 
\multicolumn{1}{c|}{\bf \em Taalam} &
\multicolumn{1}{c|}{\bf \em Sollukattu} \\ \cline{2-3} 

\multicolumn{1}{|c|}{\bf }	& 
\multicolumn{1}{c|}{\bf Name} & 
\multicolumn{1}{c|}{\bf Variants} & 
\multicolumn{1}{c|}{\bf } &
\multicolumn{1}{c|}{\bf }\\ \hline\hline

\multirow{3}{*}{1}& \multirow{3}{*}{Joining}	&  Joining 1	& \multirow{3}{*}{Adi} & Joining A \\ \cline{3-3} \cline{5-5}
&& Joining 2 &  & Joining B \\ \cline{3-3} \cline{5-5}
&& Joining 3 &  & Joining C \\ \hline

2 & Kati or Kartari	&  Kati or Kartari 1	& Roopakam	& KUMS \\ \hline

3 & Kuditta Mettu	&  Kuditta Mettu 1--4	& Adi & Kuditta Mettu \\ \hline

\multirow{3}{*}{3}& \multirow{3}{*}{Kuditta Nattal}	&  Kuditta Nattal 1--3	& \multirow{3}{*}{Adi}	& Kuditta Nattal A \\ \cline{3-3} \cline{5-5}
&& Kuditta Nattal 4--5 & & Kuditta Nattal B \\ \cline{3-3} \cline{5-5}
&& Kuditta Nattal 6 & & Kuditta Nattal A \\ \hline

5 & Kuditta Tattal	&  Kuditta Tattal 1--5	& Adi & Kuditta Tattal \\ \hline

6 & Mandi	&  Mandi 1--2	& Roopakam	& KUMS	\\ \hline

7 & Natta	&  Natta 1--8	& Adi & Natta \\ \hline

8 & Paikkal	&  Paikkal 1--3	& Adi	& Paikkal \\ \hline

9 & Pakka	&  Pakka 1--4	& Adi	& Pakka \\ \hline

10 & Sarika	&  Sarika 1--4	& Adi & Sarika	\\ \hline

11 & Sarrikkal	&  Sarrikkal 1--3	& Roopakam & KUMS \\ \hline

\multirow{7}{*}{12} & \multirow{7}{*}{Tatta} &  Tatta 1--2 & \multirow{1}{*}{Adi} & \multirow{1}{*}{Tatta A}\\ \cline{3-5}
&& Tatta 3 & Roopakam & Tatta B \\ \cline{3-5}
&& Tatta 4 & \multirow{4}{*}{Adi} & Tatta C \\ \cline{3-3} \cline{5-5}
&& Tatta 5 &  & Tatta D \\ \cline{3-3} \cline{5-5}
&& Tatta 6 &  & Tatta E \\ \cline{3-3} \cline{5-5}
&& Tatta 7 &  & Tatta F \\ \cline{3-5}
&&Tatta 8 & Roopakam & Tatta G \\ \hline

13 & Tei Tei Dhatta	&  Tei Tei Dhatta 1--3	& Adi	& Tei Tei Dhatta \\ \hline

\multirow{3}{*}{14}& \multirow{3}{*}{Tirmana}	&  Tirmana 1	& \multirow{3}{*}{Roopakam}	&  Tirmana A\\ \cline{3-3} \cline{5-5}
&& Tirmana 2 &  & Tirmana B\\ \cline{3-3} \cline{5-5} 
&& Tirmana 3 &  & Tirmana C\\ \hline

15 & Utsanga	&  Utsanga 1	& Roopakam & KUMS \\ \hline
\multicolumn{5}{l}{ } \\
\multicolumn{5}{l}{$\bullet$ KUMS stands for Kartati--Utsanga--Mandi--Sarikkal} \\
\end{tabular}
\renewcommand{\arraystretch}{1}
\end{scriptsize}
\end{table}
\renewcommand{\baselinestretch}{1.3} 
Since {\em Adavu}s are elementary units and used for training, each {\em Adavu} has a specific purpose (as shown is Table~\ref{tbl:purpose_adavus}). For example, {\em Tatta Adavu}s focus on {\em striking of the floor with foot}.  The body remains in a posture called {\em Araimandi} and the feet, by rotation, strike the floor alternately with the sole. There are 8 variants of {\em Tatta Adavu}. The features of the Variant 1 of {\em Tatta Adavu} (say, {\em Tatta} 1) are -- (a) Strike on the floor, (b) Heel to touch hip during strike, (c) No hand gesture, and (d) No movements. The {\em Sollukattu} used in {\em Tatta} 1 is {\em tei a tei} (say, {\em Tatta\_A}). This follows the {\em Adi Taalam} or 8 beats' pattern as shown in Table~\ref{tbl:tatta1_beats}. The {\em bol}s on each beat are shown in three different tempos. 


\renewcommand{\baselinestretch}{1} 
\begin{table}[!ht]
\caption[Purpose of various {\em Adavu}s]{Purpose of various {\em Adavu}s}\label{tbl:purpose_adavus}
\centering
\begin{scriptsize}
\begin{tabular}{|l|p{9.5cm}|}
 \hline
\multicolumn{1}{|c}{\bf \em Adavu} & 
\multicolumn{1}{|c|}{\bf \em Purpose of the {\em Adavu}} \\ \hline\hline
{\em Joining} & Simple connecting {\em Adavu}s to be used while building longer sequences of postures\\ \hline

{\em Kati or Kartari} & {\em Paidhal} itself includes a variety of leaps and may also be coupled with spins ({\em Bramhari}). It also includes the famous {\em Kartari} (Scissors) {\em adavu} where the movement of the hand and feet trace crisscross patterns in space.\\ \hline

{\em Kuditta Mettu} & Jumping on the toes and then striking the heels \\ \hline

{\em Kuditta Nattal} & Striking the floor by leg, jumping on toes, stretching legs and hands and also circular movement of hand \\ \hline

{\em Kuditta Tattal} & Striking the floor, jumping on toes, stretching hands, circular movements of hands, neck and head with the bending of torso and  waist and hand movements define different planes in space\\ \hline

{\em Mandi} & {\em Mandi} in some Indian languages refers to area around the thigh and knee. In some instance we can refer it to a bent knee. For example, {\em Araimandi} is where the knee is half bent. {\em Muzhumandi} or {\em Poorna Mandala} is where the knee is fully bent. In {\em Mandi} {\em adavu}s we make use of the {\em Muzhumandi} position often. Steps could vary from jumps in {\em Poorna Mandala} to jumping and touching one knee on the floor. \\ \hline

{\em Natta} & Stretching of legs \\ \hline

{\em Paikkal} & {\em Paikkal} ({\em Paidhal} or {\em Paichal}) is a Tamil term that means {\bf \em to leap}. It differs from the {\em Kuditta Mettu} in the sense, the dancer while doing the {\em Paikkal} covers space, whereas in {\em Kuditta Mettu} she / he jumps in the same spot. A very graceful step in itself, {\em Paikkal} is usually seen at the end of {Korvai} (a string of {\em Adavu}s) as part of {\em Ardhi}s. \\ \hline

{\em Pakka} & Moving towards sides \\ \hline

{\em Sarika} & {\em Sarika} means a thing of beauty or nature \\ \hline

{\em Sarrikkal} & {\em Sarrikkal} means {\bf \em to slide}. Here as one foot is lifted and placed the another foot slides towards it. \\ \hline

{\em Tatta} & Striking the floor with feet \\ \hline

{\em Tei Tei Dhatta} & Use of half and full seating, stretching legs and hand, jumping with linear and circular movements of hands\\ \hline

{\em Tirmana} & {\em Tirmana} (or {\em Teermanam} means {\bf \em to conclude} or an ending or a final stage. Thus the steps in these {\em adavu}s are used to end a dance sequence or {\em jathi}s. It is done in a set of three steps or repeated thrice. \\ \hline

{\em Utsanga} & Use of different hand position to enhance the stretching on half seating, straight standing, jump on heels, striking the floor. Also use of linear and circular movements of waist and stretching of hands.\\ \hline
\multicolumn{2}{c}{ } \\
\multicolumn{2}{c}{{\bf Source:} ~\cite{preetivasudevan} and personal communication with Debaldev Jana}
\end{tabular}
\end{scriptsize}
\end{table}
\renewcommand{\baselinestretch}{1.3}

\renewcommand{\baselinestretch}{1} 
\begin{table}[!ht]
\caption{Beat pattern of {\em Tatta} 1 ({\em Tatta Adavu} Variant 1) in {\em Adi Taalam}}\label{tbl:tatta1_beats}
\centering
\begin{scriptsize}
\begin{tabular}{|l||c|c|c|c|c|c|c|c|} \hline
\multicolumn{1}{|c||}{\bf Beats} & {\bf 1} & {\bf 2} & {\bf 3} & {\bf 4} & {\bf 5} & {\bf 6} & {\bf 7} & {\bf 8} \\ \cline{1-1}
\multicolumn{1}{|c||}{\bf Speed} &&&&&&&&  \\ \hline \hline
1$^{st}$ & tei a & tei  & tei a & tei & tei a & tei & tei a & tei \\  \hline
2$^{nd}$ & tei a tei & tei a tei & tei a tei & tei a tei & tei a tei & tei a tei & tei a tei & tei a tei \\  \hline
3$^{rd}$ & tei a tei & tei a tei & tei a tei & tei a tei & tei a tei & tei a tei & tei a tei & tei a tei \\ 
 & tei a tei & tei a tei & tei a tei & tei a tei & tei a tei & tei a tei & tei a tei & tei a tei \\ \hline
\multicolumn{9}{c}{{\bf Time Measure}: {\em Adi Taalam}} \\
\end{tabular}
\end{scriptsize}
\end{table}
\renewcommand{\baselinestretch}{1.3} 

\begin{figure}[!ht]
\centering
\begin{tabular}{ccccccc}
Right Strike  &\hspace*{0.5cm}& Left Strike &\hspace*{0.5cm}& Right Strike &\hspace*{0.5cm}& Left Strike \\
\begin{minipage}{.2\textwidth}
      \includegraphics[width=2.9cm]{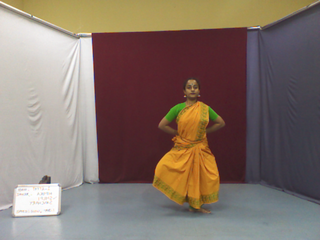}
\end{minipage}
&\hspace*{0.5cm}&
\begin{minipage}{.2\textwidth}
      \includegraphics[width=2.9cm]{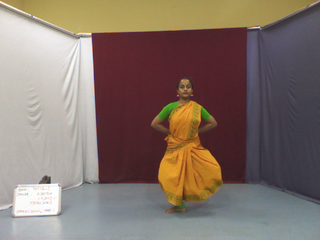}
\end{minipage}
&\hspace*{0.5cm}& 
\begin{minipage}{.2\textwidth}
      \includegraphics[width=2.9cm]{Images/Bharatanatyam_Images/tatta1.png}
\end{minipage}
&\hspace*{0.5cm}& 
\begin{minipage}{.2\textwidth}
      \includegraphics[width=2.9cm]{Images/Bharatanatyam_Images/tatta2.png}
\end{minipage}
\\ 
tei a (beat 1) &\hspace*{0.5cm}& tei (beat 2) &\hspace*{0.5cm}& tei a (beat 3) &\hspace*{0.5cm}& tei (beat 4) \\  \\
Right Strike  &\hspace*{0.5cm}& Left Strike &\hspace*{0.5cm}& Right Strike &\hspace*{0.5cm}& Left Strike \\
\begin{minipage}{.2\textwidth}
      \includegraphics[width=2.9cm]{Images/Bharatanatyam_Images/tatta1.png}
\end{minipage}
&\hspace*{0.5cm}&
\begin{minipage}{.2\textwidth}
      \includegraphics[width=2.9cm]{Images/Bharatanatyam_Images/tatta2.png}
\end{minipage}
&\hspace*{0.5cm}& 
\begin{minipage}{.2\textwidth}
      \includegraphics[width=2.9cm]{Images/Bharatanatyam_Images/tatta1.png}
\end{minipage}
&\hspace*{0.5cm}& 
\begin{minipage}{.2\textwidth}
      \includegraphics[width=2.9cm]{Images/Bharatanatyam_Images/tatta2.png}
\end{minipage}
\\ 
tei a (beat 5) &\hspace*{0.5cm}& tei (beat 6) &\hspace*{0.5cm}& tei a (beat 7) &\hspace*{0.5cm}& tei (beat 8) \\      
\end{tabular}
\caption{Example performance of {\em Tatta} 1 ({\em Tatta Adavu} Variant 1)} \label{fig:tatta1}
\end{figure}

The posture of a dancer is synchronized with the beats. The synchronized postures with beats are shown in Figure~\ref{fig:tatta1}. Here, the dancer strikes her left and right foot with the beats in rotation. 

\renewcommand{\baselinestretch}{1} 
\renewcommand{\baselinestretch}{1.3} 

\begin{figure}[!ht]
\centering
	\includegraphics[width=8.5cm]{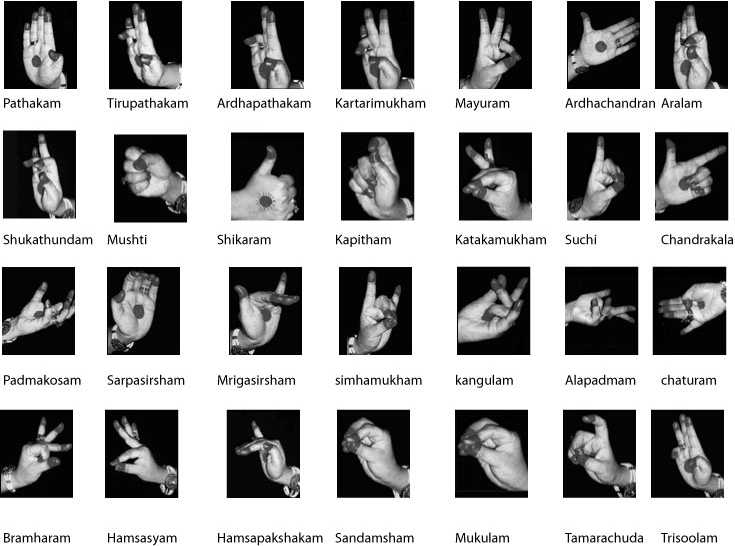}\\ 
(a) Sets of {\em Asamyutha Hasta Mudra}s\\ \ \newline
	\includegraphics[width=8.5cm]   {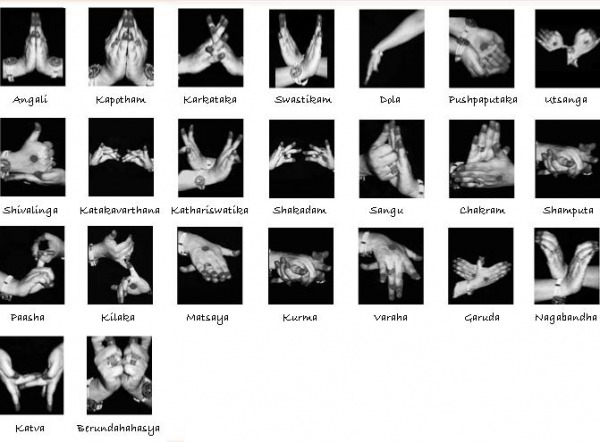}\\ 
(b) Sets of {\em Samyutha Hasta Mudra}s\\ \ \newline 
	{\tiny {\bf Image Source:} \url{https://grade1.weebly.com/theory.html}}\vspace*{-0.3cm}
\caption{{\em Hasta Mudra}s of {\em Bharatanatyam}}
\label{fig:BN_Hasta_Set}
\end{figure}

Like {\em Tatta} 1, all {\em Adavu}s are combinations of:
\begin{itemize} 
\item {\bf Position of the legs ({\em Sthanakam}) / Posture of standing ({\em Mandalam})}: {\em Adavu}s are performed in postures that are (Figure~\ref{fig:mandalams}) -- (a) {\em Samapadam} or the standing position, (b) {\em Araimandi} / {\em Ardha Mandalam} or the half sitting posture, and (c) {\em Muzhumandi} or the sitting posture.


\renewcommand{\baselinestretch}{1} 
\begin{figure}[!ht]
\centering
\begin{tabular}{ccc}
\includegraphics[width=3cm]{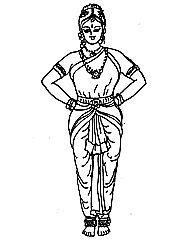} &
\includegraphics[width=3cm]{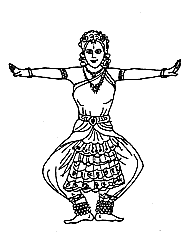} &
\includegraphics[width=3cm]{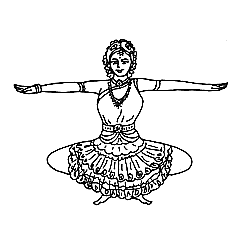} \\
{\em Samapadam} & {\em Araimandi} & {\em Muzhumandi} \\
(Standing) & (Half Sitting) & (Full Sitting) \\
\multicolumn{3}{c}{\tiny {\bf Source:} Leg Postures in Bharatanatyam by 
Nysa Dance Academy} \\
\multicolumn{3}{c}{\tiny \url{https://nysadancecom.wordpress.com/2015/09/26/leg-posture-aramandi-or-ardhamandala/}}
\end{tabular}

\caption{Three {\em Mandalam}s (types of leg support) of {\em Bharatanatyam}}
\label{fig:mandalams}
\end{figure}
\renewcommand{\baselinestretch}{1.3} 


\item {\bf Jumps ({\em Utplavana})}: Based on the mode of performances {\em Utplavana}s are classified into {\em Alaga}, {\em Kartari}, {\em Asva}, {\em Motita}, and {\em Kripalaya}. 

\item {\bf Walking Movement ({\em Chari})}: {\em Chari} are used for gaits. According to {\em Abhinayadarpana}, there are eight kinds of {\em Chari}s -- {\em Chalana}, {\em Chankramana}, {\em Sarana}, {\em Vehini}, {\em Kuttana}, {\em Luhita}, {\em Lolita}, and {\em Vishama Sanchara}. 

\item {\bf Hand Gestures ({\em Nritta Hasta}s)}: {\em Bharatanatyam} primarily uses two types\footnote{Few other types like {\em Nritya Hasta} are used at times.} of {\em Hasta Mudra}s (Figure~\ref{fig:BN_Hasta_Set}) that play a significant role in communication --  28 single hand gestures ({\em Asamyutha Hasta}) and 23 combined (both) hand gestures ({\em Samyutha Hasta}). There are twelve major hand gesture for {\em Adavu}s -- {\em Pataka}, {\em Tripataka}, {\em Ardhachandra}, {\em Kapittha}, {\em Katakamukha}, {\em Suchi Musthi}, {\em Mrigasirsha}, {\em Alapadma}, {\em Kaetarimukha}, {\em Shikhara}, and {\em Dola}.

%

\end{itemize}

In {\em Bharatanatyam}, {\bf \em Adavu} is used in dual sense. It either denotes just the dance part (postures and movements) or the dance and the accompanying music together. To maintain clarity of reference, in this paper, we refer to the dance simply by {\em Adavu} and the composite of dance and music by {\em Bharatanatyam Adavu}.

\renewcommand{\baselinestretch}{1} 
\renewcommand{\baselinestretch}{1.3}

%
%
%



\section{Object-based Modeling of {\em Adavu}s}\label{sec:ontological_models}
To express the ontology we follow an extended object-based modeling framework comprising a set of classes (Table~\ref{tbl:ontology_classes}), a set of instances (Table~\ref{tbl:ontology_instances}), and a set of relations (Table~\ref{tbl:ontology_relations}). Classes are used to represent generic as well as specific concepts. These can be {\em Abstract} or {\em Concrete}. A concrete class has one or more instances while an abstract class has one or more specialized classes. Relations are usually binary and are defined between two classes, between a class and an instance, or between two instances.

\renewcommand{\baselinestretch}{1} 
\begin{table}[!ht]
\centering
\caption{List of Classes for the ontology of {\em Bharatanatyam Adavu} \label{tbl:ontology_classes}}
\begin{scriptsize}
\begin{tabular}{|p{4.5cm}|l|p{6cm}|} 
\hline
\multicolumn{1}{|c}{\bf Class} & \multicolumn{1}{|c}{\bf Type} & \multicolumn{1}{|c|}{\bf Remarks} \\ \hline\hline
$\bullet$ {\em Sollukattu} & Concrete & The music (audio) of {\em Adavu}s (Table~\ref{tbl:bn_Bols})\\
$\bullet$ {\em Adavu} & Abstract & The movements (video) of {\em Adavu}s (Table~\ref{tbl:adavus_sollukattu}) \\
$\bullet$ {\em Tatta Adavu}, {\em Natta Adavu}, $\cdots$ & Concrete & Types of {\em Adavu}s (Table~\ref{tbl:adavus_sollukattu}) \\
$\bullet$ Carnatic Music & Abstract & The style of {\em Bharatanatyam} music \\
$\bullet$ Sequence & Abstract & Ordered list of elements of one kind\\
$\bullet$ Beat & Abstract & Basic unit of time -- an instance on timescale\\
$\bullet$ {\em Bol} & Concrete & Mnemonic syllable or vocal utterances (Table~\ref{tbl:bol_vocabs}) \\
$\bullet$ Posture & Abstract & Standing or sitting position of a dancer \\
$\bullet$ {\em Taalam} & Concrete & Rhythmic pattern of beats \\
$\bullet$ Tempo & Concrete & Beats per minute -- defines speed\\
$\bullet$ Instrumental Strike & Concrete & Beating of a percussion\\
$\bullet$ Position (Time Stamp) & Concrete & Instant of time \\
$\bullet$ Key Posture & Concrete & Momentarily stationary posture (Figure~\ref{fig:tatta1}) \\
$\bullet$ Transition Posture & Abstract & Non-stationary posture\\
$\bullet$ Trajectorial Transition Posture & Concrete & Transitions along a well-defined trajectory\\
$\bullet$ Natural Transition Posture & Concrete & Natural posture transitions by the dancer\\
$\bullet$ Leg Support ({\em Mandalam}) & Concrete & Ways to support the body (Figure~\ref{fig:mandalams}) \\
$\bullet$ Legs Position ({\em Pada Bheda}) & Concrete & Positions of both legs in {\em Bharatanatyam}  \\
$\bullet$ Arms Position ({\em Bahu Bheda}) & Concrete & Positions of both arms in {\em Bharatanatyam}  \\
$\bullet$ Head Position ({\em Shiro Bheda}) & Concrete & Positions of head in {\em Bharatanatyam}  \\
$\bullet$ Neck Position ({\em Griba Bheda}) & Concrete & Positions of neck in {\em Bharatanatyam} \\
$\bullet$ Eyes Position ({\em Drishti Bheda}) & Concrete & Eye movements depicting {\em navarasa} \\
$\bullet$ Hands Position ({\em Hasta Mudra}) & Abstract & Positions of both hands in {\em Bharatanatyam}  \\
$\bullet$ Single Hand Gesture & Concrete & {\em Asamyukta Hasta Mudra}s (Figure~\ref{fig:BN_Hasta_Set})\\
$\bullet$ Double Hand Gesture & Concrete & {\em Samyukta Hasta Mudra}s (Figure~\ref{fig:BN_Hasta_Set})\\
$\bullet$ Left Leg (Formation) & Concrete & Left leg in {\em Pada Bheda} (Table~\ref{tbl:leg_vocabs})\\
$\bullet$ Right Leg (Formation) & Concrete & Right leg in {\em Pada Bheda} (Table~\ref{tbl:leg_vocabs})\\
$\bullet$ Left Arm (Formation) & Concrete & Left arm in {\em Bahu Bheda} (Table~\ref{tbl:arm_vocabs})\\
$\bullet$ Right Arm (Formation) & Concrete & Right arm in {\em Bahu Bheda} (Table~\ref{tbl:arm_vocabs})\\
$\bullet$ Left Hand (Formation) & Concrete & Left hand in {\em Hasta Mudra} (Table~\ref{tbl:hand_vocabs})\\
$\bullet$ Right Hand (Formation) & Concrete & Right hand in {\em Hasta Mudra} (Table~\ref{tbl:hand_vocabs})\\ 
\hline
\end{tabular}
\end{scriptsize}
\end{table}
\renewcommand{\baselinestretch}{1.3} 

\renewcommand{\baselinestretch}{1} 
\begin{table}[!ht]
\centering
\caption{List of Instances for the ontology of {\em Bharatanatyam Adavu} \label{tbl:ontology_instances}}
\begin{scriptsize}
\begin{tabular}{|p{7cm}l|p{5cm}|} 
\hline
\multicolumn{1}{|c}{\bf Class:Instance} && \multicolumn{1}{|c|}{\bf Remarks} \\ \hline\hline
$\bullet$ {\em Sollukattu}: {\em Tatta\_A}, $\cdots$, {\em Tatta\_G}, {\em Natta}, {\em Kuditta Mettu} && 23 types of {\em Sollukattu}s (Table~\ref{tbl:bn_Bols})\\
$\bullet$ {\em Adavu}: {\em Tatta 1}, $\cdots$, {\em Tatta 8}, {\em Natta}, {\em Kuditta Mettu} && 58 types of {\em Adavu}s (Table~\ref{tbl:adavus_sollukattu})\\
$\bullet$ {\em Bol}: {\em tei, yum, tat, $\cdots$, } && 31 types of {\em Bol}s (Table~\ref{tbl:bol_vocabs})\\
$\bullet$ {\em Taalam}: {\em Adi Taalam}, {\em Roopakam Taalam} && 2 of the 7 types of {\em Taalam}s\\
$\bullet$ Tempo ({\em Laya}): {\em Vilambit Laya}, {\em Madhay Laya, Drut Laya} && 3 types of {\em Laya}s (speed) or tempo\\
$\bullet$ :Spinal Bending (boolean) && Spine may or may not be bent \\
$\bullet$ Key Postures: {\em Natta1P1}, {\em Natta1P2}, {\em Natta1P3}, $\cdots$ && Key Postures of {\em Natta Adavu} Variant 1 \\ 
$\bullet$ Leg Support: {\em Samapadam} (Standing),  {\em Araimandi} (Half-Sitting),  {\em Muzhumandi} (Full Sitting) && 3 types of leg support\\
$\bullet$ Legs Position: {\em Aayata} {\bf [S]}, {\em Prenkhanam} {\bf [M]}, $\cdots$ && Types of both legs positions \\
$\bullet$ Arms Position: {\em Natyarambhe} {\bf [S]}, {\em Natyarambhe} {\bf [M]}, $\cdots$ && Types of both arma positions \\
$\bullet$ Head Position: {\em Samam}, {\em Left Paravrittam}, {\em Right} \newline {\em Paravrittam}, $\cdots$ && Types of head positions (Table~\ref{tbl:head_vocabs})\\
$\bullet$ Hands Position: {\em Tripataka} {\bf [S]}, $\cdots$ && Types of both handa gestures \\
$\bullet$ Left / Right Leg (Formation): {\em Aayata}, {\em Anchita}, $\cdots$ && Types of single leg formations (Table~\ref{tbl:leg_vocabs})\\
$\bullet$ Left / Right Arm (Formation): {\em Natyarambhe}, {\em Kunchita} \newline {\em Natyarambhe}, $\cdots$ && Types of single arm formations (Table~\ref{tbl:arm_vocabs}) \\
$\bullet$ Left / Right Hand (Formation): {\em Tripataka}, $\cdots$ && Types of single hand gestures  \\
\hline
\multicolumn{3}{c}{ } \\
\multicolumn{3}{l}{$\bullet$ {\bf [S]}: Denotes symmetric ({\bf [S]}) positions between left and right limbs} \\
\multicolumn{3}{l}{$\bullet$ {\bf [M]}: Denotes asymmetric positions between left and right limbs and its mirror ({\bf [M]})} \\
\multicolumn{3}{l}{$\bullet$ {\em Instances of Neck Position, Eyes Position, and Double Hand Gestures are not considered}} \\
\end{tabular}
\end{scriptsize}
\end{table}
\renewcommand{\baselinestretch}{1.3} 

\renewcommand{\baselinestretch}{1} 
\begin{table}[!ht]
\centering
\caption{List of (Binary) Relations for the ontology of {\em Bharatanatyam Adavu} \label{tbl:ontology_relations}}
\begin{scriptsize}
\begin{tabular}{|l|l|p{1.4cm}|p{7cm}|} 
\hline
\multicolumn{1}{|c}{\bf Relation} & \multicolumn{1}{|c}{\bf Domain} & \multicolumn{1}{|c}{\bf Co-} & \multicolumn{1}{|c|}{\bf Remarks} \\
\multicolumn{1}{|c}{\bf } & \multicolumn{1}{|c}{\bf } & \multicolumn{1}{|c}{\bf Domain} & \multicolumn{1}{|c|}{\bf } \\ \hline \hline
{\em is\_a} & Class & Class & Specialization / Generalization or {\em is\_a} hierarchy of Object-based Modeling. This is used to build the {\em taxonomy}. For example, {\em Tatta Adavu} {\em is\_a} {\em Adavu}. \\ \hline

{\em has\_a} & Class & Class / \newline Instance & Composition or {\em has\_a} hierarchy of Object-based Modeling. This is used to build the {\em partonomy}. For example, {\em Sollukattu} {\em has\_a} {\em Taalam}. \\ \hline

{\em isInstanceOf} & Instance & Class & Distinct instances of a class \\ \hline

{\em isAccompaniedBy} & Class & Class & {\em isAccompaniedBy} captures the association between video and audio streams. Hence, {\em Adavu} {\em isAccompaniedBy} {\em Sollukattu}. \\ \hline

{\em isSyncedWith} & Class & Class & Expresses high-level synchronization -- between audio and video streams. Every {\em Adavu} {\em isSyncedWith} a unique {\em Sollukattu}. \\ 
\hline

{\em isSequenceOf} & Class & Class & {\em isSequenceOf} builds a sequence from elements of the same type. For example, every {\em Sollukattu} ({\em Adavu}) {\em has\_a} a sequence of beats (postures) constructed from beat (postures) by {\em isSequenceOf} relation. {\em isFollowedBy} is a dual of this relation. \\ \hline

{\em isAccentedBy} & Class & Class & A beat {\em isAccentedBy} a {\em bol}. \\ \hline

{\em isFollowedBy} & Class & Class & Ordering of audio events (like {\em beats}) or video events (like {\em postures}) -- Event $E_1$ is {\em isFollowedBy} event $E_2$. {\em isSequenceOf} is a dual of this relation. \\ \hline

{\em triggers} & Instance & Instance & Expresses low-level synchronization -- between audio and video events. Hence, a beat {\em triggers} a posture as the dance is driven by the music. \\ \hline
{\em repeats} & Class & Class & Once a {\em taalam} completes a bar, it may repeat itself. \\ \hline
\end{tabular}
\end{scriptsize}
\end{table}
\renewcommand{\baselinestretch}{1.3} 

\subsection{Ontology of {\em Bharatanatyam Adavu}s -- Top Level}
\begin{figure}[!ht]
\centering
\includegraphics[width=7.5cm]{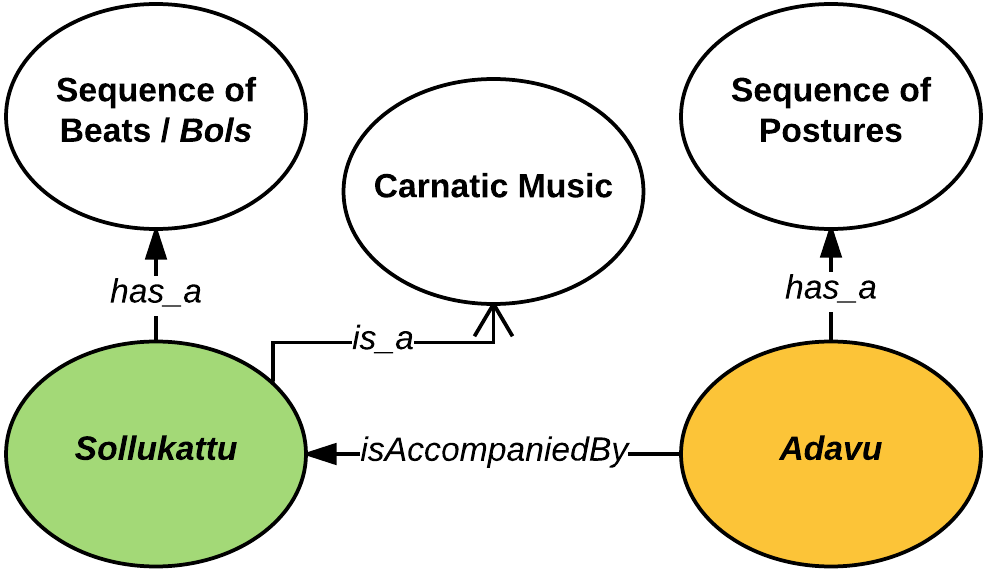} \vspace*{-0.3cm}
\caption{Ontology of {\em Bharatanatyam Adavu} at the abstract level\label{fig:Abstract Concept of BN Adavu}}
\end{figure}

At the top level, a {\em Bharatanatyam Adavu} can be expressed simply as a dance ({\em Adavu}) accompanied and driven by ({\em isAccompaniedBy}) music ({\em Sollukattu}) (Figure~\ref{fig:Abstract Concept of BN Adavu}). In other words, the musical {\em meter}\footnote{The {\em meter} of music is its rhythmic structure.} of an {\em Adavu} is called a {\em Sollukattu} which is a sequence of beats / {\em bol}s. An {\em Adavu} is a sequence of postures. We also note that {\em Sollukattu} is a form of Carnatic Music. 

\renewcommand{\baselinestretch}{1} 
\renewcommand{\baselinestretch}{1.3} 

\begin{figure}[!ht]
\centering
\includegraphics[width=\textwidth]{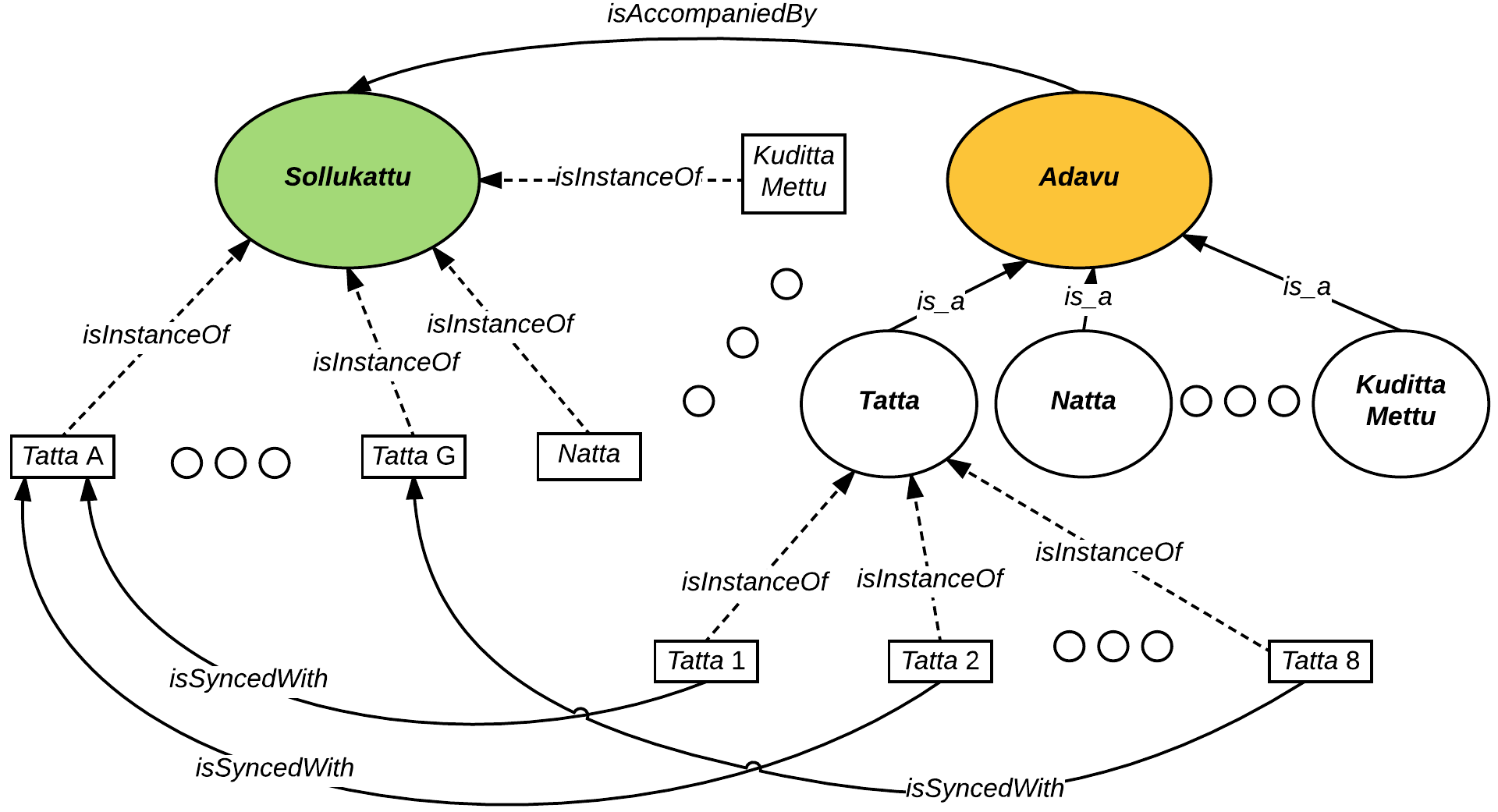} \vspace*{-0.3cm}
\caption[Ontology of {\em Bharatanatyam Adavu} with specializations and instances]{Ontology of {\em Bharatanatyam Adavu} with specializations and instances
\label{fig:Concept of BN Adavu – Expanded}}
\end{figure}

Elaborating on the basic concept of {\em Adavu}s, we show in Figure~\ref{fig:Concept of BN Adavu – Expanded} that there are several specializations of {\em Adavu}s like {\em Tatta Adavu} or {\em Natta Adavu} having instances {\em Tatta Adavu 1}, $\cdots$, {\em Tatta Adavu 8} etc. and there are several instances of {\em Sollukattu}s like {\em Tatta A}, $\cdots$ {\em Kuditta Mettu}, etc. Specifically, every {\em Adavu} is synchronized with ({\em isSyncedWith}) a unique {\em Sollukattu}. 

\renewcommand{\baselinestretch}{1} 
\renewcommand{\baselinestretch}{1.3} 

\subsection{Ontology of {\em Sollukattu}s}
\begin{figure}[!ht]
\centering
\includegraphics[width=\textwidth]{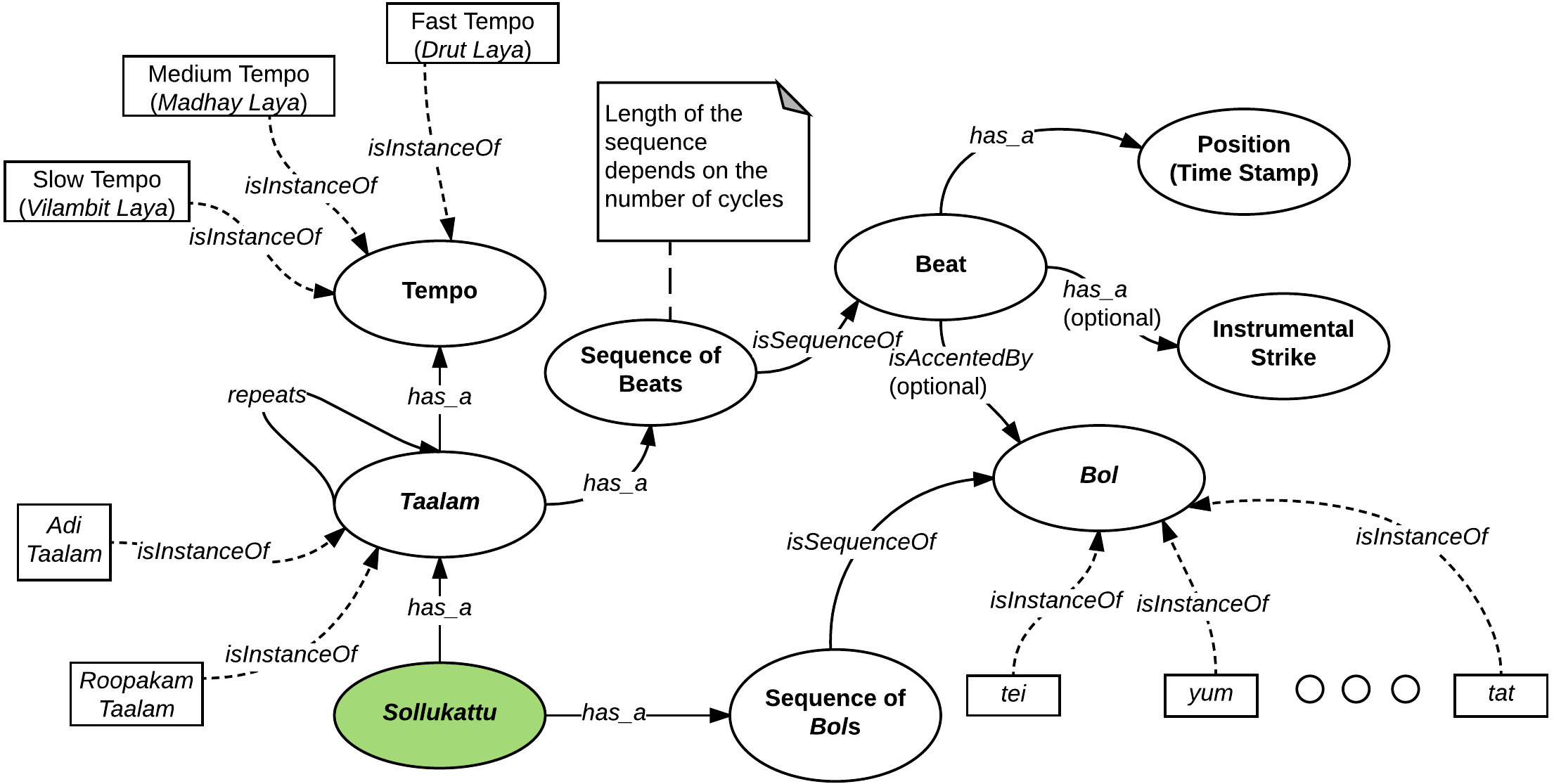} \vspace*{-0.3cm}
\caption{Ontology of {\em Sollukattu}s \label{fig:Concept of BN Audio Structure}}
\end{figure}

Next, we elaborate the ontology of a {\em Sollukattu} (Section~\ref{sec:music}) in Figure~\ref{fig:Concept of BN Audio Structure}. A {\em Sollukattu} is performed in a {\em Taalam} that designates a specific pattern of rhythm. A {\em Taalam} is composed of a sequence of beats ({\em isSequenceOf}) going at a certain tempo (speed). At the end of the sequence of beats (or the bar), the {\em Taalam} {\em repeats} itself. A tempo corresponds to the  speed of the rhythm which may  be carried out  in one of the three speeds ({\em Laya}) -- slow, medium, and fast. {\em Adi Taalam} and {\em Roopakam Taalam} are the typical rhythms used in {\em Bharatanatyam}.

\renewcommand{\baselinestretch}{1} 
\renewcommand{\baselinestretch}{1.3} 
A beat is an instant in time that may be marked by beating of a stick and optionally accented by a {\em bol}. Hence, it {\em has\_a} temporal position (time stamp), an instrumental strike (for example, beating of {\em Tatta Kazhi}), and a {\em bol} like {\em tei}, {\em yum}, {\em tat}, $\cdots$ (vocabulary of {\em Bol}s by {\em Bharatanatyam} experts is given in Table~\ref{tbl:bol_vocabs}). 


\renewcommand{\baselinestretch}{1} 
\begin{table}[!ht]
\centering
\caption{{\em Bol} vocabulary of {\em Sollakattu}s \label{tbl:bol_vocabs}}
\begin{scriptsize}
\begin{tabular}{|r|l||r|l||r|l|} 
\hline
\multicolumn{1}{|c|}{\bf Sl. \#}	&	\multicolumn{1}{c||}{\bf Bol}	&	\multicolumn{1}{c|}{\bf Sl. \#}	&	\multicolumn{1}{c||}{\bf Bol}	&	\multicolumn{1}{c|}{\bf Sl. \#}	&	\multicolumn{1}{c|}{\bf Bol}	\\ \hline \hline
1		&		a		&	12		&		ha		&		23		&		tak		 \\ \hline
2		&		da		&		13		&		hat		&	24		&		tam		 \\ \hline
3		&		dha		&		14		&		hi		&		25		&		tan		 \\ \hline
4		&		dhat 	&		15		&		jag		&		26		&		tat		 \\ \hline
5		&		dhi		&		16		&		jham 	&		27		&		tei		 \\ \hline
6		&		dhin 	&		17		&		ka		&		28		&		tom		 \\ \hline
7		&		dhit 	&		18		&		ki		&		29		&		tta		 \\ \hline
8		&		ding 	&		19		&		ku		&		30		&		ya		 \\ \hline
9		&		e		&		20		&		na		&		31		&		yum		 \\ \hline
10		&		gadu 	&		21		&		ri		&		32		&	 {\em Stick Beat}		 \\ \hline
11		&		gin		&		22		&		ta		&			&		 \\ \hline
\multicolumn{6}{c}{} \\

\multicolumn{6}{p{5.5cm}}{\scriptsize {\em Stick Beat} is treated as a pseudo-{\em bol}. The {\em bols} shown in the table are typical as {\em Bharatanatyam} does not follow a strictly fixed set of {\em bol}.}

\end{tabular}
\end{scriptsize}
\end{table}
\renewcommand{\baselinestretch}{1.3}


\subsection{Ontology of {\em Adavu}s}
\begin{figure}[!ht]
\centering
\includegraphics[width=\textwidth]{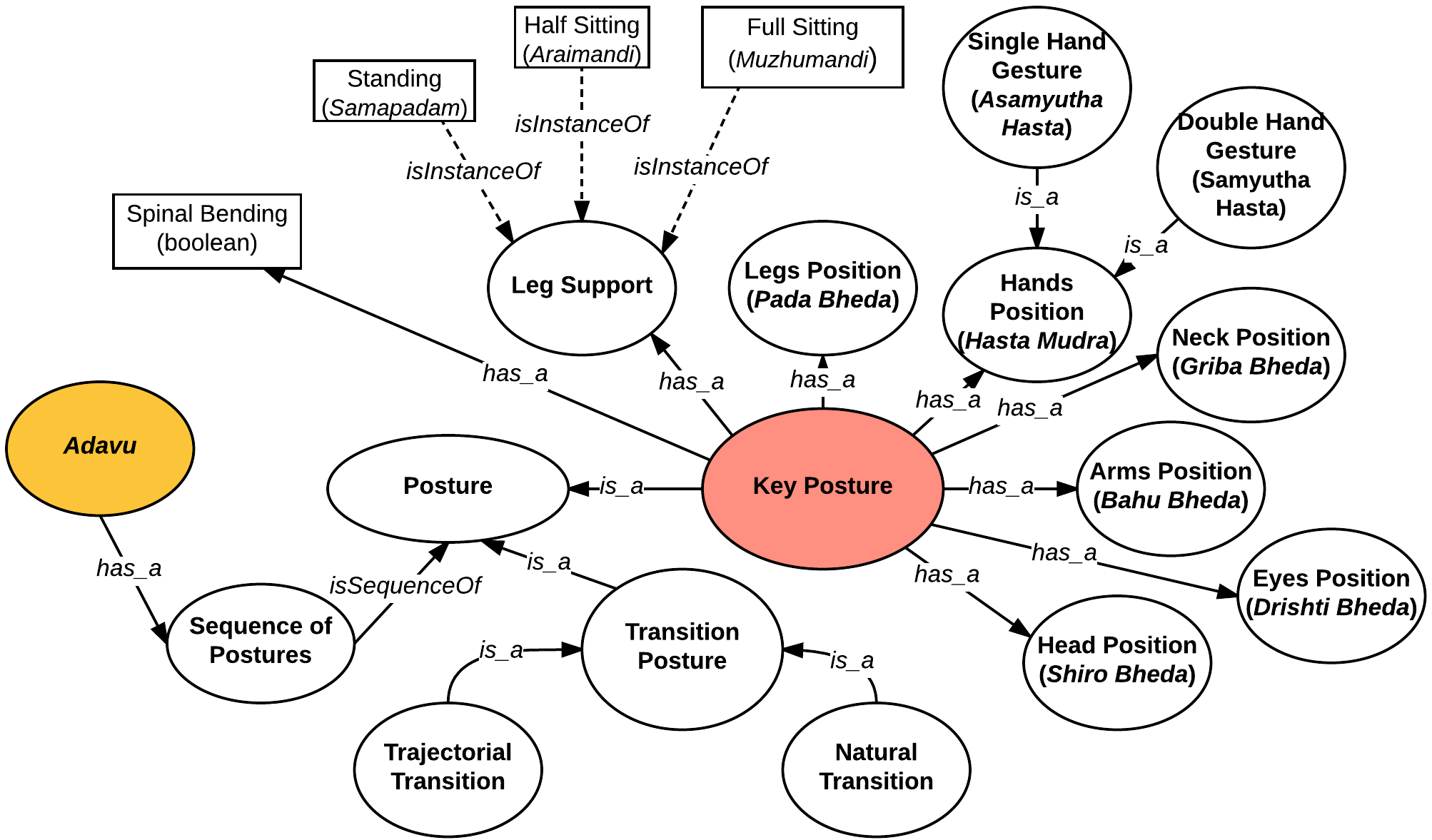} \vspace*{-0.3cm}
\caption{Ontology of {\em Adavu}s\label{fig:Concept of BN Video Structure}}
\end{figure}

We elaborate the ontology of an {\em Adavu} (Section~\ref{sec:adavus}) in Figure~\ref{fig:Concept of BN Video Structure}. An {\em Adavu} is created by a sequence of {\em Postures} and intervening {\em Movements} like {\em Utplavana} (Jumps), {\em Chari} (Walking), or {\em Karana}\footnote{{\em Karana}s (`{\em doing}' in Sanskrit) are the 108 key transitions described in {\em Natya Shastra}.} (synchronized movement of hands and feet). A posture may be a {\em Key Posture} or a {\em Transition Posture}. A Key Posture is defined as a momentarily stationary pose taken by the dancer with well-defined positions for the {\em Legs} ({\em Pada Bheda}), the {\em Arms} ({\em Bahu Bheda}), the {\em Head} ({\em Shiro Bheda}), the {\em Neck} ({\em Griba Bheda}), the {\em Eyes} ({\em Drishti Bheda}), and the {\em Hands}  ({\em Hasta Mudra}). Every Key Posture is also defined with a specific {\em Leg Support} and {\em Spinal Bending} to support and balance the body. A Transition Posture, in turn, is a transitory pose (ill-defined, at times) between two consecutive Key Postures in a sequence or a pose assumed as a part of a movement. It may be {\em Trajectorial} or {\em Natural}. While a {\em Trajectorial Transition Posture} occurs in a well-defined trajectory path of body parts, a {\em Natural Transition Posture} may be suitably chosen by a dancer to move from one Key Posture to the next.

\renewcommand{\baselinestretch}{1} 
\renewcommand{\baselinestretch}{1.3} 

In the current work, we focus only on Key Postures and do not model and / or analyze movements and transitions. Hence, we do not elaborate the ontology for {\em Transition Postures} or movements. However, the concept of Key Postures are detailed in Figure~\ref{fig:Concept of Key Postures}.

\subsubsection{Vocabulary of Positions and Formations}\label{sec:vocab_postures}
To elaborate the ontology for a Key Posture, we introduce the notions of {\em positions} and {\em formations} of constituent limbs or body parts. A {\em formation} describes the specific manner in which a body part is posed in the posture. For body parts that occur in pair (like leg, arm, hand, eye), the combined formation of the individual (left and right) parts define a {\em position}. For the rest (like head, neck) {\em position} and {\em formation} are taken to be synonymous. Accepted nomenclature (as identified by the experts) exists for many positions / formations of most of the body parts in {\em Bharatanatyam}. Naturally, we adopt those. For the rest, we assign names based on crisp descriptors of the positions. 
We observe that the postures mostly are distinguishable based on the four major body parts -- leg, arm, head, and hand. Hence, we have not considered the eyes and the neck in building the posture ontology.

In Table~\ref{tbl:leg_vocabs}, we list the vocabulary for formations of left and right legs as well as their combined legs positions. Some of the positions are asymmetric in which the left and the right leg assume different formations. For example, if the left leg is in {\em Anchita} formation and the right leg is in {\em Samapadam} formation, the combined legs position is named as {\em Ardha Prenkhanam}. Naturally, every asymmetric position has a position which is a mirror image of the other one, marked by {\bf [M]} (Mirror), where the formations of the legs are swapped. That is, in {\em Ardha Prenkhanam} {\bf [M]}, the right leg is in {\em Anchita} formation and the left leg is in {\em Samapadam} formation. In the table, we have listed only one of these mirrored positions. Remaining leg positions are symmetric in which both legs assume the same formation. In such cases, the position is marked with an {\bf [S]} (Symmetric) and the same name is used for the formation and the position. Hence in {\em Aayata} {\bf [S]} position, both legs are in {\em Aayata} formation.
\renewcommand{\baselinestretch}{1} 
\begin{table}[!ht]
\centering
\caption{Vocabulary of formations and positions of legs ({\em Pada Bheda})\label{tbl:leg_vocabs}}
\begin{scriptsize}
\begin{tabular}{|l|l|l|} 
\hline
\multicolumn{1}{|c}{\em Left Leg Formation}	&	\multicolumn{1}{|c}{\em Right Leg Formation}	&	\multicolumn{1}{|c|}{\em Leg Position}	\\ \hline\hline
\multicolumn{3}{|c|}{\bf Asymmetric Positions} \\ \hline\hline
$\bullet$ {\em 	Anchita	} & $\bullet$ {\em 	Samapadam	} & $\bullet$ {\em 	Ardha Prenkhanam	} \\
$\bullet$ {\em 	Aayata	} & $\bullet$ {\em 	Back Swastikam	} & $\bullet$ {\em 	Back Swastikam	} \\
$\bullet$ {\em 	Agratala Sanchara	} & $\bullet$ {\em 	Samapadam	} & $\bullet$ {\em 	Chalan Chari	} \\
$\bullet$ {\em 	Aayata 	} & $\bullet$ {\em 	Diagona Anchita	} & $\bullet$ {\em 	Diagonal Prenkhanam	} \\
$\bullet$ {\em 	Bend On Knee	} & $\bullet$ {\em 	Support	} & $\bullet$ {\em 	Ekapadam	} \\
$\bullet$ {\em 	Aayata 	} & $\bullet$ {\em 	Front Anchita	} & $\bullet$ {\em 	Front Prenkhanam	} \\
$\bullet$ {\em 	Aayata	} & $\bullet$ {\em 	Front Swastikam	} & $\bullet$ {\em 	Front Swastikam	} \\
$\bullet$ {\em 	Aayata	} & $\bullet$ {\em 	Prerita	} & $\bullet$ {\em 	Prerita	} \\
$\bullet$ {\em 	Parsasuchi	} & $\bullet$ {\em 	Bisamasuchi	} & $\bullet$ {\em 	Garudamandalam	} \\
$\bullet$ {\em 	Aayata	} & $\bullet$ {\em 	Forward / Side Low	} & $\bullet$ {\em 	Lolita Chari	} \\
$\bullet$ {\em 	Aayata	} & $\bullet$ {\em 	Anchita	} & $\bullet$ {\em 	Prenkhanam	} \\
$\bullet$ {\em 	Aayata	} & $\bullet$ {\em 	Side Middle / Low	} & $\bullet$ {\em 	Prenkhanam Above Floor	} \\ 
$\bullet$ {\em 	Aayata	} & $\bullet$ {\em 	Kunchita	} & $\bullet$ {\em 	Aaleeda} ({\bf [M]} = {\em Pratyaaleeda}) \\
$\bullet$ {\em 	Kunchita	} & $\bullet$ {\em 	Aayata	} & $\bullet$ {\em 	Pratyaaleeda} \\ \hline \hline
\multicolumn{3}{|c|}{\bf Symmetric Positions} \\ \hline\hline
$\bullet$ {\em 	Aayata	} & $\bullet$ {\em 	Anchita	} & $\bullet$ {\em 	Ekapadam Bhramari	} \\
$\bullet$ {\em 	Samapadam	} & $\bullet$ {\em 	Motita Mandal	} & $\bullet$ {\em 	Side Chankramanang	} \\
$\bullet$ {\em 	Muzmandi	} & $\bullet$ {\em 	Slip With Left Knee	} & $\bullet$ {\em 	Chankramanang	} \\
$\bullet$ {\em 	Kuttana	} & $\bullet$ {\em 	Slip With Right Knee	} & $\bullet$ {\em 	Back Chankramanang	} \\
$\bullet$ {\em 	Parswa Aayata	} &  &  \\ 
\hline
\end{tabular}
\end{scriptsize}
\end{table}

\renewcommand{\baselinestretch}{1.3} 
In Table~\ref{tbl:arm_vocabs}, we list the vocabulary for the formations of the arms. Either arm can assume any of these formations. In case of arms, no specific names are used for combined arms positions. Hence, they are referred to with the names of both the formations if they are different. For example, if the left arm is in {\em Kunchita Natyarambhe} formation and the right arm is in {\em Natyarambhe} formation, the combined arms position is named as {\em Natyarambhe--Kunchita Natyarambhe}. If, however, both formations are same, we name the position with an {\bf [S]}. Hence {\em Natyarambhe} {\bf [S]} has {\em Natyarambhe} formation for both arms. In Table~\ref{tbl:head_vocabs}, we list the vocabulary for the formations of the head. Naturally, there is no position descriptor here. Next we list the vocabulary for the formations of the hands ({\em hasta mudra}) in Table~\ref{tbl:hand_vocabs}. Like arms, these are also denoted with formations of single hands only and combined hands position is similarly named. It may be noted that the vocabulary listed here is a subset of {\em Asamyutha Hasta} or single hand gestures as commonly observed in the {\em Adavu}s. We do not consider {\em Samyutha Hasta} or combined (both) hand gestures in building the vocabulary. 

\renewcommand{\baselinestretch}{1} 
\begin{table}[!ht]
\centering
\caption{Vocabulary of formations of arms ({\em Bahu Bheda})\label{tbl:arm_vocabs}}
\begin{scriptsize}
\begin{tabular}{|l|l|l|} 
\hline
$\bullet$ {\em 	Above Head Natyarambhe	} & $\bullet$ {\em 	Diagonal High	} & $\bullet$ {\em 	Kunchita Natyarambhe} \\
$\bullet$ {\em 	Above Head Natyarambhe } & $\bullet$ {\em 	Diagonal Middle	} & $\bullet$ {\em 	Left Diagonal High 	} \\
({\em 	Joined	}) & $\bullet$ {\em 	Elbow Down Anchita	} & $\bullet$ {\em 	Natyarambhe	} \\
$\bullet$ {\em 	Anchita	} & $\bullet$ {\em 	Forward High	} & $\bullet$ {\em 	Right Diagonal High 	} \\
$\bullet$ {\em 	Anchita Above Left Ear	} & $\bullet$ {\em 	Forward High Above Head	} & $\bullet$ {\em 	Right Diagonal Middle	} \\
$\bullet$ {\em 	Anchita Above Right Ear	} & $\bullet$ {\em 	Forward Low	} & $\bullet$ {\em 	Side High	} \\
$\bullet$ {\em 	Ardha Vithi	} & $\bullet$ {\em 	Forward Middle	} & $\bullet$ {\em 	Side High Natyarambhe	} \\
$\bullet$ {\em 	Backward High	} & $\bullet$ {\em 	Front Natyarambhe	} & $\bullet$ {\em 	Side Low	} \\
$\bullet$ {\em 	Backward Low	} & $\bullet$ {\em 	Katyang Behind Waist	} & $\bullet$ {\em 	Side Middle	} \\
$\bullet$ {\em 	Backward Middle 	} & $\bullet$ {\em 	Kunchita} & $\bullet$ {\em 	Utsanga	} \\
$\bullet$ {\em 	Cross Kunchita	} & $\bullet$ {\em 	Kunchita Above Shoulder	} &		\\ \hline
\end{tabular}
\end{scriptsize}
\end{table}
\renewcommand{\baselinestretch}{1.3}



\subsubsection{Ontology of Key Postures}\label{sec:ontology_key_postures}
\begin{figure}[!ht]
\centering
\includegraphics[width=\textwidth]{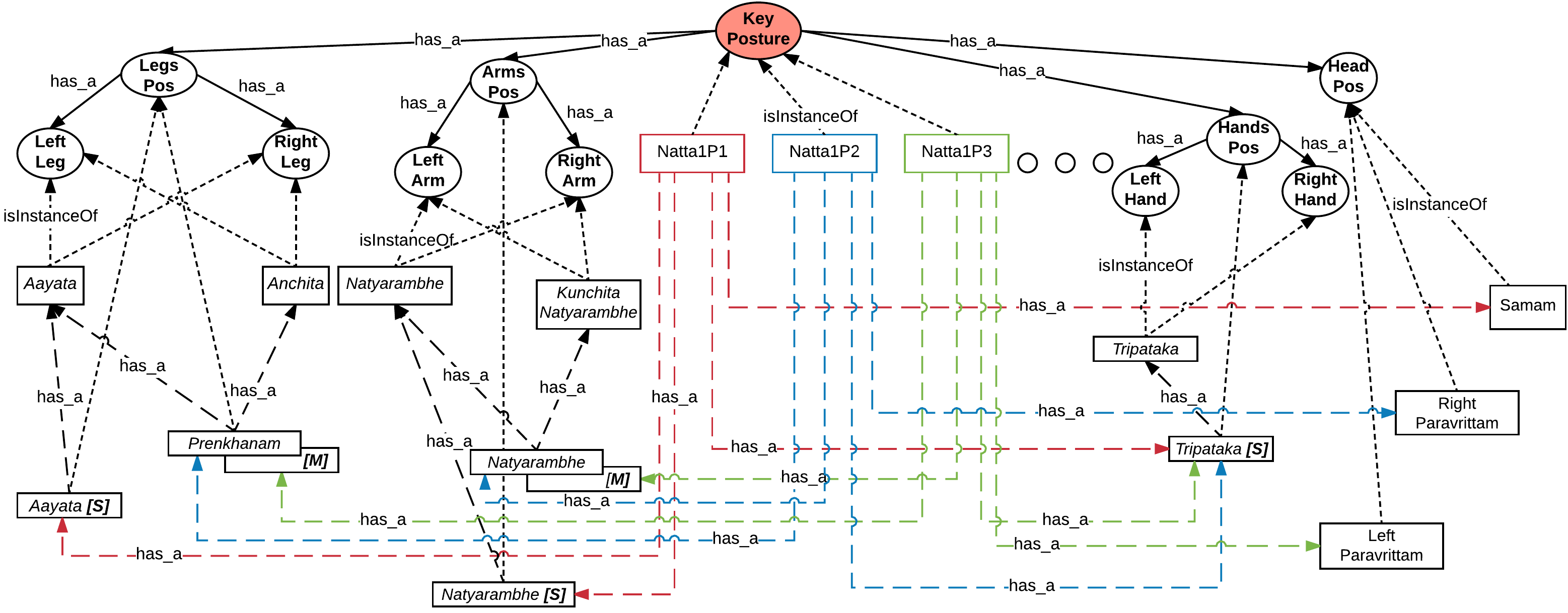} 

\begin{tabular}{ccc} 
\multicolumn{3}{l}{\small $\bullet$ Dotted lines denote {\em isInstanceOf} between an instance and a class} \\
\multicolumn{3}{l}{\small $\bullet$ Dashed lines denote {\em has\_a} between two an instances} \\
\multicolumn{3}{l}{ } \\
\includegraphics[width=3.5cm]{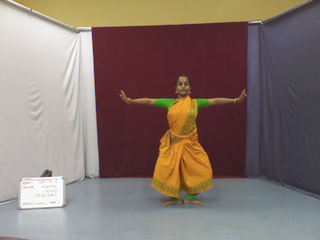} & 
\includegraphics[width=3.5cm]{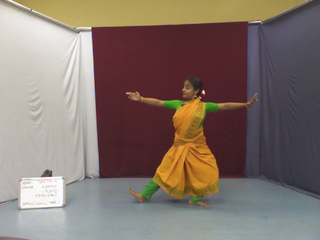} & 
\includegraphics[width=3.5cm]{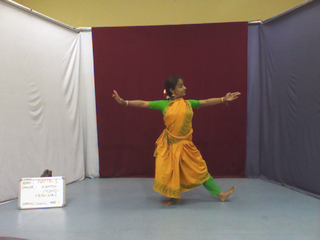} \\
{\small Natta1P1} & {\small  Natta1P2} & {\small Natta1P3} \\
\end{tabular}

\caption{Ontology of Key Postures
\label{fig:Concept of Key Postures}}
\end{figure}

We elaborate the ontology of Key Postures in Figure~\ref{fig:Concept of Key Postures}.
Consider the {\em Legs Positions}. For the {\em Prenkhanam} Legs Position in Natta1P2 in the figure, the left leg makes the {\em Aayata} (bent at knee) and the right leg makes {\em Anchita} formation (straight and stretched). {\em Prenkhanam} {\bf [M]} is a mirror image position of {\em Prenkhanam} where the formations of the two legs are swapped. Natta1P3 is a mirrored posture of Natta1P2 and has {\em Prenkhanam} {\bf [M]} for the legs positions. With symmetry Natta1P1 has {\em Aayata} {\bf [S]} legs position. 

Consider instances of 3 key postures -- Natta1P1, Natta1P2, and Natta1P3 -- of {\em Natta 1 Adavu}. For example, for instance Natta1P1, we have {\em Legs Position} = {\em Aayata} {\bf [S]}, {\em Arms Position} = {\em Natyarambhe} {\bf [S]}, {\em Hands Position} = {\em Tripataka} {\bf [S]}, and {\em Head Position} = {\em Samam}.

\renewcommand{\baselinestretch}{1} 
\begin{table}[!ht]
\centering
\caption{Vocabulary of positions / formations of head ({\em Shiro Bheda})\label{tbl:head_vocabs}}
\begin{scriptsize}
\begin{tabular}{|l|l|l|} 
\hline
$\bullet$ {\em 	Samam	} &	$\bullet$ {\em 	Left Adhomukham	} &	$\bullet$ {\em 	Right Adhomukham	} \\
$\bullet$ {\em 	Adhomukham	} &	$\bullet$ {\em 	Left Ardha Paravrittam	} &	$\bullet$ {\em 	Right Ardha Paravrittam	} \\
$\bullet$ {\em 	Back Paravrittam	} &	$\bullet$ {\em 	Left Paravrittam	} &	$\bullet$ {\em 	Right Paravrittam	} \\
$\bullet$ {\em 	Udvahitam	} &	$\bullet$ {\em 	Left Utshiptam	} &	$\bullet$ {\em 	Right Utshiptam	} \\
$\bullet$ {\em 	Ardha Aalolitam	} &			&			\\ \hline
\end{tabular}
\end{scriptsize}
\end{table}
\renewcommand{\baselinestretch}{1.3} 


\renewcommand{\baselinestretch}{1} 
\begin{table}[!ht]
\centering
\caption{Vocabulary of formations of hands ({\em Hasta Mudra})\label{tbl:hand_vocabs}}
\begin{scriptsize}
\begin{tabular}{|l|l|l|l|} 
\hline
$\bullet$ {\em Alapadma} & $\bullet$ {\em Kartarimukha} & $\bullet$ {\em 	Mushti} & 
$\bullet$ {\em Suchi} \\
$\bullet$ {\em Avahitya} & $\bullet$ {\em Katakamukha} & $\bullet$ {\em Pataka} & $\bullet$ {\em Tripataka} \\
$\bullet$ {\em Dola} & $\bullet$ {\em Mrigashirsha} & $\bullet$ {\em Shikhara} & \\ \hline
\end{tabular}
\end{scriptsize}
\end{table}
\renewcommand{\baselinestretch}{1.3}


We identify 361 distinct postures and 48 distinct movements in the 58 {\em Adavu}s.

\renewcommand{\baselinestretch}{1} 
\renewcommand{\baselinestretch}{1.3} 

\subsection{Ontology of Audio-Visual Sync between {\em Sollukattu} \& {\em Adavu}}
\begin{figure}[!ht]
\centering
\includegraphics[width=\textwidth]{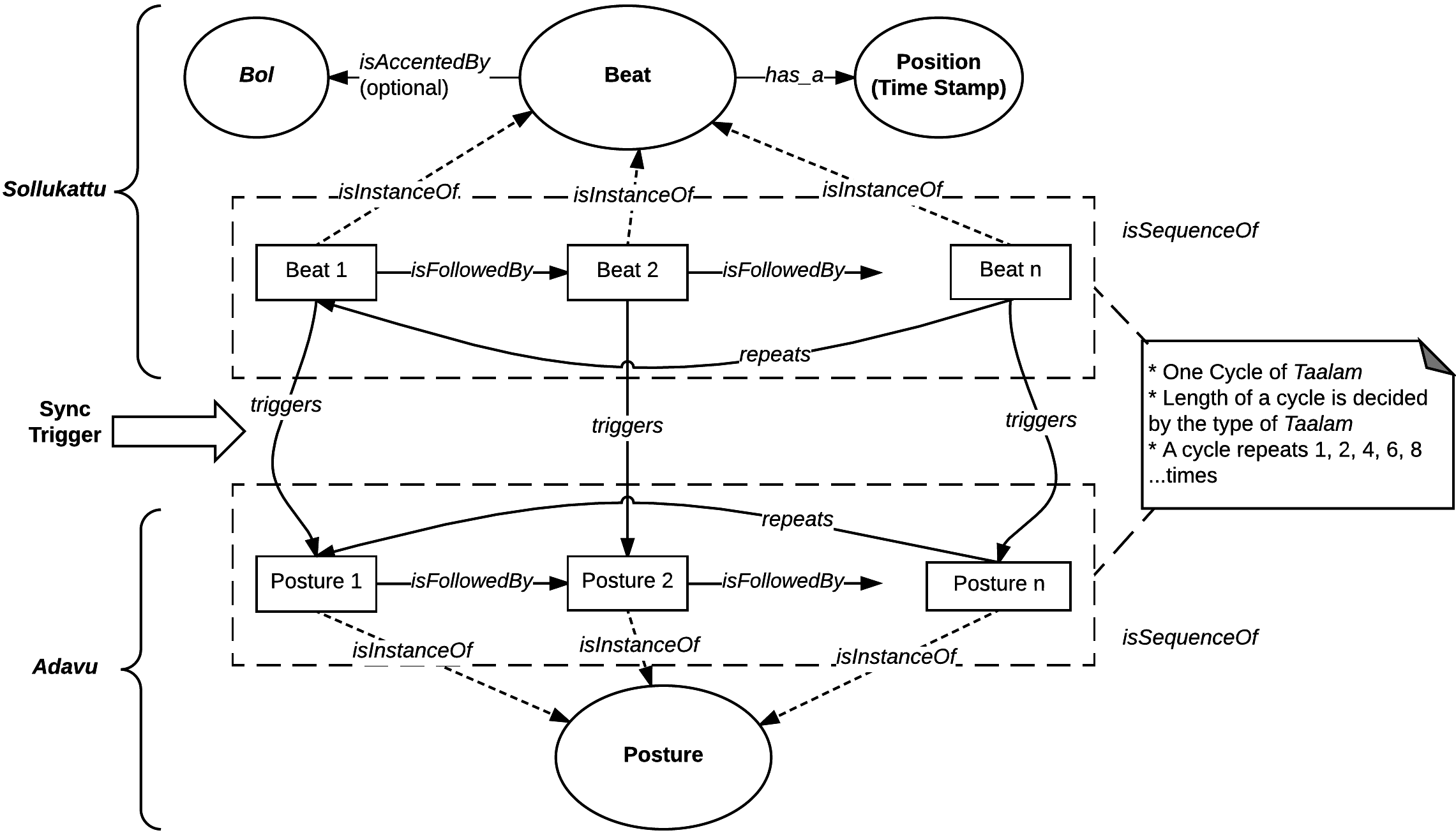} \vspace*{-0.3cm}
\caption{Ontology of Audio-Visual Sync in {\em Bharatanatyam} \label{fig:Concept of BN Audio-Visual Sync}}
\end{figure}

With the ontology of music ({\em Sollukattu}) and (visual) sequence of postures ({\em Adavu}) of {\em Bharatanatyam}, we next capture the synchronization of the events. As the postures are driven by and are synchronized with the beats of the music, and as the performance repeats after a bar of the rhythm, we capture the ontology of synchronization between an {\em Adavu} and its {\em Sollukattu} as in  Figure~\ref{fig:Concept of BN Audio-Visual Sync}. Here specific instances of beats -- Beat 1, Beat 2, $\cdots$, Beat $n$ -- form the sequence of beats in a {\em Sollukattu}. So we expresses that Beat 1 {\em isFollowedBy} Beat 2, Beat 2 {\em isFollowedBy} Beat 3, and so on. Finally, after Beat $n$, the bar {\em repeats}, and hence, Beat $n$ {\em isFollowedBy} Beat 1. Similarly, instances of key postures -- Posture 1, Posture 2, $\cdots$, Posture $n$ -- form the sequence of postures in an {\em Adavu} that also {\em repeats}. Being driven by music, every beat {\em triggers} the corresponding posture. In the figure, we show only one cycle (bar) of the {\em Taalam}. In an {\em Adavu}, usually 1, 2, 4, 6, 8, or more number of repetitions are performed by the dancer. Explicit instances of {\em bol}s and time instants are omitted on the diagram for better clarity.

\renewcommand{\baselinestretch}{1} 
\renewcommand{\baselinestretch}{1.3} 

In this section, we have captured the central concepts of {\em Bharatanatyam Adavu}s in terms of a set of object-based ontological models. These models identify the key items with their interrelationships and help the annotation of data sets 
for training as well as testing. Naturally, they lead to algorithms for the analysis and recognition of various items (like {\em bol}s and postures). However, these models are structural, and hence, are limited in their temporal specification.

\section{Event-based Modeling of {\em Adavu}s}\label{sec:characterize}
The framework used so far is good for taxonomical and partonomical representation but lacks the expressibility in temporal terms. But Dance is multimedia in nature with music driving the steps.
In order to capture dynamic association between music and video, we first tried to use the concept of triggers to model synchronization of events. The progression of time is captured by simple sequences ({\em isFollowedBy}) of occurrences of {\em bol}s and beats . This approach is illustrated in Figure~\ref{fig:Concept of BN Audio-Visual Sync}. Since a simple sequence of {\em bol}s and beats  misses actual quantum of time slice, it  cannot deal with triggers between 
beat and posture actions, and cannot ensure equal time gap between beats. Temporal behavioral models are necessary to analyze and recognize such temporal and synchronization details in depth. Hence, we introduce an event-based modeling framework that, on one hand, can relate to the key concepts as introduced above and is defined in terms of temporal relationships on the other. 

This event-based framework treats a performance as a multimedia stream and takes the models closer to the structure of the data that we capture later by Kinect. A {\em Bharatanatyam Adavu}, therefore, consists of (1) Composite {\bf Audio Stream} ({\em Sollakattu}) containing -- (a) {\em Instrumental Sub-stream} as generated by instrumental strikes and (b) {\em Vocal Sub-stream} as generated by vocalizations or {\em bol}s; (2) {\bf Video Stream} of frames  containing either -- (a) Key Posture (called, {\em K-Frame}), or (b) Transition Posture (called, {\em T-Frame}); and (3) {\bf Synchronization (Sync)} of Position, Posture, Movement, and Gesture of an {\em Adavu} as performed in synchronization among themselves, and in synchronization with the rhythm of the music. In {\em Instrumental} and {\em Vocal Sub-streams} of a {\em Sollukattu}, beating and {\em bol}s are usually generated in sync. The rules or structure of synchronization have been defined for every {\em Sollakattu} in {\em Bharatanatyam}. 

\subsection{Events of {\em Adavu}s}
An {\em Event} denotes the occurrence of an activity (called {\em Causal Activity}) in the audio or the video stream of an {\em Adavu}. Further, sync events are defined between multiple events based on temporal constraints. Sync events may be defined jointly between audio and video streams. An event is described by:

\renewcommand{\baselinestretch}{1} 
\begin{table}[!ht]
\caption{List of Events of {\em Adavu}s\label{tbl:events}}
\centering
\begin{scriptsize}
\begin{tabular}{|l|l|l|l|} \hline
\multicolumn{1}{|c|}{\bf Event} & \multicolumn{1}{c|}{\bf Event} & \multicolumn{1}{c|}{\bf Event} & \multicolumn{1}{c|}{\bf Event} \\
\multicolumn{1}{|c|}{\bf Category} & \multicolumn{1}{c|}{\bf Type} & \multicolumn{1}{c|}{\bf Description} & \multicolumn{1}{c|}{\bf Label} \\ \hline \hline
&&&\\
Audio & $\alpha^{fb}$ & Full beat$^1$ with {\em bol} & {\em bol}$^2$, downbeat$^3$, upbeat$^4$ \\
Audio & $\alpha^{hb}$ & Half beat$^5$ with {\em bol} & {\em bol} \\
Audio & $\alpha^{qb}$ & Quarter beat$^6$ with {\em bol} & {\em bol} \\
Audio & $\alpha^{fn}$ & Full beat having no {\em bol}& upbeat  \\
Audio & $\alpha^{hn}$ & Half beat having no {\em bol}&  \\
Audio & $\beta$ & {\em bol} is vocalized & {\em bol} \\ 
&&&\\

Video & $\nu^{nm}$ & No motion$^7$ & Range of Frames$^8$, Key Posture$^9$ \\ 
Video & $\nu^{tr}$ & Transition Motion$^{10}$ & Range of Frames \\ 
Video & $\nu^{tj}$ & Trajectory Motion$^{11}$ & Range of Frames, Trajectory\\
&&&\\

Sync & $\phi^{fb}$ & {\em bol} @ Full beat & {\em bol} \\ 
Sync & $\phi^{hb}$ & {\em bol} @ Half beat & {\em bol} \\ 
Sync & $\psi^{fb}$ & No motion @ Full beat$^{12}$ & Key Posture \\ 
Sync & $\psi^{hb}$ & No motion @ Half beat & Key Posture \\ \hline
\end{tabular}

\begin{tabular}{rp{10.5cm}}
1: & A (full) beat is the basic unit of time -- an instance on the timescale \\
2: & Vocalized {\em bol}s accompany some beats \\
3: & The first beat of a bar \\
4: & The last beat in the previous bar which immediately precedes the downbeat \\
5: & Half beats are soft strikes at the middle of a tempo period \\
6: & Quarter beats strike at the middle of a Full-to-Half or a Half-to-Full beat  \\
7: & Frames over which the dancer does not move (assumes a Key Posture) \\
8: & Sequence of consecutive frames over which the events spreads \\ 
9: & A Key Posture is a well-defined and stationery posture \\
10: & Transitory motion to change from one Key Posture to the next \\
11: & Motion that follows a well-defined trajectory of movement for limbs \\
12: & $\alpha^{fb}$ and $\nu^{nm}$ in sync. That is, $\tau(\alpha^{fb}) \cap \tau(\nu^{nm}) \neq \phi$ \\ \hline
\end{tabular}
\end{scriptsize}
\end{table}
\renewcommand{\baselinestretch}{1.3} 

\begin{enumerate}
\item {\em Category}: The nature of the event based on its origin ({\em audio}, {\em video} or {\em sync}).

\item {\em Type}: Type relates to the causal activity of an event in a given category. Event types are listed in Table~\ref{tbl:events} with brief description.

\item {\em Time-stamp / range}: The time of occurrence of the causal activity of the event. This is elapsed time from the beginning of the stream and is marked by a function $\tau(.)$. Often a causal activity may spread over an interval $[\tau_s, \tau_e]$ which will be associated with the event. For video events, we use range of video frame numbers $[\eta_s, \eta_e]$ as the temporal interval. Since the video has a fixed rate of 30 fps, for any event we interchangeably use $[\tau_s, \tau_e]$ or $[\eta_s, \eta_e]$ as is appropriate in a context.

\item {\em Label}: Optional labels may be attached to an event for annotating details.
 
\item {\em ID}: Every instance of an event in a stream is distinguishable. These are sequentially numbered in the temporal order of their occurrence (Table~\ref{tbl:Sollukattu_bols}).

\end{enumerate}

The list of events are given in Table~\ref{tbl:events} and characterized in the next sections.

\subsection{Characterization of Audio Events}\label{sec:characterize_audio}
A {\em Sollukattu} is the musical {\em meter} of an {\em Adavu}. Traditionally, a {\em Tatta Palahai} (wooden stick) is periodically struck on a {\em Tatta Kozhi} (wooden block) in the rhythmic pattern of {\em Adi} or {\em Roopakam} {\em Taalam}s to produce the periodic beats (or $\alpha^{fb}$ events in Table \ref{tbl:Sollukattu_bols_Kuditta Mettu}). Usually beats repeat in a {\em bar}\footnote{A {\em bar} (or {\em measure}) is a segment of time corresponding to a specific $\Lambda$ number of beats. {\em Sollukattu}s also use longer bars (12, 16, 24, or 32).} of $\Lambda$ = 6 or 8. The {\em tempo} of a meter is measured by beats per minute ($bpm$) and can be slow, medium or fast. We use {\em Tempo Period} or {\em Period} $T = (60 / bpm)$ or the time interval between two consecutive beats in secs as the temporal measure for a meter. 

In the current work we use only the slow tempo. While there is no fixed definition for the {\em bpm} of a slow tempo (medium and fast progressively doubles relative to the slow one), it is typically found to be between 75 (period = 0.8 sec.) and 30 (period = 2 sec.) in most of the performances. Theoretically, the tempo period should not vary during the performance of a specific {\em Sollukattu} or across {\em Sollukattu}s. However, in reality it does vary depending on the skill of the beat player. Naturally, the event model needs to take care of such variations.

Next let us consider two consecutive beats $\alpha^{fb}_i$ and $\alpha^{fb}_{i+1}$ in a bar of length $\Lambda$, where $i$ denotes the $i^{th}$ ($1 \le i < \Lambda$) period. The time-stamps of the respective events are then related as $\tau(\alpha^{fb}_{i+1}) - \tau(\alpha^{fb}_i) \approx T$. Further the bar repeats after an equal time interval of $T$. That is, $\tau(\alpha^{fb}_{\Lambda*i+1}) - \tau(\alpha^{fb}_{\Lambda*i}) \approx T$, $i \ge 1$. We refer to such beats as {\em full beats} and hence the superscript {\em fb} in $\alpha^{fb}$ events. The first beat $\alpha^{fb}_0$ (last beat $\alpha^{fb}_{\Lambda-1}$) of a bar is referred to as a {\em downbeat} ({\em upbeat}). We mark these on the events as labels. In many {\em Sollukattu}s beating is also performed at the middle of a period. These are called {\em half beats} and produce the $\alpha^{hb}_i$ events in the $i^{th}$ period. Naturally, $\tau(\alpha^{hb}_i) - \tau(\alpha^{fb}_i) \approx \tau(\alpha^{fb}_{i+1}) - \tau(\alpha^{hb}_i) \approx T/2$.


Often in a {\em Sollukattu} the beat player (an accomplice of the dancer) also utters {\em bol}s. These are done in sync with a full beat or a half beat. We represent {\em bol}s as labels of the respective $\alpha^{fb}$ or $\alpha^{hb}$ events. A {\em bol} is {\em optional} for an event. 

It may be noted that a beat is actually an instant of time that occurs in every $T$ secs. So it is possible that a beat has no beating (and obviously no {\em bol}). Such cases, however, are not in the scope of the present study and we always work with a beating at a beat. 




There are 23 {\em Sollukattu}s. We illustrate a few here to understand various meters. All {\em Sollukattu}s are shown in slow tempo or {\em Vilambit Laya}.
\begin{enumerate}
\item {\bf \em Kuditta Mettu} ($T \approx $ 1.2 secs, $\Lambda$ = 8): We show two bars in Tables~\ref{tbl:Sollukattu_bols_Kuditta Mettu} with {\em bol}s and time-stamps. In Figure~\ref{fig:Kuditta_Mettu_Annotations}, we illustrate the signal for a {\em Kuditta Mettu} recording highlighting various events, time-stamps, and {\em bol}s. While this {\em Sollukattu} has only $\alpha^{fb}$ events by definition, some incidental $\alpha^{hn}$ 
events can still be seen in the signal. These will need to be later removed.

Table~\ref{tbl:Sollukattu_bols} shows its relationship with the {\em Adavu}.

\renewcommand{\baselinestretch}{1} 
\begin{table}[!ht]
\caption[Pattern of {\em Kuditta Mettu Sollukattu}]{Pattern of {\em Kuditta Mettu Sollukattu} (Figure~\ref{fig:Kuditta_Mettu_Annotations} (a)) \label{tbl:Sollukattu_bols_Kuditta Mettu}}
\centering
{\renewcommand{\arraystretch}{1.5}%
\begin{scriptsize}
\begin{tabular}{|l|r|r||l|r|r|}\hline
\multicolumn{1}{|c|}{\bf Event}& \multicolumn{1}{c|}{\bf Time} & \multicolumn{1}{c||}{\bf Beat Offset} & \multicolumn{1}{c|}{\bf Event} & \multicolumn{1}{c|}{\bf Time} & \multicolumn{1}{c|}{\bf Beat Offset}	\\ 
\multicolumn{1}{|c|}{\bf }& \multicolumn{1}{c|}{\bf (sec.)} & \multicolumn{1}{c||}{\bf (sec.)} & \multicolumn{1}{c|}{\bf } & \multicolumn{1}{c|}{\bf (sec.)} & \multicolumn{1}{c|}{\bf (sec.)}	\\ 
\multicolumn{1}{|c|}{\bf }& \multicolumn{1}{c|}{\bf ($\tau(\alpha)$)} & \multicolumn{1}{c||}{\bf ($\tau(\alpha_{i+1}) - \tau(\alpha_i)$)} & \multicolumn{1}{c|}{\bf } & \multicolumn{1}{c|}{\bf ($\tau(\alpha)$)} & \multicolumn{1}{c|}{\bf ($\tau(\alpha_{i+1}) - \tau(\alpha_i)$)}	\\ \hline \hline
\textcolor{blue}{$\alpha^{fb}_1$(tei)}	&	2.681	&		&\textcolor{blue}{$\alpha^{fb}_9$(tei)}	&	12.271	&	1.207	\\ \hline
\textcolor{red}{$\alpha^{fb}_2$(hat)}	&	3.912	&	1.231	&\textcolor{red}{$\alpha^{fb}_{10}$(hat)}	&	13.386	&	1.115	\\ \hline
\textcolor{blue}{$\alpha^{fb}_3$(tei)}	&	5.108	&	1.196	&\textcolor{blue}{$\alpha^{fb}_{11}$(tei)}	&	14.512	&	1.126	\\ \hline
\textcolor{red}{$\alpha^{fb}_4$(hi)}	&	6.269	&	1.161	&\textcolor{red}{$\alpha^{fb}_{12}$(hi)}	&	15.603	&	1.091	\\ \hline
\textcolor{blue}{$\alpha^{fb}_5$(tei)}	&	7.523	&	1.254	&\textcolor{blue}{$\alpha^{fb}_{13}$(tei)}	&	16.764	&	1.161	\\ \hline
\textcolor{red}{$\alpha^{fb}_6$(hat)}	&	8.742	&	1.219	&\textcolor{red}{$\alpha^{fb}_{14}$(hat)}	&	17.902	&	1.138	\\ \hline
\textcolor{blue}{$\alpha^{fb}_7$(tei)}	&	9.891	&	1.149	&\textcolor{blue}{$\alpha^{fb}_{15}$(tei)}	&	19.028	&	1.126	\\ \hline
\textcolor{red}{$\alpha^{fb}_8$(hi)}	&	11.064	&	1.173	&\textcolor{red}{$\alpha^{fb}_{16}$(hi)}	&	20.178	&	1.150	\\ \hline
\multicolumn{6}{c}{$\bullet$ $T = 1.2$ sec., $\Lambda = 8$} \\
\end{tabular}
\end{scriptsize}
}
\end{table}
\renewcommand{\baselinestretch}{1.3} 

\renewcommand{\baselinestretch}{1} 
\begin{figure}[!ht]
\centering
\includegraphics[width=\textwidth]{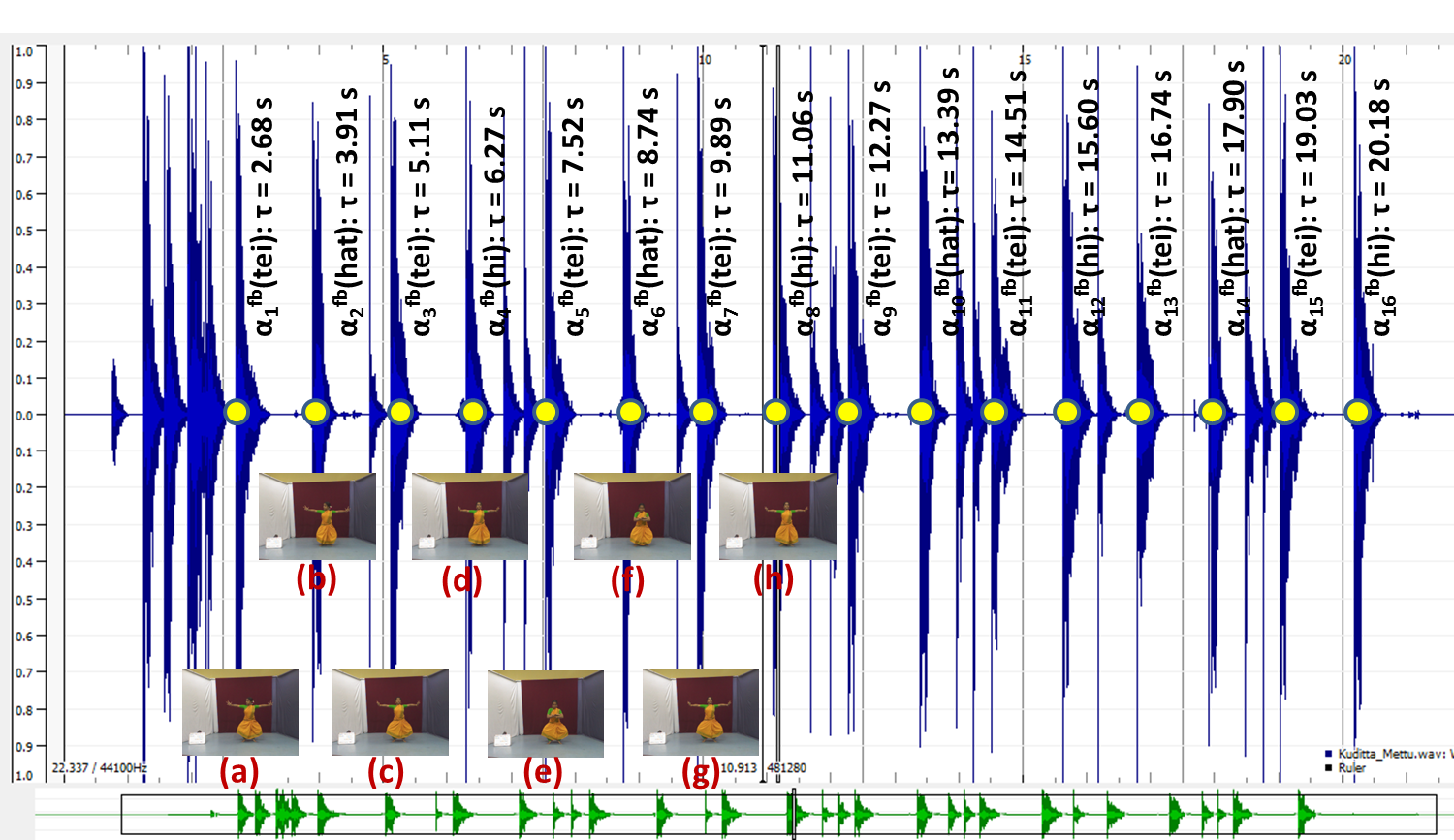} 

\begin{scriptsize}
\begin{tabular}{p{12cm}}
$\bullet$ {\bf Parameters}: No. of bars = 2, $\Lambda$ = 8 and $T = 1.16$ sec. \\
$\bullet$ Full beat ($\alpha^{fb}$) event positions are highlighted (yellow blobs) and corresponding {\em bol}s and time-stamps are shown (Table~\ref{tbl:Sollukattu_bols_Kuditta Mettu}). Note that several $\alpha^{hn}$
 are visible in the signals. These are rather incidental and not intended in the {\em Sollukattu}. Also, the beatings before the downbeat ($\alpha^{fb}_1$) are ignored. Right-sided {\em Key Postures} (Figure~\ref{fig:Kuditta_Mettu_Postures}) are also shown for the first 8 beats. Left-sided {\em Key Postures} are performed for the next 8 beats. 
\end{tabular}
\end{scriptsize}
\caption{Marking of beats and annotations of {\em bol}s for {\em Kuditta Mettu Sollukattu}  
\label{fig:Kuditta_Mettu_Annotations}}
\end{figure}
\renewcommand{\baselinestretch}{1.3} 

\item {\bf \em Tatta\_C} ($T \approx $ 1.6 secs, $\Lambda$ = 8): It has $\alpha^{fb}$ as well as $\alpha^{hb}$ events (Table~\ref{tbl:Sollukattu_bols_Tatta_C} and Figure~\ref{fig:Tatta_C_Annotations}).

\renewcommand{\baselinestretch}{1} 
\begin{table}[!ht]
\caption[Patterns of {\em Tatta\_C Sollukattu}]{Patterns of {\em Tatta\_C Sollukattu} (Figure~\ref{fig:Kuditta_Mettu_Annotations} (b)) \label{tbl:Sollukattu_bols_Tatta_C}}
\centering
{\renewcommand{\arraystretch}{1.5}%
\begin{scriptsize}
\begin{tabular}{|l|r|r|r|} \hline
\multicolumn{1}{|c|}{\bf Event}	& \multicolumn{1}{c|}{\bf Time} & \multicolumn{1}{c|}{\bf Beat Offset} & \multicolumn{1}{c|}{\bf 1/2--Beat Offset} \\ 
\multicolumn{1}{|c|}{\bf }	& \multicolumn{1}{c|}{\bf (sec.)} & \multicolumn{1}{c|}{\bf (sec.)} & \multicolumn{1}{c|}{\bf (sec.)} \\ 
\multicolumn{1}{|c|}{\bf }& \multicolumn{1}{c|}{\bf ($\tau(\alpha)$)} & \multicolumn{1}{c|}{\bf ($\tau(\alpha^{fb}_{i+1}) - \tau(\alpha^{fb}_i)$)} & \multicolumn{1}{c|}{\bf ($\tau(\alpha^{hb}_i) - \tau(\alpha^{fb}_i)$)} 	\\ \hline \hline
\textcolor{blue}{$\alpha^{fb}_1$(tei)}	&	6.571	&\multicolumn{1}{r|}{}&\\ \cline{1-4}
\textcolor{blue}{$\alpha^{hb}_1$(ya)}&	7.395	&\multicolumn{1}{r|}{}&	0.82	\\ \cline{1-4} 
\textcolor{red}{$\alpha^{fb}_2$(tei)}	&	8.185	&	\multicolumn{1}{r|}{1.61} & \\ \cline{1-4}
\textcolor{red}{$\alpha^{hb}_2$(ya)}	&	8.962	&\multicolumn{1}{r|}{}&	0.78	\\ \cline{1-4}
\textcolor{blue}{$\alpha^{fb}_3$(tei)}	&	9.752	&	\multicolumn{1}{r|}{1.57} &\\ \cline{1-4}
\textcolor{blue}{$\alpha^{hb}_3$(ya)}	&	10.565	&\multicolumn{1}{r|}{}&	0.81	\\ \cline{1-4}
\textcolor{red}{$\alpha^{fb}_4$(tei)}	&	11.366	&	\multicolumn{1}{r|}{1.61} &\\ \cline{1-4}
\textcolor{blue}{$\alpha^{fb}_5$(tei)}	&	13.003	&	\multicolumn{1}{r|}{1.64} &\\ \cline{1-4}
\textcolor{blue}{$\alpha^{hb}_5$(ya)}	&	13.815	&\multicolumn{1}{r|}{}&	0.81	\\ \cline{1-4}
\textcolor{red}{$\alpha^{fb}_6$(tei)} &	14.628	&	\multicolumn{1}{r|}{1.63} &\\ \cline{1-4}
\textcolor{red}{$\alpha^{hb}_6$(ya)} 	&	15.441	&\multicolumn{1}{r|}{}&	0.81	\\ \cline{1-4}
\textcolor{blue}{$\alpha^{fb}_7$(tei)} &	16.184	&	\multicolumn{1}{r|}{1.56} &\\ \cline{1-4}
\textcolor{blue}{$\alpha^{hb}_7$(ya)}	&	17.031	&\multicolumn{1}{r|}{}&	0.85	\\ \cline{1-4}
\textcolor{red}{$\alpha^{fb}_8$(tei)} &	17.809	&	\multicolumn{1}{r|}{1.63} &\\ \cline{1-4}
\multicolumn{4}{c}{$\bullet$ $T = 1.6$ sec., $\Lambda = 8$} \\
\end{tabular}
\end{scriptsize}
}
\end{table}
\renewcommand{\baselinestretch}{1.3} 

\renewcommand{\baselinestretch}{1} 
\begin{figure}[!ht]
\centering
\includegraphics[width=.7\textwidth]{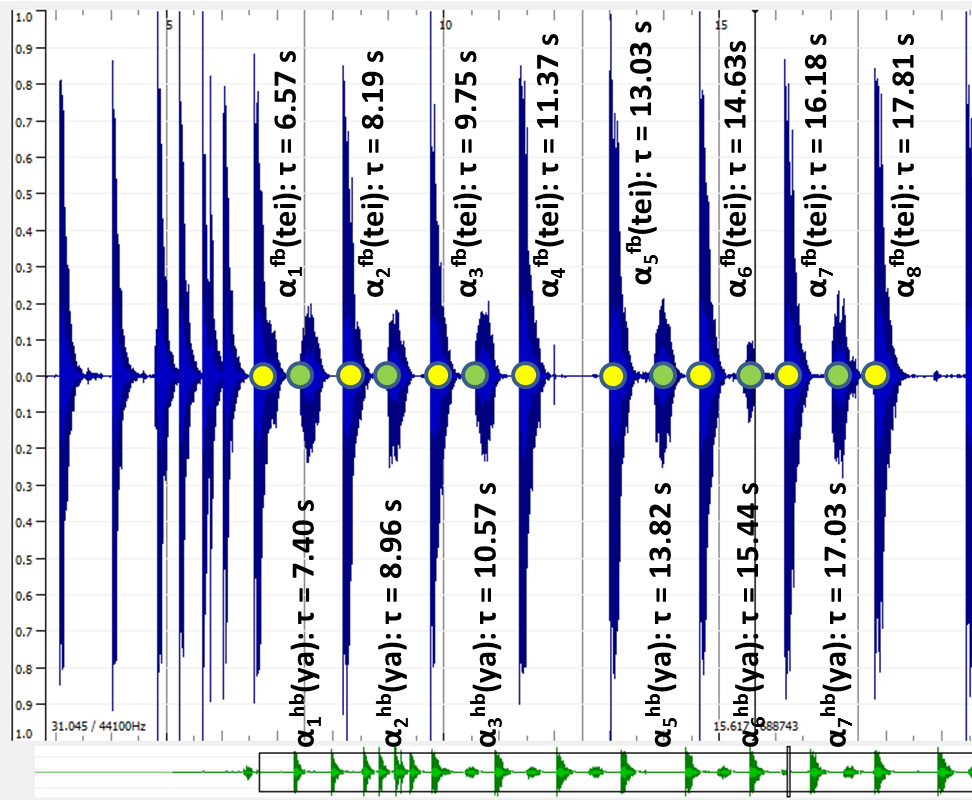} 

\centering
\begin{scriptsize}
\begin{tabular}{p{12cm}}
$\bullet$ {\bf Parameters}: No. of bars = 2, $\Lambda$ = 8 and $T = 1.56$ sec. \\
$\bullet$ Full beat ($\alpha^{fb}$) (yellow blobs) and half beat ($\alpha^{hb}$) (green blobs) event positions are highlighted and corresponding {\em bol}s and time-stamps are shown (Table~\ref{tbl:Sollukattu_bols_Tatta_C}). 
\end{tabular}
\end{scriptsize}
\caption{Marking of beats and annotations of {\em bol}s for {\em Tatta\_C Sollukattu}
\label{fig:Tatta_C_Annotations}}
\end{figure}
\renewcommand{\baselinestretch}{1.3} 

\item {\bf \em Kuditta Nattal\_A} \& {\bf \em Tatta\_E} ($T \approx $ 1.0 secs, $\Lambda$ = 8):  In addition to $\alpha^{fb}$, $\alpha^{fn}$ and $\alpha^{hn}$ events are also found (Table~\ref{tbl:Sollukattu_bols}) where there is only beating and no {\em bol}.

\item {\bf \em Joining\_B} ($T \approx $ 1.5 secs, $\Lambda$ = 8): As such it uses only $\alpha^{fb}$s (Table~\ref{tbl:Sollukattu_bols}). 
\end{enumerate}

All {\em Sollukattu}s in terms of the {\em Bol}s are listed in Table~\ref{tbl:bn_Bols}.

\renewcommand{\baselinestretch}{1} 
\begin{table}[!ht]
\caption{Variations in the patterns of {\em Sollukattu}s with {\em Adavu}s \label{tbl:Sollukattu_bols}}
\centering
{\renewcommand{\arraystretch}{1.5}%
\begin{tabular}{|l|l|} \hline
\multicolumn{1}{|c|}{\bf {\em Sollukattu}}	& \multicolumn{1}{c|}{\bf Description of Bol / {\em Adavu}s}	\\ \hline
{\em Kuditta}	&	
	\textcolor{blue}{$\alpha^{fb}_1$(tei)}
	\textcolor{red}{$\alpha^{fb}_2$(hat)} 
	\textcolor{blue}{$\alpha^{fb}_3$(tei)}
	\textcolor{red}{$\alpha^{fb}_4$(hi)} \\ 
{\em Mettu}	&	
	\textcolor{blue}{$\alpha^{fb}_5$(tei)}
	\textcolor{red}{$\alpha^{fb}_6$(hat)}
	\textcolor{blue}{$\alpha^{fb}_7$(tei)}
	\textcolor{red}{$\alpha^{fb}_8$(hi)} \\ \cdashline{2-2}
		& {\em Adavu}: {\em Kuditta\_Mettu 1, 2, 3, 4} \\ \hline

{\em Kuditta}	&	
	\textcolor{blue}{$\alpha^{fb}_1$(tat)} 
	\textcolor{red}{$\alpha^{fb}_2$(tei) $\alpha^{hn}_2$}
	\textcolor{blue}{$\alpha^{fb}_3$(tam)}
	\textcolor{red}{$\alpha^{fn}_4$ $\alpha^{hn}_4$} \\  
{\em Nattal A}	&	
	\textcolor{blue}{$\alpha^{fb}_5$(dhit)}
	\textcolor{red}{$\alpha^{fb}_6$(tei) $\alpha^{hn}_6$} 
	\textcolor{blue}{$\alpha^{fb}_7$(tam)}
	\textcolor{red}{$\alpha^{fn}_8$ $\alpha^{hn}_8$}  \\ \cdashline{2-2}
		& {\em Adavu}: {\em Kuditta\_Nattal 1, 2, 3, 6} \\ \hline

{\em Tatta E}	&	
	\textcolor{blue}{$\alpha^{fb}_1$(tei)}
	\textcolor{red}{$\alpha^{fb}_2$(tei)}
	\textcolor{blue}{$\alpha^{fb}_3$(tam)}
	\textcolor{red}{$\alpha^{fn}_4$ $\alpha^{hn}_4$}  \\ 
			&	
	\textcolor{blue}{$\alpha^{fb}_5$(tei)}
	\textcolor{red}{$\alpha^{fb}_6$(tei)}
	\textcolor{blue}{$\alpha^{fb}_7$(tam)}
	\textcolor{red}{$\alpha^{fn}_8$ $\alpha^{hn}_8$} \\ \cdashline{2-2}
 		& {\em Adavu}: {\em Tatta 6}  \\ \hline
 
{\em Joining B}	&	
	\textcolor{blue}{$\alpha^{fb}_1$(dhit)}
	\textcolor{red}{$\alpha^{fb}_2$(dhit)}
	\textcolor{blue}{$\alpha^{fb}_3$(tei)}
	\\ 
			&
	\textcolor{blue}{$\alpha^{fb}_4$(dhit)}
	\textcolor{red}{$\alpha^{fb}_5$(dhit)}
	\textcolor{blue}{$\alpha^{fb}_6$(tei)}
	\\ 
	\cdashline{2-2}
 		& {\em Adavu}: {\em Joining 2} \\ \hline
\end{tabular}}
\end{table}
\renewcommand{\baselinestretch}{1.3} 

\subsection{Characterization of Video Events}\label{sec:characterize_video}
While performing an {\em Adavu} the dancer closely follows the beats of the accompanying music.
At a beat, the dancer assumes a {\em Key Posture} and holds it for a little while before quickly changing to the next {\em Key Posture} at the next beat. Consequently, while the dancer holds the key posture, she stays almost stationary and there is no or very slow motion in the video. This leads to $\nu^{nm}$ no-motion events. Further, while the dancer changes to the next key posture, we observe the $\nu^{tr}$ (transition) or $\nu^{tj}$ (trajectory) motion events. Since a frame is an atomic observable unit in a video, we can classify the frames of the video of an {\em Adavu} into 2 classes:

\renewcommand{\baselinestretch}{1} 
\begin{figure}[!ht]
\centering
\begin{tabular}{llllllll}
{\includegraphics[width=3cm]{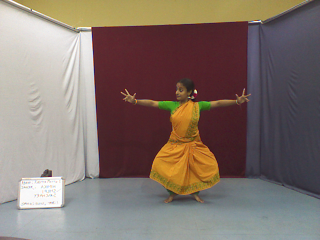}} & 
{\includegraphics[width=3cm]{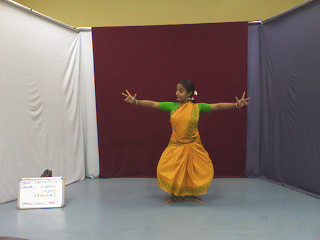}} & 
{\includegraphics[width=3cm]{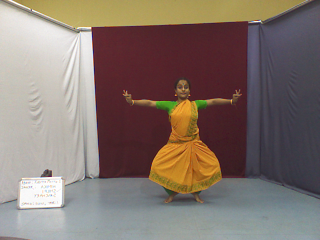}} & 
{\includegraphics[width=3cm]{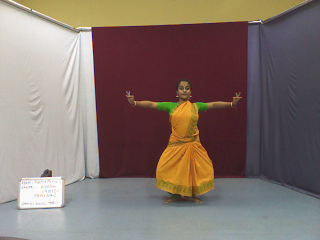}} \\
\multicolumn{1}{c}{(a) $\nu^{nm}_1$, $\alpha^{fb}_1$(tei)} & 
\multicolumn{1}{c}{(b) $\nu^{nm}_2$, $\alpha^{fb}_2$(hat)} & 
\multicolumn{1}{c}{(c) $\nu^{nm}_3$, $\alpha^{fb}_3$(tei)} & 
\multicolumn{1}{c}{(d) $\nu^{nm}_4$, $\alpha^{fb}_4$(hi)} \\
{\includegraphics[width=3cm]{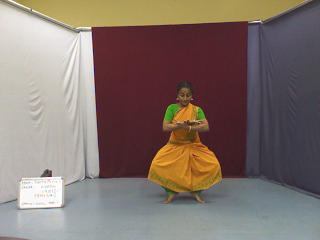}} & 
{\includegraphics[width=3cm]{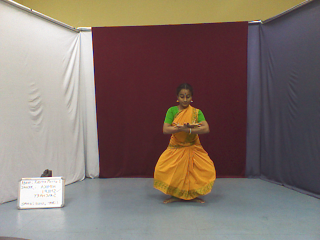}} & 
{\includegraphics[width=3cm]{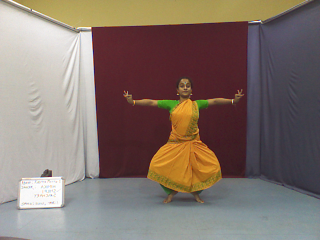}} & 
{\includegraphics[width=3cm]{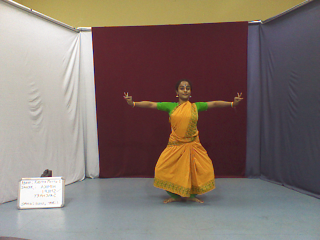}} \\ 
\multicolumn{1}{c}{(e) $\nu^{nm}_5$, $\alpha^{fb}_5$(tei)} & 
\multicolumn{1}{c}{(f) $\nu^{nm}_6$, $\alpha^{fb}_6$(hat)} & 
\multicolumn{1}{c}{(g) $\nu^{nm}_7$, $\alpha^{fb}_7$(tei)} & 
\multicolumn{1}{c}{(h) $\nu^{nm}_8$, $\alpha^{fb}_8$(hi)} 
\end{tabular}

\begin{scriptsize}
\begin{tabular}{p{10.5cm}} \\
$\bullet$ {\em Sollukattu = Kuditta Mettu}) with {\em bol}s for Bar 1. \\
$\bullet$ From a {\em tei} to the next {\em hat} or {\em hi} the dancer sharply lowers her raised feet. \\
$\bullet$ Further, 8 left-sided Key Postures are performed for the next 8 beats in Bar 2. \\
\end{tabular}
\end{scriptsize}
\caption{Right-sided Key Postures of {\em Kuditta Mettu Adavu} (Variant 2)}
\label{fig:Kuditta_Mettu_Postures}
\end{figure}
\renewcommand{\baselinestretch}{1.3} 

\begin{enumerate}
\item {\bf {\em K-frame}s or Key Frames}: These frames contain key postures where the dancer {\em holds} the Posture. Evidently, a $\nu^{nm}$ has the sequence of {\em K-frames} as labels. All {\em K-frames} of an $\nu^{nm}$ contain the same key posture.

\item {\bf {\em T-frame}s of Transition Frame}: These are transition frames between two {\em K-frames} while the dancer is rapidly changing posture to assume the next key posture from the previous one. {\em T-frame}s contain {\em Natural Transition Postures} (leading to $\nu^{tr}$ events) or {\em Trajectorial Transition Postures} (leading to $\nu^{tj}$ events). A $\nu^{tr}$ or $\nu^{tj}$ event has the corresponding sequence of {\em T-frames} as labels. In the current work, we do not deal with movements and transitions. Hence, we ignore {\em T-frames}.
\end{enumerate} 

In Figure~\ref{fig:Kuditta_Mettu_Postures} we show the key postures of {\em Kuditta Mettu Adavu} at every beat of the first bar of {\em Kuditta Mettu Sollukattu}. The corresponding video and audio events are marked in Table~\ref{tbl:kuditta_mettu_v_events} with {\em K-/T-Frames}. These are also marked on the {\em Sollukattu} in Figure~\ref{fig:Kuditta_Mettu_Annotations}. Note that only the right-sided half of the postures are shown in both figures. 


\renewcommand{\baselinestretch}{1} 
\begin{table}[!ht]
\caption[Patterns of {\em Kuditta Mettu Adavu}]{Patterns of {\em Kuditta Mettu Adavu} (Figure~\ref{fig:Kuditta_Mettu_Postures})\label{tbl:kuditta_mettu_v_events}}
\centering

{\renewcommand{\arraystretch}{1.5}%
\begin{tabular}{|l|r|r||l|r|r|} \hline
\multicolumn{1}{|c|}{\bf Events}	&\multicolumn{2}{c||}{\bf {\em K-/T-Frames}} & \multicolumn{1}{c|}{\bf Events} & \multicolumn{2}{c|}{\bf {\em K-/T-Frames}} \\ \cline{2-3} \cline{5-6}
\multicolumn{1}{|c|}{\bf }	& \multicolumn{1}{c|}{\bf Range} & \multicolumn{1}{c||}{\bf \# of} & \multicolumn{1}{c|}{\bf }	& \multicolumn{1}{c|}{\bf Range} & \multicolumn{1}{c|}{\bf \# of} \\ \hline \hline
\textcolor{blue}{$\nu^{nm}_1$} [$\alpha^{fb}_1$(tei)]& 70--99&30&\textcolor{blue}{$\nu^{nm}_9$} [$\alpha^{fb}_9$(tei)]&359--386&28 \\ \hline
\textcolor{red}{$\nu^{tr}_1$}& 100--103&4&\textcolor{red}{$\nu^{tr}_9$}    &387--390&4 \\ \hline
\textcolor{blue}{$\nu^{nm}_2$} [$\alpha^{fb}_2$(hat)]&104--124&21&\textcolor{blue}{$\nu^{nm}_{10}$} [$\alpha^{fb}_{10}$(hat)]&391--410&20 \\ \hline
\textcolor{red}{$\nu^{tr}_2$}& 125--145&21&\textcolor{red}{$\nu^{tr}_{10}$} &411--429&19 \\ \hline
\textcolor{blue}{$\nu^{nm}_3$} [$\alpha^{fb}_3$(tei)]&146--172&27&\textcolor{blue}{$\nu^{nm}_{11}$} [$\alpha^{fb}_{11}$(tei)]&430--451&22 \\ \hline
\textcolor{red}{$\nu^{tr}_3$}& 173--176&4&\textcolor{red}{$\nu^{tr}_{11}$} &452--455&4 \\ \hline
\textcolor{blue}{$\nu^{nm}_4$} [$\alpha^{fb}_4$(hi)]&177--191&15&\textcolor{blue}{$\nu^{nm}_{12}$} [$\alpha^{fb}_{12}$(hi)]&456--470&15 \\ \hline
\textcolor{red}{$\nu^{tr}_4$}& 192--214&23&\textcolor{red}{$\nu^{tr}_{12}$} &471--492&22 \\ \hline
\textcolor{blue}{$\nu^{nm}_5$} [$\alpha^{fb}_5$(tei)]&215--245&31&\textcolor{blue}{$\nu^{nm}_{13}$} [$\alpha^{fb}_{13}$(tei)]&493--521&29 \\ \hline
\textcolor{red}{$\nu^{tr}_5$}& 246--249&4&\textcolor{red}{$\nu^{tr}_{13}$} &522--525&4 \\ \hline
\textcolor{blue}{$\nu^{nm}_6$} [$\alpha^{fb}_6$(hat)]&250--262&13&\textcolor{blue}{$\nu^{nm}_{14}$} [$\alpha^{fb}_{14}$(hat)]&526--542&17 \\ \hline
\textcolor{red}{$\nu^{tr}_6$}& 263--287&25&\textcolor{red}{$\nu^{tr}_{14}$} &543--564&22 \\ \hline
\textcolor{blue}{$\nu^{nm}_7$} [$\alpha^{fb}_7$(tei)]&288--314&27&\textcolor{blue}{$\nu^{nm}_{15}$} [$\alpha^{fb}_{15}$(tei)]&565--587&23 \\ \hline
\textcolor{red}{$\nu^{tr}_7$}& 315--317&3&\textcolor{red}{$\nu^{tr}_{15}$} &588--590&3 \\ \hline
\textcolor{blue}{$\nu^{nm}_8$} [$\alpha^{fb}_8$(hi)]&318--345&28&\textcolor{blue}{$\nu^{nm}_{16}$} [$\alpha^{fb}_{16}$(hi)]&591--620&30 \\ \hline
\textcolor{red}{$\nu^{tr}_8$}& 346--358&13&\textcolor{red}{$\nu^{tr}_{16}$} &621-- & -- \\ \hline
\multicolumn{4}{c}{} \\
\end{tabular}}
\end{table}
\renewcommand{\baselinestretch}{1.3} 

\subsection{Characterization of Synchronization}\label{sec:sync}
A {\em Bharatanatyam} dancer intends to perform the key postures of an {\em Adavu} in synchronization with the beats. Hence various audio events like $\alpha^{fb}$ and corresponding video events like $\nu^{nm}$ should be in sync. Every {\em Adavu} has a well-defined set of rules that specifies this synchronization based on its associated {\em Sollukattu}. For example, in Figure~\ref{fig:Kuditta_Mettu_Postures}, we show how different key postures of {\em Kuditta Mettu Avadu} should be assumed at every beat of the {\em Kuditta Mettu Sollukattu}. That is, how the $\alpha^{fb}$s of a bar in the audio should sync with the $\nu^{nm}$s of the video. Other {\em Adavu}s require several other forms of synchronization between the audio-video events including sync between beats and trajectory-based body movements $\nu^{tj}$.

We assert a sync event $\psi^{fb}$ if a key posture ($\nu^{nm}$) should sync with a corresponding (full) beat ($\alpha^{fb}$). In simple terms, a $\psi^{fb}$ occurs if the time intervals of $\alpha^{fb}$ and $\nu^{nm}$ events overlap. That is, $\tau(\alpha^{fb}) \cap \tau(\nu^{nm}) \neq \phi$. Similar sync events may be defined between other audio and video events according to the rules of {\em Adavu}s.

Perfect synchronization is always intended and desirable for a performance. However, we often observe the lack of it due to various reasons. The beating instrument, vocal {\em bol}s, and body postures each has a different latency. If a posture is assumed {\em after hearing} the beat, $\nu^{nm}$ will lag $\alpha^{fb}$. If the dancer assumes the posture in {\em anticipation}, $\nu^{nm}$ may lead $\alpha^{fb}$. Lack of sync may also arise due to imperfect performance of the dancer, the beater or the vocalist. Hence, analysis and estimation of sync is critical for processing {\em Adavu}.

While sync between the audio and video streams is fundamental to the choreography, there are a variety of other synchronization issues that need to be explored. These include sync between beating (instrumental) beats and (vocalized) {\em bol}s, uniformity of time gap between consecutive beats, sync between different body limbs while changing from one key posture to the next, and so on.
 


%

\section{Ontology of Events and Streams}
We have captured the structural models of {\em Sollukattu}s and {\em Adavu}s in Section~\ref{sec:ontological_models} and then, the temporal behavioral models in Section~\ref{sec:characterize} based on these structures. Now, we would like to relate these to the actual recording data of the performances. For the current work we capture the performances of {\em Bharatanatyam Adavu}s using Kinect XBox 360 (Kinect 1.0) sensor. So in this section, we model the relationships between the events and the Kinect streams to facilitate the formulation of the algorithms later.

Kinect 1.0 is an RGBD sensor that captures a multi-channel audio stream with 3 video streams -- RGB, Depth, and Skeleton in its data file. The video streams are captured at 30 frames per second (fps). The RGB stream comprises frames containing color intensity images. The depth stream comprises frames containing depth images. And the skeleton stream comprises frames containing 20-joints skeleton images of human beings in the view. The video streams are synchronized between themselves. Hence for any RGB frame, the corresponding depth and skeleton frames carry the same frame number. The audio is also synchronized with the video by the same clock. Hence, any time $t$ on the audio stream corresponds to an RGB (depth, skeleton) frame by $t/30$. 

We now present a combined ontology for the events (as introduced in the last section) and the streams (of a Kinect data file), and capture their interrelationships. For this we identify sets of classes (Table~\ref{tbl:event_ontology_classes}), instances (Table~\ref{tbl:event_ontology_instances}), and relations (Table~\ref{tbl:event_ontology_relations}).

\renewcommand{\baselinestretch}{1} 
\begin{table}[!ht]
\centering
\caption{List of Classes for the ontology of Events and Streams\label{tbl:event_ontology_classes}}
\begin{scriptsize}
\begin{tabular}{|l|l||l|l|} \hline
\multicolumn{1}{|c}{\bf Class} & \multicolumn{1}{|c}{\bf Type} & \multicolumn{1}{||c}{\bf Class} & \multicolumn{1}{|c|}{\bf Type} \\ \hline\hline
$\bullet$ Kinect Data File 	&	 Concrete 	&	$\bullet$ Audio-Event Stream 	&	 Concrete 	 \\ 
$\bullet$ Audio Stream 	&	 Concrete 	&	$\bullet$ Video-Event Stream 	&	 Concrete 	 \\ 
$\bullet$ Video Stream 	&	 Concrete 	&	$\bullet$ Audio Event 	&	 Abstract 	 \\ 
$\bullet$ RGB Stream 	&	 Concrete 	&	$\bullet$ Video Event 	&	 Abstract 	 \\ 
$\bullet$ Depth Stream 	&	 Concrete 	&	$\bullet$ Beat Event 	&	 Abstract 	 \\ 
$\bullet$ Skeleton Stream 	&	 Concrete 	&	$\bullet$ {\em Bol} Event 	&	 Concrete 	 \\ 
$\bullet$ RGB Frame 	&	 Concrete 	&	$\bullet$ Full Beat Event 	&	 Concrete 	 \\ 
$\bullet$ Depth Frame 	&	 Concrete 	&	$\bullet$ Half Beat Event 	&	 Concrete 	 \\ 
$\bullet$ Skeleton Frame 	&	 Concrete 	&	$\bullet$ Full Beat with {\em bol} (FB+B) Event 	&	 Concrete 	 \\ 
$\bullet$ K-Frame 	&	 Concrete 	&	$\bullet$ Half Beat with {\em bol} (HB+B) Event 	&	 Concrete 	 \\ 
$\bullet$ T-Frame 	&	 Concrete 	&	$\bullet$ No-Motion Event 	&	 Concrete 	 \\ 
	&		&	$\bullet$ Transition Event 	&	 Concrete 	 \\ \hline
\end{tabular}
\end{scriptsize}
\end{table}
\renewcommand{\baselinestretch}{1.3} 

\renewcommand{\baselinestretch}{1} 
\begin{table}[!ht]
\centering
\caption{List of Instances for the ontology of Events and Streams \label{tbl:event_ontology_instances}}
\begin{scriptsize}
\begin{tabular}{|p{5cm}|p{6.5cm}|} \hline
\multicolumn{1}{|c}{\bf Class:Instance} & \multicolumn{1}{|c|}{\bf Remarks} \\ \hline\hline
$\bullet$ Full Beat Event: FBB1, FBB2, $\cdots$ & Instances of full beat with {\em bol} events\\
$\bullet$ Half Beat Event: HBB1, HBB2, $\cdots$ & Instances of half beat with {\em bol} events\\
$\bullet$ No-Motion Event: NM1, NM2, $\cdots$ & Instances of {\em no motion} events\\
$\bullet$ K-Frame: $I.p$, $I\cdots$, $I.p+u$, $\cdots$ & Intensity (RGB) image frames from no. $p$ to $p+u$\\
$\bullet$ T-Frame: $I.q+1$, $I\cdots$, $I.q+v$, $\cdots$ & Intensity image frames from no. $q+1$ to $q+v$, where $q=p+u$\\
$\bullet$ :$D.p$, $D\cdots$, $\cdots$ & Depth image frames from number $p$ \\
$\bullet$ :$S.p$, $S\cdots$, $\cdots$ & Skeleton image frames from number $p$ \\
\hline
\end{tabular}
\end{scriptsize}
\end{table}
 
\begin{table}[!ht]
\centering
\caption{List of Relations for the ontology of Events and Streams \label{tbl:event_ontology_relations}}
\begin{scriptsize}
\begin{tabular}{|l|l|p{1.8cm}|p{6cm}|} \hline
\multicolumn{1}{|c}{\bf Relation} & \multicolumn{1}{|c}{\bf Domain} & \multicolumn{1}{|c}{\bf Co-Domain} & \multicolumn{1}{|c|}{\bf Remarks} \\ \hline \hline
{\em is\_a} & Class & Class & As in Table~\ref{tbl:ontology_relations}\\ \hline

{\em has\_a} & Class & Class & As in Table~\ref{tbl:ontology_relations} \\ \hline

{\em isInstanceOf} & Instance & Class & As in Table~\ref{tbl:ontology_relations} \\ \hline

{\em isSyncedWith} & Instance & Instance & Expresses low-level synchronization -- between audio / video events and video frames. For example, an audio event FBB1 (instance of 'full-beat with bol') {\em isSyncedWith} a unique K-Frame. \\ \hline

{\em isSequenceOf} & Class & Class & As in Table~\ref{tbl:ontology_relations} \\ \hline

{\em isExtractedFrom} & Class & Class & An event {\em isExtractedFrom} Kinect video\\ \hline

{\em isInSync} & \multicolumn{2}{l|}{Relation over 3 Instances} & Expresses the inherent synchronization in data -- between audio and multiple video streams -- RGB, Depth and Skeleton. Every {\em RGB Frame} {\em isInSync} with a corresponding {\em Depth Frame} or {\em Skeleton Frame}. \\ \hline
\multicolumn{4}{c}{ } \\
\multicolumn{4}{l}{\em All relations, with the exception of 'isInSync', are binary} \\
\end{tabular}
\end{scriptsize}
\end{table}

The ontology is presented in Figure~\ref{fig:Concept of Kinect Data}. The following points about the ontology may be noted:

\begin{itemize}
\item The event-side is shown in blue and the stream-side is shown in black. 
\item A {\em K-frame} is a semantic notion that is instantiated as a triplet of an RGB, Depth and Skeleton frames. Also, it actually represents a sequence of consecutive frames in the video having {\em no-motion}. {\em T-frame}s are treated similarly. 
\item {\em isExtractedFrom} represents the processes of extraction (or detection, estimation etc.) of audio (video) events from audio (video) streams. These are not directly available from the Kinect streams and need to be computationally determined. Specific algorithms required include:
\begin{itemize}
\item Beat detection to produce $\alpha^f$ or $\alpha^h$ 
\item {\em Bol} recognition to produce $\alpha^{fb}(<bol>)$ or $\alpha^{hb}(<bol>)$ 
\item {\em No-Motion} detection to produce $\nu^{nm}$ events 
\end{itemize} 
\item {\em isInSync} represents the fact that streams in Kinect are synchronized by the sensor. \item In contrast {\em isSyncedWith} denotes the explicit attempt of the dancer to synchronize her / his moves and postures with the beats and {\em bol}s. These are $\psi^{fb}$ or $\psi^{hb}$ events. To estimate {\em isSyncedWith}, {\em K-Frame}s and {\em T-frame}s need to be extracted.
\end{itemize}

\begin{figure}[!ht]
\centering
\includegraphics[width=\textwidth]{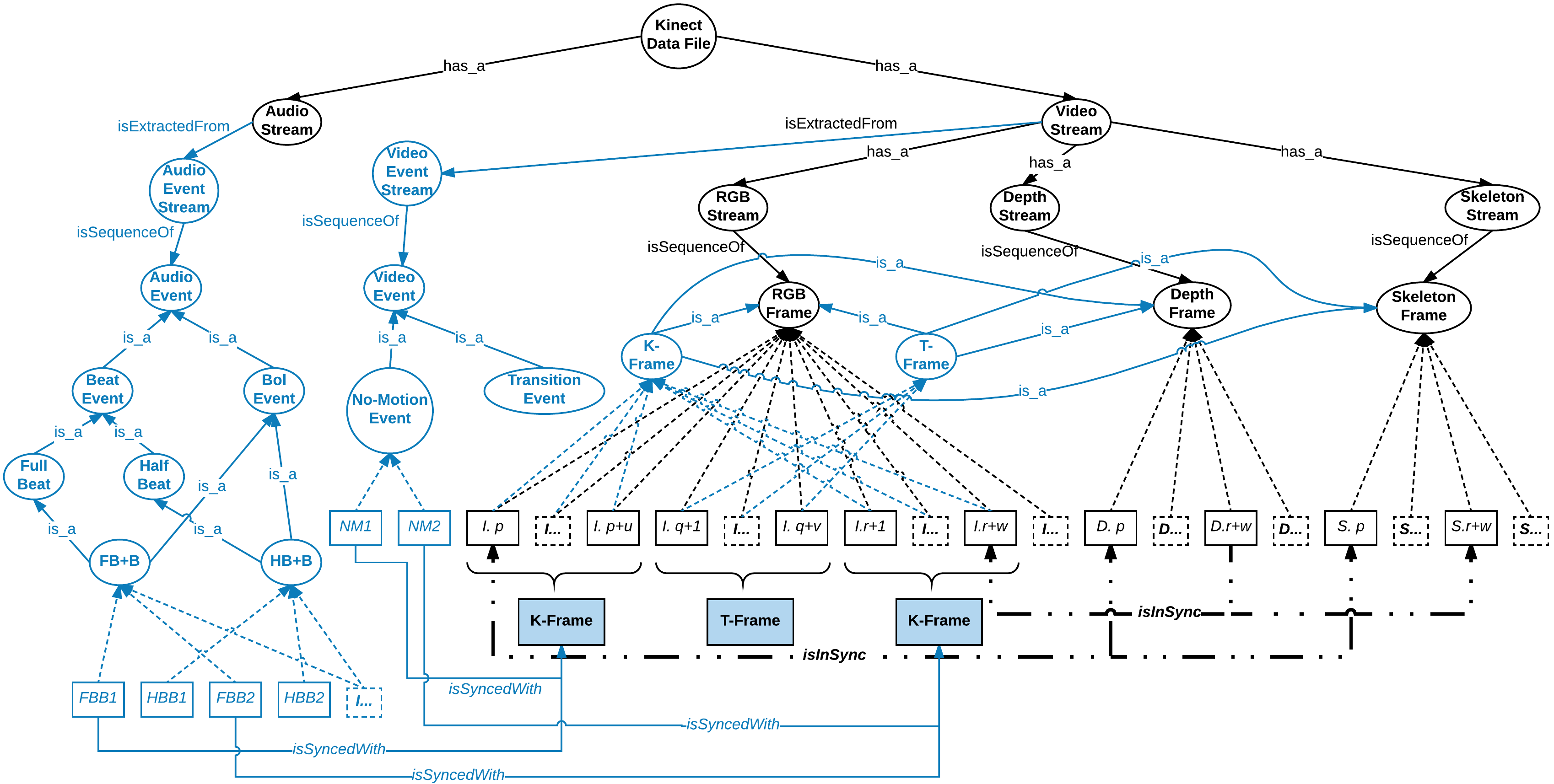} \vspace*{-0.3cm}
\caption{Ontology of Kinect Data File, Streams and Audio-Video Events
\label{fig:Concept of Kinect Data}}
\end{figure}

\section{Representation of {\em Adavu}s in Labanotation}
\begin{figure}[!ht]
\centering
\begin{tabular}{cc}
\includegraphics[width=.55\textwidth]{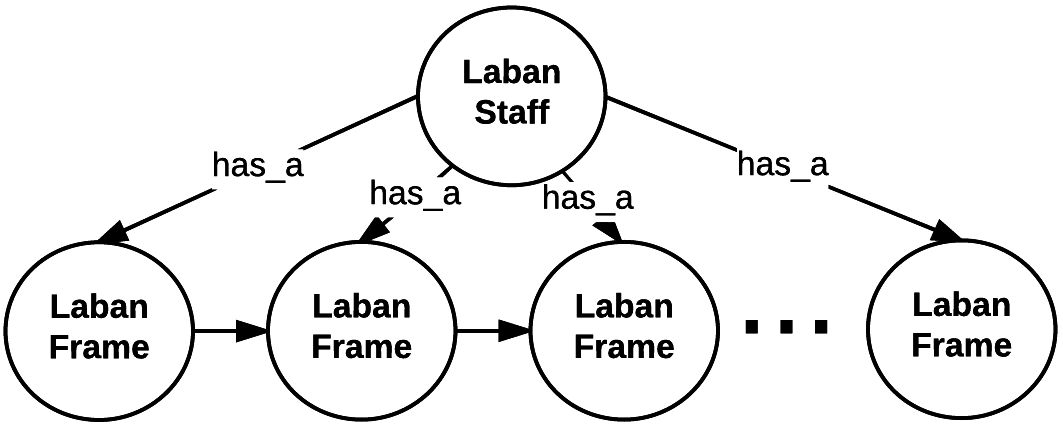} & 
\includegraphics[width=5cm]{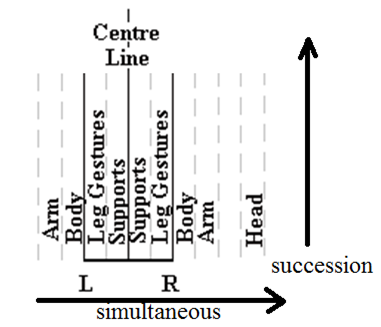} \\
(a) & (b)
\end{tabular}
\caption{Ontology of Labanotation
\label{fig_frame_seq}}
\end{figure}

We intend to represent {\em Bharatantyam} ontology according to the ontology of a parse-able standard notation. Labanotation ~\cite{guest2005labanotation} (often referred to as {\em Laban Encoding} or simply {\em Laban}) is a standard notation system used for recording human movements. To record a movement the Laban system symbolizes {\em space}, {\em time}, {\em energy}, and {\em body parts}. Here, we introduce a limited set of symbols that are particularly used for representing posture of {\em Bharatantyam Adavu}s. A Posture are encoded in laban is called frame and the laban frames are stack in laban staff as shown in Figure~\ref{fig_frame_seq}. When there is a sequence of postures or gestures changing over time, we stack their symbols on  the staff vertically to show the progression over time. The center line of the staff indicates the time. The symbols are read from the bottom to the top of the staff. 


The {\em Staff} represents the body. The {\em Center Line} divides the body into two parts -- Left and Right. The immediate next to the center line are {\em Support Columns}. The symbols placed in these columns indicate the body parts which carry the weight of the body. Other columns are represent the gestures of other body parts such as {\em Leg}, {\em Body} (torso), {\em Arm}, and {\em Head}. Except head, other body parts have left and right columns. Labanotation captures the movements of the human body parts in terms of the {\em directions} and {\em levels} of the movement. The direction symbols are used to indicate in which direction in the space the movements occur and in any direction can have three different levels, namely, {\em upward or high}, {\em horizontal or middle}, and {\em downwards or low}. Every body part can be expressed in terms of the direction and level by placing respective symbols in the designated columns.

\begin{figure}[!ht]
\centering
\includegraphics[width=0.9\textwidth]{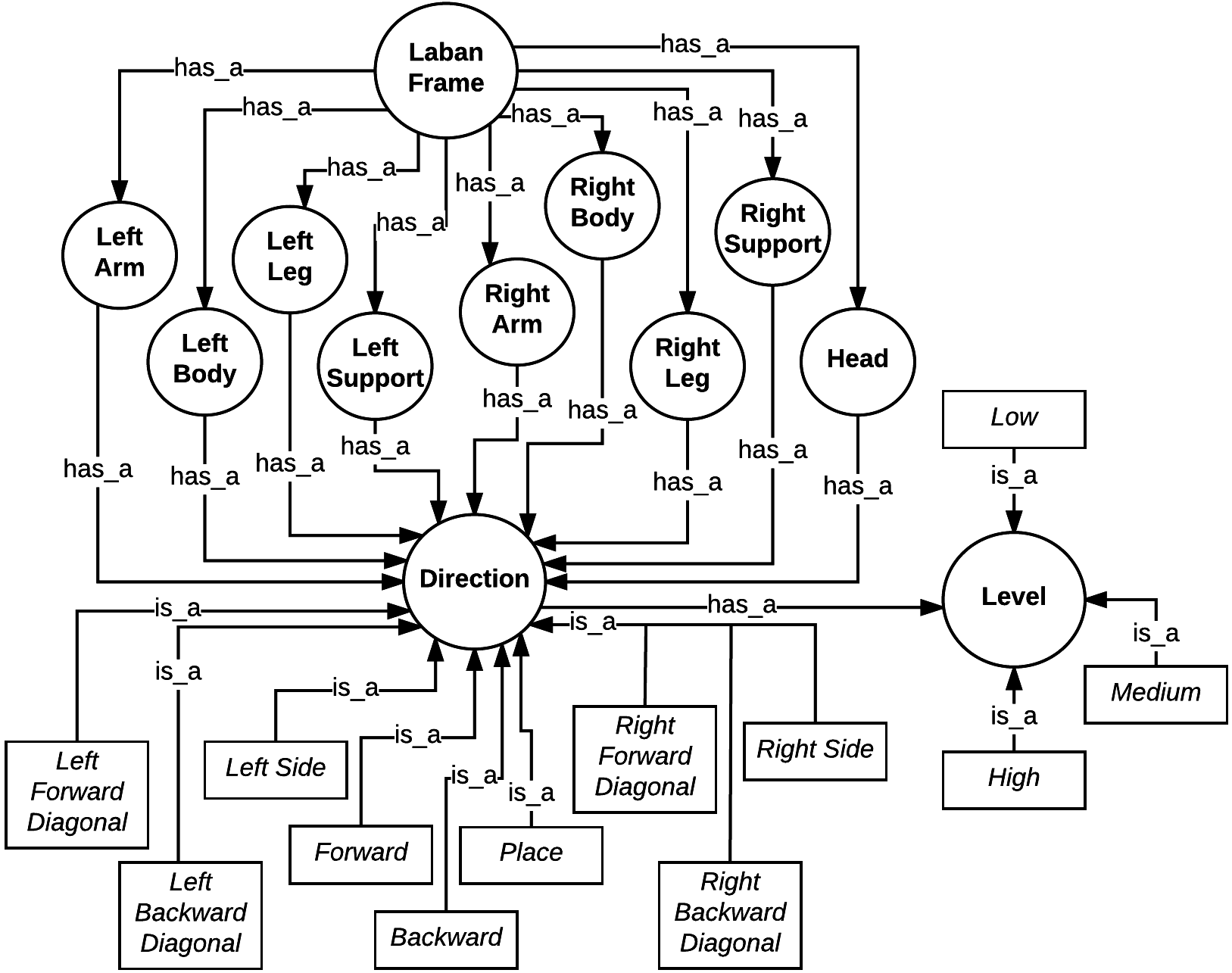} \vspace*{-0.2cm}
\caption{Ontology of Labanotation
\label{fig:Concept of Kinect Data}}
\end{figure}

\begin{figure}[!ht]
\centering
\includegraphics[width=0.9\textwidth]{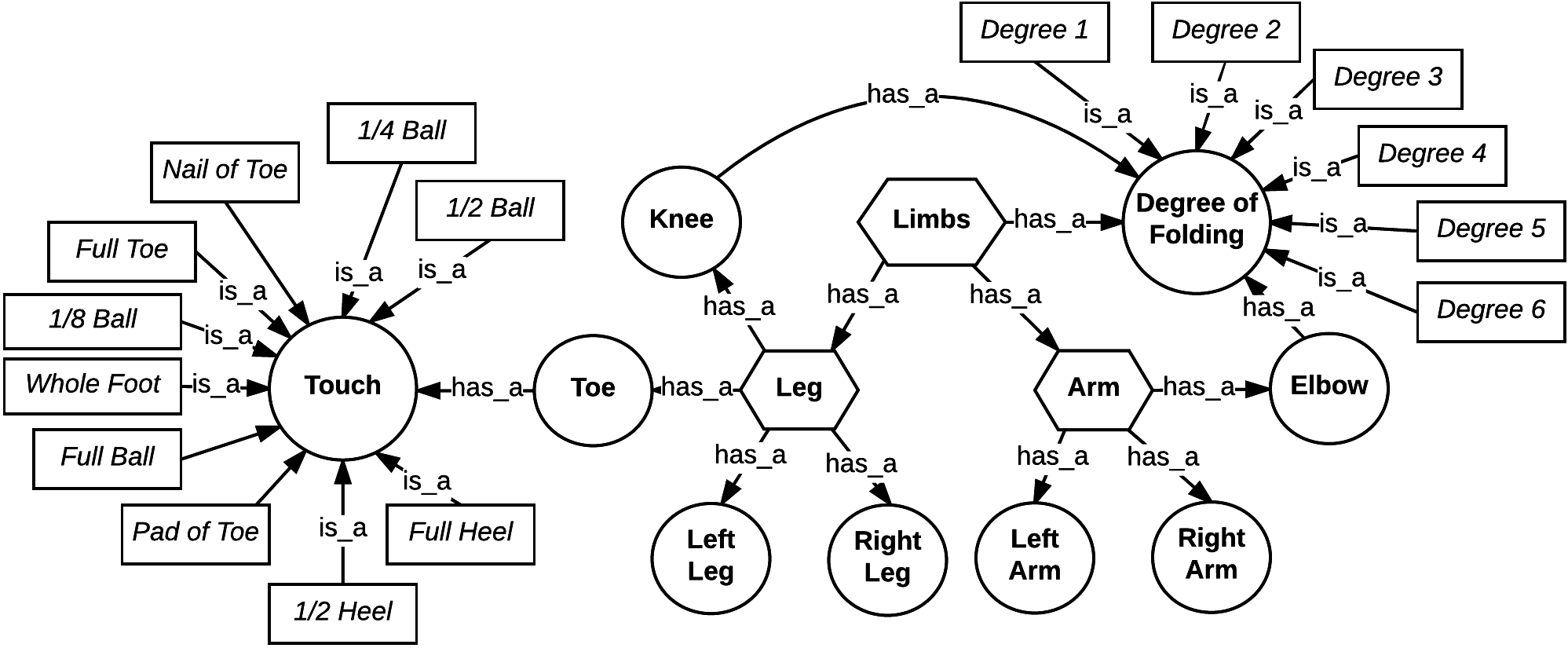} \vspace*{-0.3cm} 
\caption{Ontology of Labanotation
\label{fig_folding_touch}}
\end{figure}

The arms and the legs do not always remain straight while performing an {\em Adavu}. Few joints of the body like knee and elbow can get folded. Hence {\em degree of folding} is useful for these joints. There are a total of six degrees of folding. {\em Bharatanatyam} also involves a lot of foot work. Hence, we need to encode the type of touch between the foot and the ground and also which part of the foot is in contact with the ground. Labanotation system has symbols to diagrammatically illustrate the specific part of the foot that contact the ground. This attribute is called {\em touch} in Labanotation. There are 11 parts of foot that can touch the ground. The concepts are shown in Figure~\ref{fig_folding_touch}.

\begin{figure}[!ht]
\centering
\includegraphics[width=0.9\textwidth]{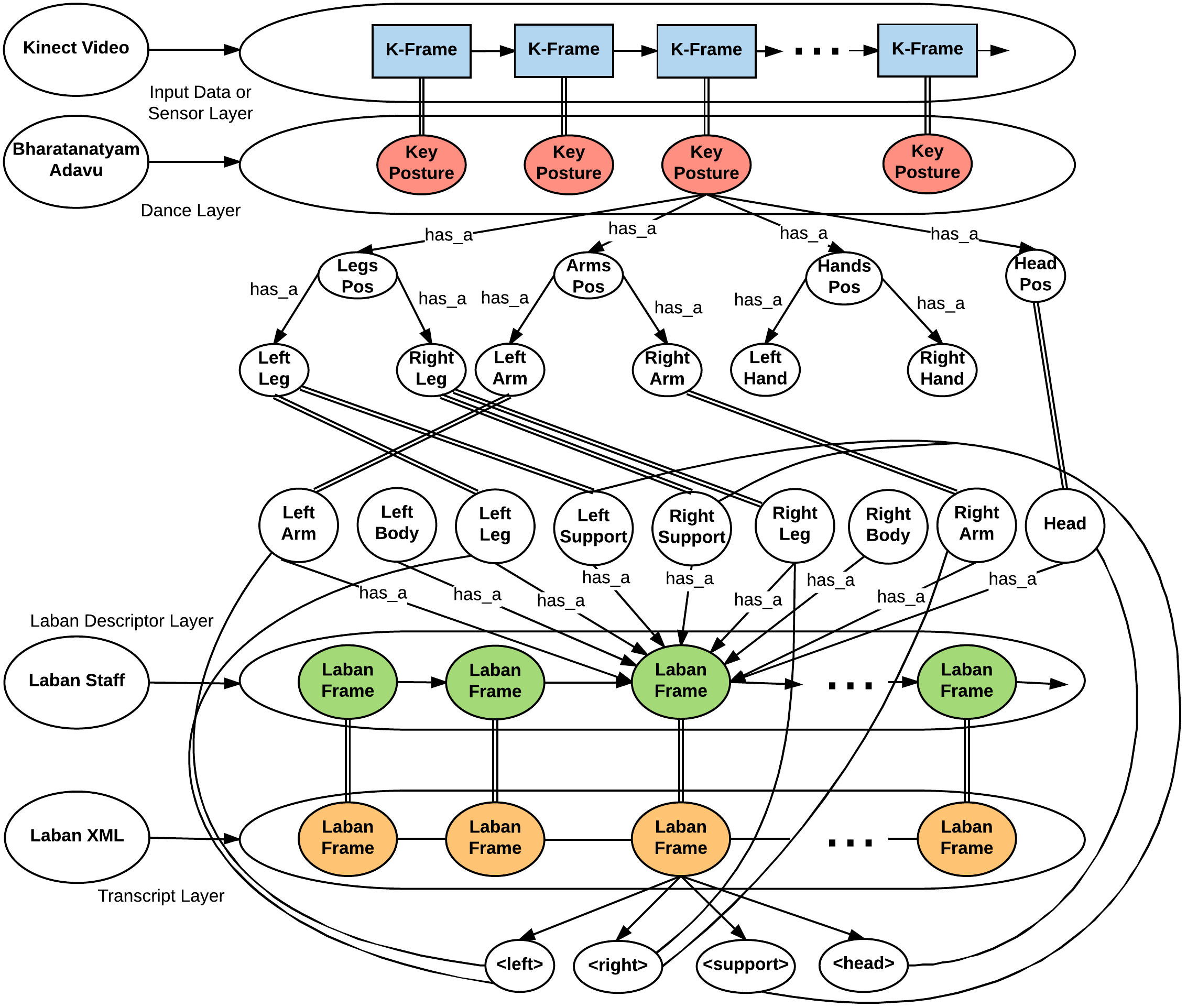} \vspace*{-0.3cm} 
\caption{Ontology of Transcription/ Sensor to parser representation 
\label{fig_kinect_xml}}
\end{figure}

We want to use concepts of Labanotation to transcript the data captured by the sensor into machine parse-able form. We map the kinect data to the concepts {\em Bharatanatyam} in Figure~\ref{fig:Concept of Kinect Data}. Now, we intend to map the concept of {\em Bharatanatyam} to the concept of Labanotation as our goal is generate a parse-able XML descriptor of {\em Bharatanatyam Adavu}. The ontology is shown in Figure~\ref{fig_kinect_xml}. There are 4 layers in the ontology--
\begin{enumerate}
\item {\bf Input or Sensor Layer}: This layers contains the data captured by the sensor. We capture the video of the dance using Kinect sensor. The data contains the {\em K-frames} as well as {\em T-frames}. Here we intend to transcript only the {\em K-frames}.
\item {\bf Dance Layer}: According to the ontology shown in Figure~\ref{fig:Concept of Kinect Data}, the {\em K-frames} contains No-Motions events. The No-Motions events are nothing but the Key postures of {\em Bharatanatyam Adavu} as shown in Figure~\ref{fig:Concept of Key Postures}. Here, we map the key posture in terms of direction, level, degree of folding and touch concept of Labanotation. The leg, arm and head of a key posture are get mapped into the Laban concept. 
\item {\bf Laban Descriptor Layer}: Each key posture has corresponding Laban frame in the Laban staff. The legs described in terms of leg and support of Labanotation. Arm and head have one to one mapping between {\em Bharatanatyam} ontology to Laban ontology. 
\item {\bf Transcript Layer}: Finally, the Laban ontology gets encoded into a parse-able XML format so that the Labanotation can get visualized or animation can get generated from the XML.
\end{enumerate}

\section{Laban Encoding of an {\em Adavu} Posture}\label{sec:Natta1_posture_trans}
We represent {\em Adavu} as a sequence of key postures. To transcribe an {\em Adavu}, we need to transcribe every key posture that occur in the {\em Adavu}. For the purpose of use, we encode the symbols of direction and level in Table~\ref{tbl:dir_level}, the degree of folding in Table~\ref{tbl:degre_folding}, and the touch attribute in Table~\ref{tbl:touch}.. 

\begin{table}[!ht]
\centering
\caption{Encoding of Directions and Levels in Labanotation\label{tbl:dir_level}}
\resizebox{\textwidth}{!}{%
\begin{tabular}{|c|c|c|c|lll} \hline
\textbf{Direction}             & \textbf{Place}                                                      & \textbf{Left Side}                                                              & \textbf{Right Side}                                                              & \multicolumn{1}{c|}{\textbf{Left Forward}}                                                       & \multicolumn{1}{c|}{\textbf{Right Forward}}                                                       & \multicolumn{1}{c|}{\textbf{Left Backward}} \\ \hline
Encoding                       & 1                                                                   & 2                                                                          & 3                                                                           & \multicolumn{1}{c|}{4}                                                                           & \multicolumn{1}{c|}{5}                                                                            & \multicolumn{1}{c|}{6}                      \\ \hline

\multicolumn{7}{c}{ } \\ \cline{1-6}

\textbf{Direction}             & \textbf{\begin{tabular}[c]{@{}c@{}}Right\\   Backward\end{tabular}} & \textbf{\begin{tabular}[c]{@{}c@{}}Left\\   Forward Diagonal\end{tabular}} & \textbf{\begin{tabular}[c]{@{}c@{}}Right\\   Forward Diagonal\end{tabular}} & \multicolumn{1}{c|}{\textbf{\begin{tabular}[c]{@{}c@{}}Left\\   Backward Diagonal\end{tabular}}} & \multicolumn{1}{c|}{\textbf{\begin{tabular}[c]{@{}c@{}}Right\\   Backward Diagonal\end{tabular}}} &                                             \\ \cline{1-6}
\multicolumn{1}{|l|}{Encoding} & \multicolumn{1}{l|}{7}                                              & \multicolumn{1}{l|}{8}                                                     & \multicolumn{1}{l|}{9}                                                      & \multicolumn{1}{l|}{10}                                                                          & \multicolumn{1}{l|}{11}                                                                           &                                             \\ \cline{1-6}

\multicolumn{7}{c}{ } \\ \cline{1-4}

\textbf{Level}                 & \textbf{HIGH}                                                       & \textbf{MID}                                                               & \textbf{LOW}                                                                &                                                                                                  &                                                                                                   &                                             \\ \cline{1-4}
Encoding                       & 1                                                                   & 2                                                                          & 3                                                                           &                                                                                                  &                                                                                                   &                                             \\ \cline{1-4}
\end{tabular}
}
\end{table}

\begin{table}[!ht]
\centering
\caption{Degree of Folding\label{tbl:degre_folding}}
\resizebox{\textwidth}{!}{
\begin{tabular}{|l|l|l|l|l|l|l|l|} \hline
\multicolumn{1}{|c|}{\textbf{Degree of Folding}} & \multicolumn{1}{c|}{\textbf{No Fold}} & \multicolumn{1}{c|}{\textbf{\begin{tabular}[c]{@{}c@{}}Fold \\ Degree 1\end{tabular}}} & \multicolumn{1}{c|}{\textbf{\begin{tabular}[c]{@{}c@{}}Fold \\ Degree 2\end{tabular}}} & \multicolumn{1}{c|}{\textbf{\begin{tabular}[c]{@{}c@{}}Fold \\ Degree 3\end{tabular}}} & \multicolumn{1}{c|}{\textbf{\begin{tabular}[c]{@{}c@{}}Fold \\ Degree 4\end{tabular}}} & \multicolumn{1}{c|}{\textbf{\begin{tabular}[c]{@{}c@{}}Fold \\ Degree 5\end{tabular}}} & \multicolumn{1}{c|}{\textbf{\begin{tabular}[c]{@{}c@{}}Full \\ Fold \end{tabular}}} \\ \hline
Encoding                                   & 0                                     & 1                                                                                      & 2                                                                                      & 3                                                                                      & 4                                                                                      & 5                                                                                      & 6                                                                                  \\ \hline
\end{tabular}
}
\end{table}

\begin{table}[!ht]
\centering
\caption{Type of touch with floor\label{tbl:touch}}
\resizebox{\textwidth}{!}{ 
\begin{tabular}{|l|l|l|l|l|l|l} 
\hline
\multicolumn{1}{|c|}{\textbf{\begin{tabular}[c]{@{}c@{}}Foot\\   Parts\end{tabular}}} & \multicolumn{1}{c|}{\textbf{Full heel}} & \multicolumn{1}{c|}{\textbf{One half heel}} & \multicolumn{1}{c|}{\textbf{Whole foot}} & \multicolumn{1}{c|}{\textbf{\begin{tabular}[c]{@{}c@{}}One eighth \\ ball\end{tabular}}} & \multicolumn{1}{c|}{\textbf{\begin{tabular}[c]{@{}c@{}}One fourth \\ ball\end{tabular}}} & \multicolumn{1}{c|}{\textbf{\begin{tabular}[c]{@{}c@{}}One half \\ ball\end{tabular}}} \\ \hline
Encoding                                                                              & 1                                       & 2                                           & 3                                        & 4                                                                                        & 5                                                                                        & \multicolumn{1}{l|}{6}                                                                 \\ \hline
\multicolumn{7}{c}{ } \\ \cline{1-6}
\textbf{Foot Parts}                                                                   & \textbf{Full ball}                      & \textbf{Pad of toe}                         & \textbf{Full toe}                        & \textbf{Nail of toe}                                                                     & \textbf{No touch}                                                                        &                                                                                        \\ \cline{1-6}
Encoding                                                                              & 7                                       & 8                                           & 9                                        & 10                                                                                       & 0                                                                                        &                                                                                        \\ \cline{1-6}
\end{tabular}
}
\end{table}

\renewcommand{\baselinestretch}{1} 
\begin{table}[!ht]
\centering
\caption{Annotation of the video of {\em Natta Adavu Variation 1}}
\label{tbl:annotated_data}
\begin{small}
\begin{tabular}{|l|r|r|r|l|} \multicolumn{5}{c}{} \\ \hline
\multicolumn{1}{|c}{\bf Posture}	&	\multicolumn{1}{|c}{\bf Start}	&	\multicolumn{1}{|c}{\bf End}	&	\multicolumn{1}{|c}{\bf Beat}	&	\multicolumn{1}{|c|}{\bf Bols}	\\ 
\multicolumn{1}{|c}{\bf Name}	&	\multicolumn{1}{|c}{\bf Frame}	&	\multicolumn{1}{|c}{\bf Frame}	&	\multicolumn{1}{|c}{\bf Number}	&	\multicolumn{1}{|c|}{\bf }	\\ 
\multicolumn{1}{|c}{\bf (a)}	&	\multicolumn{1}{|c}{\bf (b)}	&	\multicolumn{1}{|c}{\bf (c)}	&	\multicolumn{1}{|c}{\bf (d)}	&	\multicolumn{1}{|c|}{\bf (e)}	\\ \hline \hline
Natta1P1	&	70	&	89	&	0	&	No Bol	\\ \hline
Natta1P2	&	101	&	134	&	1	&	{\em tei yum}	\\ \hline
Natta1P1	&	144	&	174	&	2	&	{\em tat ta}	\\ \hline
Natta1P3	&	189	&	218	&	3	&	{\em tei yum}	\\ \hline
Natta1P1	&	231	&	261	&	4	&	{\em ta}	\\ \hline
\end{tabular}
\end{small}
\end{table}
\renewcommand{\baselinestretch}{1.3} 

\begin{figure}[!ht]
\centering
\includegraphics[width=0.6\textwidth]{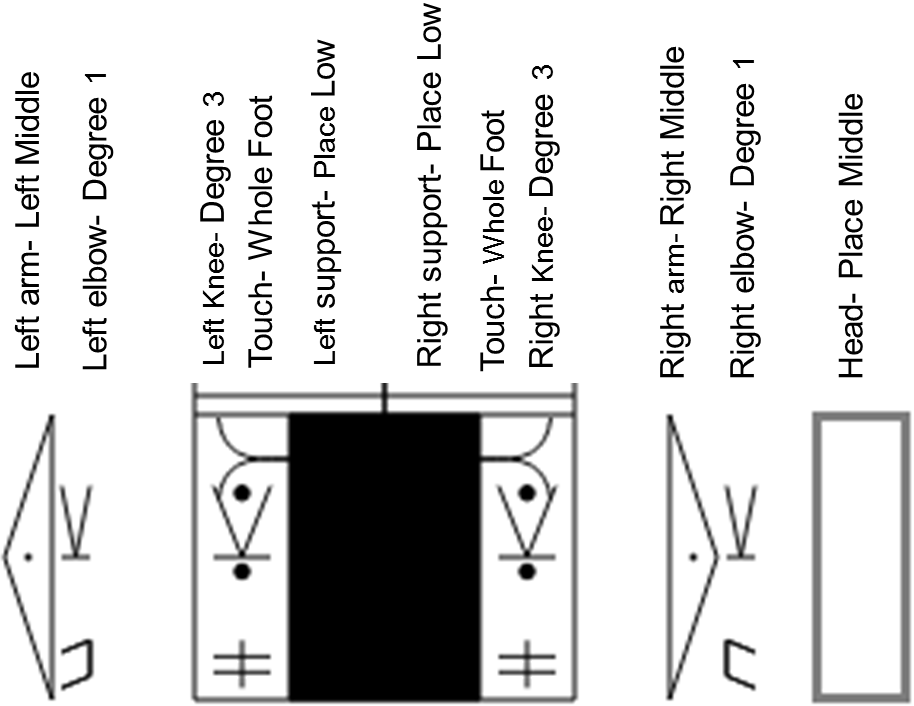} \vspace*{-0.3cm}
\caption{Labanotation of Leg, Arm and Head of Posture = Natta1P1\label{fig:BN_Laban}}
\end{figure}

The sequence of key postures occurring in the first 4 beats of {\em Natta Adavu Variation 1} are shown in Table~\ref{tbl:annotated_data}. A posture is described in terms of legs, arms, head, and hands using the vocabulary (Section~\ref{sec:vocab_postures}) for the annotation of the limbs. 
Now, we want to transcribe the posture. Hence, we need to encode the body parts from {\em Bharatanatyam} terminology to Labanotation descriptor. For example, consider key posture Natta1P1 {\em Natta Adavu Variation 1}. Let us describe the posture using the Labanotation symbols. The posture is shown in Figure~\ref{fig:BN_Laban}. The different body parts of the posture are marked in different colors like arm is marked as yellow. Annotation of the body parts of postures Natta1P1 is given in Table~\ref{tbl:Natta1P1_posture_anno} (We exclude {\em Hasta Mudra} from the transcription work).


\begin{figure}[!ht]
\centering
\includegraphics[width=5cm]{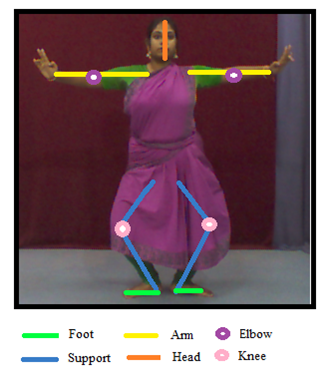}
\caption{Natta1P1 Posture annotated for Laban encoding\label{posture_Natta1P1}}
\end{figure}

\renewcommand{\baselinestretch}{1} 
\begin{table}[!ht]
\centering 
\caption{Annotation of the body parts of postures in {\em Natta Adavu 1}\label{tbl:Natta1P1_posture_anno}}
\begin{footnotesize} 
\begin{tabular}{|l|l|l|l|l|} 
\hline
\multicolumn{1}{|c|}{\bf Body}	&	\multicolumn{1}{c|}{\bf Position} 	&	\multicolumn{2}{c|}{\bf Formation}		&	\multicolumn{1}{c|}{\bf Vocab}	\\ \cline{3-4}
\multicolumn{1}{|c|}{\bf Part}	&		&	\multicolumn{1}{c|}{\bf Left}	&	\multicolumn{1}{c|}{\bf Right}	&		\\ \hline \hline
\multicolumn{5}{|c|}{\bf Posture = Natta1P1} \\ \hline\hline

Leg	& {\em 	Aayata {\bf [S]}} & {\em 	Aayata} & {\em 	Aayata}	& Table~\ref{tbl:leg_vocabs}	\\ \hline
Arm	& {\em 	Natyarambhe {\bf [S]}} & {\em Natyarambhe} & {\em Natyarambhe} & Table~\ref{tbl:arm_vocabs}	\\ \hline
Head& {\em 	Samam} & & & Table~\ref{tbl:head_vocabs}	\\ \hline
\end{tabular}
\end{footnotesize}
\end{table}
\renewcommand{\baselinestretch}{1.3} 

The next challenge is to map the {\em Bharatanatyam} ontology to the Labanotation ontology. As an example, we encode the posture Natta1P1 (Figure~\ref{posture_Natta1P1}, Table~\ref{tbl:Natta1P1_posture_anno}) to Laban in Table~\ref{tbl:laban_encoding_Natta1P1} and Figure~\ref{fig:BN_Laban}.

\begin{enumerate}
\item {\bf Leg}: The leg is in {\em Aayata} position which means: 
\begin{itemize}
\item The weight of the body is on both legs. So the legs are in support (as both leg are taking the weight of the body). The Support Direction and Support Level are encoded accordingly in Table~\ref{tbl:laban_encoding_Natta1P1}. 
\item The left (right) foot is in left (right) direction. The legs are not stretched in any direction, so the legs are in place. 
\item The folding of the legs indicate that the level of the leg is low. 
\item The legs are not crossing each other, so the Leg Crossing = 0. 
\item We mark the symmetric position of both the legs using Mirror = 1. If Mirror = 1, then the direction of the right leg will just be in the opposite of the left leg. 
\item The body weight is not on hip so Hip Support = 0.
\item Both legs are folded at the knee. The Knee in folding in around $90^\circ$, so Knee Folding = 3 (Figure~\ref{tbl:degre_folding}). 
\item The whole feet are touching the ground, so Touch = 3 (Figure~\ref{tbl:touch}). 
\end{itemize}

\item {\bf Arm}: The arms are in {\em Natyarambhe} which means:
\begin{itemize}
\item The hands are stretched in left and right side of the body at the shoulder level and are slightly folded at elbow. So, Arm Direction = 2, Arm Level = 2 and Elbow Folding = 1. 
\item Arm is not occluding with the body (Body Inclusion = 0) and 
\item Both arms are similar (Mirror = 1). 
\end{itemize}

\item {\bf Head}: The head is in {\em Samam} which means:
\begin{itemize}
\item The head is straight and forward (Head Direction = 1 and Level = Middle). 
\end{itemize}
\end{enumerate}

The complete Laban encoding for Natta1P1 is shown in Table~\ref{tbl:laban_encoding_Natta1P1}. In a similar manner we have encodes the other postures used in {\em Bharatanatyam Adavu}. This has been done with the help of the experts.

\renewcommand{\baselinestretch}{1} 
\begin{table}[!ht]
\centering
\caption{Laban Encoding of Leg, Arm and Head of Posture = Natta1P1
\label{tbl:laban_encoding_Natta1P1}}
\resizebox{\textwidth}{!}{ 
\begin{tabular}{c|c|c|cccl} 
\hline
\multicolumn{1}{|l|}{\bf Leg Vocab}  & \begin{tabular}[c]{@{}l@{}}Support\\   Direction\end{tabular} & Support Level   & \multicolumn{1}{l|}{Leg Direction} & \multicolumn{1}{l|}{Leg Level}         & \multicolumn{1}{l|}{Leg Crossing}   & \multicolumn{1}{l|}{Mirror} \\ \hline
\multicolumn{1}{|l|}{Aayata}      & 1                                                             & 3                & \multicolumn{1}{l|}{0}              & \multicolumn{1}{l|}{0}                  & \multicolumn{1}{l|}{0}               & \multicolumn{1}{l|}{1}      \\ \hline

\multicolumn{7}{c}{ } \\ \cline{2-4} 
                                  & Hip Support                                                  & Knee Folding & \multicolumn{1}{l|}{Touch}          &                                         &                                      &                             \\ \cline{2-4}                         
                                  & 0                                                             & 3                & \multicolumn{1}{l|}{3}              &                                         &                                      &                             \\ \cline{2-4}
                                  \multicolumn{7}{c}{ } \\ \hline       
\multicolumn{1}{|l|}{\bf Arm Vocab}  & Arm Direction                                                & Arm Level       & \multicolumn{1}{l|}{Arm Crossing}  & \multicolumn{1}{l|}{Elbow Folding} & \multicolumn{1}{l|}{Body Inclusion} & \multicolumn{1}{l|}{Mirror} \\ \hline
\multicolumn{1}{|l|}{Natyarambhe} & 2                                                             & 2                & \multicolumn{1}{l|}{0}              & \multicolumn{1}{l|}{1}                  & \multicolumn{1}{l|}{0}               & \multicolumn{1}{l|}{1}      \\ \hline

\multicolumn{7}{c}{ } \\ \cline{1-3}
\multicolumn{1}{|l|}{\bf Head Vocab} & Direction                                                     & Level            &                                     &                                         &                                      &                             \\ \cline{1-3}
\multicolumn{1}{|l|}{Samam}       & 1                                                             & 2                &                                     &                                         &                                      &                             \\ \cline{1-3}
\end{tabular}
}
\end{table}
\renewcommand{\baselinestretch}{1.3} 

\subsection{LabanXML}\label{sec:laban_xml}
While the graphical symbolization of Laban and our encoding in tabular formats as above are both forms of transcription, neither is amenable to machine processing. To visualize the postures and to build further applications based on the transcripts, we need a searchable and parseable representation. So we adopt LabanXML~\cite{nakamura2006xml} -- an {\em eXtensible Markup Language} (XML) design for Labanotation. LabanXML bundles columns of the staff in four groups -- left, right, support and head. Left and right, in turn, contains arm and leg.
 

The tags of LabanXML are as follows:
\begin{itemize}
\item \texttt{<laban>}: This is root tag which includes \texttt{<attribute>} and \texttt{<notation>} tags.
\item	\texttt{<attribute>}: This includes tag \texttt{<title>} used to name the XML file.
\item	\texttt{<notation>}: This includes tag \texttt{<measure>}.
\item	\texttt{<measure>}: Which gives position of current pose on time line.
\item	\texttt{<left>}: Contains tags for columns appearing on left side of Labanotation.
\item	\texttt{<right>}: Contains tags for columns appearing on right side of Labanotation.
\item	\texttt{<support>}: Describes the support element in Labanotation columns. It has attribute {\em side} having value {\em left} or {\em right} indicating the side of the support.
\item	\texttt{<arm>, <leg>, <foot>, <head>}, and \texttt{<support>}: These tags include \\ \texttt{<direction>} and \texttt{<level>} tags of the respective limb.
\item	\texttt{<elbow>, <knee>}: These tags include \texttt{<degree>} for degree of folding.
\item	\texttt{<touch>}: This tag is included in \texttt{<support>} tag and \texttt{<leg>} tag. It describes how foot is hooked to floor.
\end{itemize}

Using the above tags, we represent the information from Table~\ref{tbl:laban_encoding_Natta1P1} in XML format in Table~\ref{tbl:Natta1P1_XML}. The graphical representation of Laban encoding is shown in Figure~\ref{fig:BN_Laban}. The symbols described earlier are used to write the XML tags in Laban staff. 

\renewcommand{\baselinestretch}{1}
\begin{table}[!ht]
\caption{LabanXML  of Posture Natta1P1}\label{tbl:Natta1P1_XML}
\begin{scriptsize}
\begin{tabular}{|l|l|} 
\hline
\begin{minipage}{.45\textwidth}
\begin{verbatim}

-<laban>
   -<attribute>
      <title>natta_1</title>
   </attribute>
   -<notation>
      -<measure num="0">
         -<left>
            -<arm duration="1">
               <direction>2</direction>
               <level>2</level>
             </arm>
            -<elbow duration="1">
               <Degree>1</Degree>
             </elbow>
            -<foot>
               <touch>3</touch>
             </foot>
            -<knee duration="1">
               <Degree>3</Degree>
             </knee>
          </left>
         -<right>
            -<arm duration="1">
               <direction>3</direction>
               <level>2</level>
             </arm>
             
\end{verbatim}
\end{minipage}
&
\begin{minipage}{.45\textwidth}
\begin{verbatim}

            -<elbow duration="1">
               <Degree>1</Degree>
             </elbow>
            -<foot>
               <touch>3</touch>
             </foot>
            -<knee duration="1">
               <Degree>3</Degree>
             </knee>
          </right>
         -<support side="left">
            <direction>1</direction>
            <level>3</level>
          </support>
         -<support side="right">
            <direction>1</direction>
            <level>3</level>
          </support>
         -<head>
            <direction>1</direction>
            <level>2</level>
            </head>
          </measure>          
    </notation>
</laban>


\end{verbatim}
\end{minipage} \\ \hline
\end{tabular}
\end{scriptsize}
\end{table}
\renewcommand{\baselinestretch}{1.3}


\subsection{Tool Overview}
\begin{figure}[!ht]
\centering
\includegraphics[width=.75\textwidth]{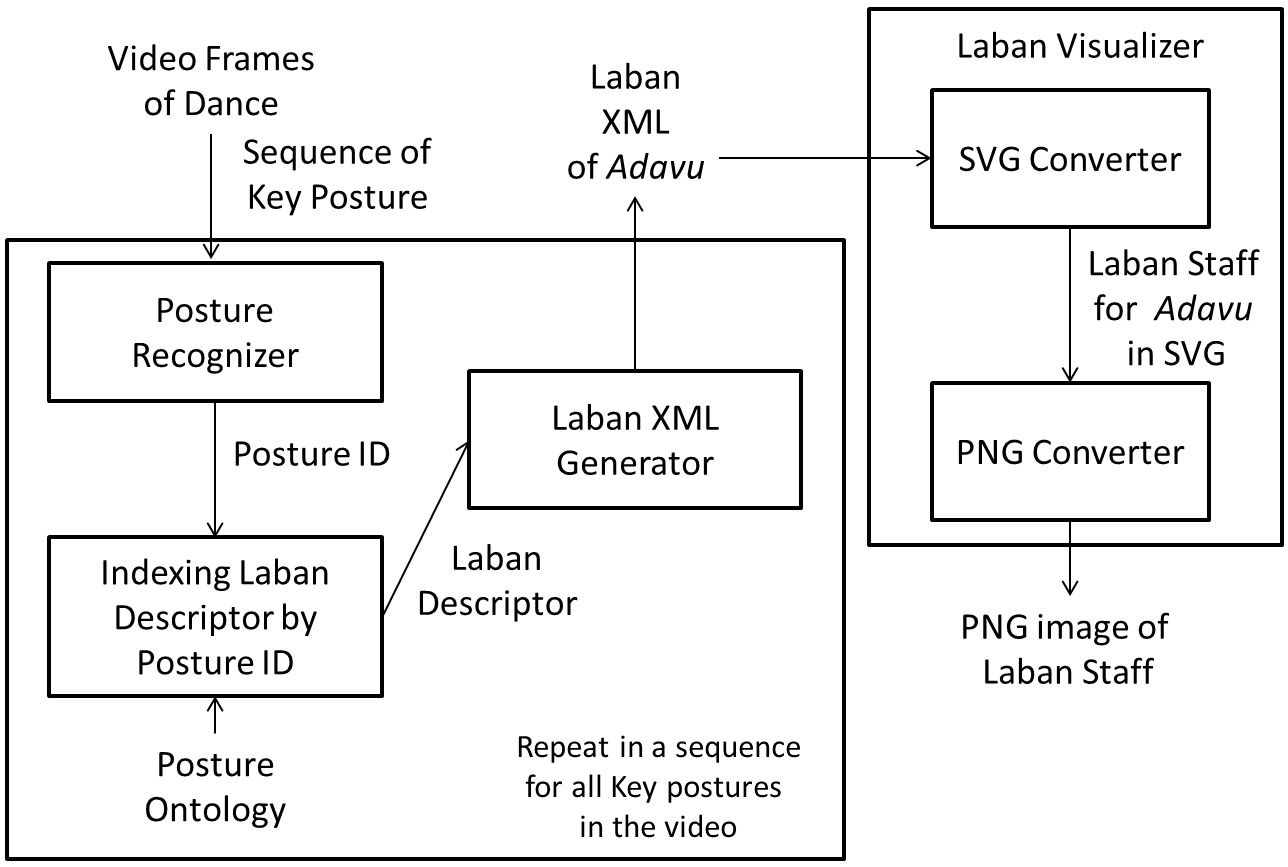}\vspace*{-0.3cm}
\caption{Architecture of the {\em Adavu} Transcription Tool\label{fig:trans_system}}
\end{figure}

To build the {\em Adavu} Transcription Tool, we first encode our ontological models of {\em Adavu}s, especially the key postures and their sequences, and the video annotations in Laban ontology following the approach as illustrated in Section~\ref{sec:Natta1_posture_trans}. This cross-ontology of concepts (called {\em Posture Ontology}) are then represented in a mapping database indexed by the posture ID. This is used by the  {\em Adavu} Transcription Tool as given in Figure~\ref{fig:trans_system}. We explain the modules below.

\begin{figure}[!ht]
\centering
\begin{scriptsize}
\begin{tabular}{cccc}
\begin{minipage}{.2\textwidth}
 \includegraphics[width=2.5cm]{Images/Avavu_Images/2nP1.png}
\end{minipage}
&
\begin{minipage}{.2\textwidth}
 \includegraphics[width=2.5cm]{Images/Avavu_Images/2nP2.png}
\end{minipage}
& 
\begin{minipage}{.2\textwidth}
 \includegraphics[width=2.5cm]{Images/Avavu_Images/2nP3.png}
\end{minipage}
& 
\begin{minipage}{.2\textwidth}
 \includegraphics[width=2.5cm]{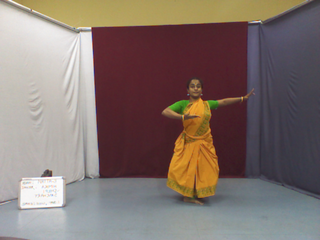}
\end{minipage}
\\ 
(C01) & (C02) & (C03) & (C04) \\ 
\begin{minipage}{.2\textwidth}
 \includegraphics[width=2.5cm]{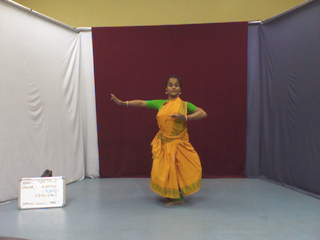}
\end{minipage}
&
\begin{minipage}{.2\textwidth}
 \includegraphics[width=2.5cm]{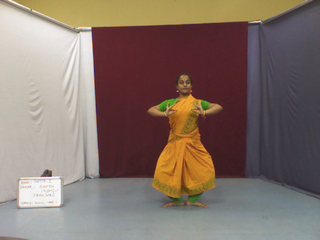}
\end{minipage}
& 
\begin{minipage}{.2\textwidth}
 \includegraphics[width=2.5cm]{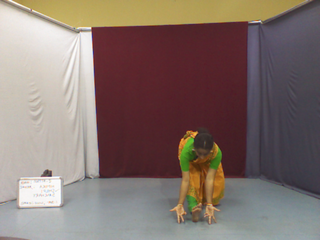}
\end{minipage}
& 
\begin{minipage}{.2\textwidth}
 \includegraphics[width=2.5cm]{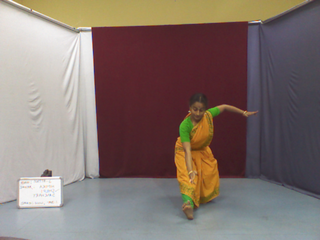}
\end{minipage} \\
(C05) & (C06) & (C07) & (C08) \\ 
\begin{minipage}{.2\textwidth}
 \includegraphics[width=2.5cm]{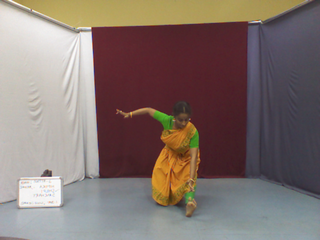}
\end{minipage}
&
\begin{minipage}{.2\textwidth}
 \includegraphics[width=2.5cm]{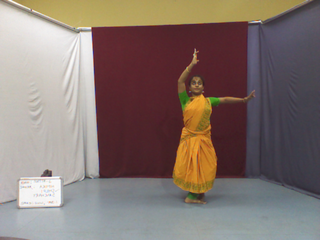}
\end{minipage}
& 
\begin{minipage}{.2\textwidth}
 \includegraphics[width=2.5cm]{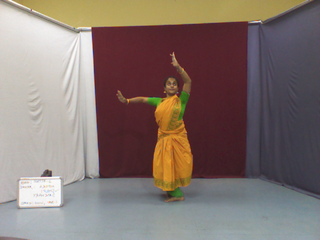}
\end{minipage}
& 
\begin{minipage}{.2\textwidth}
 \includegraphics[width=2.5cm]{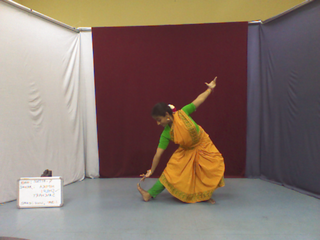}
\end{minipage} \\
(C09) & (C10) & (C11) & (C12) \\ 
\begin{minipage}{.2\textwidth}
 \includegraphics[width=2.5cm]{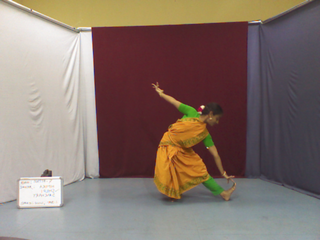}
\end{minipage}
&
\begin{minipage}{.2\textwidth}
 \includegraphics[width=2.5cm]{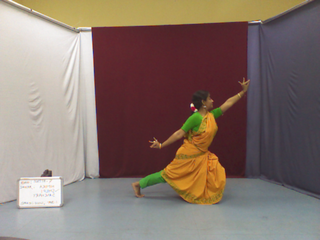}
\end{minipage}
& 
\begin{minipage}{.2\textwidth}
 \includegraphics[width=2.5cm]{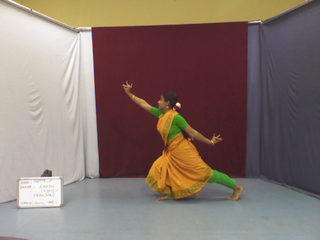}
\end{minipage}
& 
\begin{minipage}{.2\textwidth}
 \includegraphics[width=2.5cm]{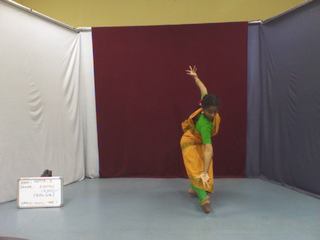}
\end{minipage} \\
(C13) & (C14) & (C15) & (C16) \\ 

\begin{minipage}{.2\textwidth}
 \includegraphics[width=2.5cm]{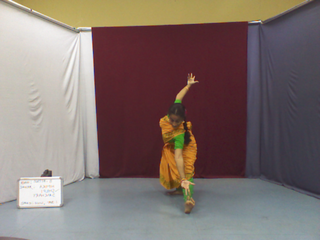}
\end{minipage}
&
\begin{minipage}{.2\textwidth}
 \includegraphics[width=2.5cm]{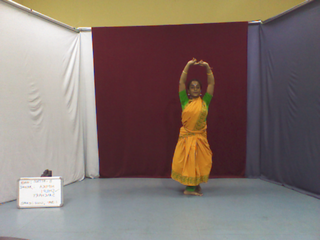}
\end{minipage}
& 
\begin{minipage}{.2\textwidth}
 \includegraphics[width=2.5cm]{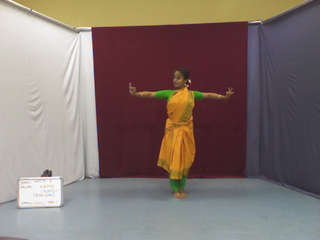}
\end{minipage}
& 
\begin{minipage}{.2\textwidth}
 \includegraphics[width=2.5cm]{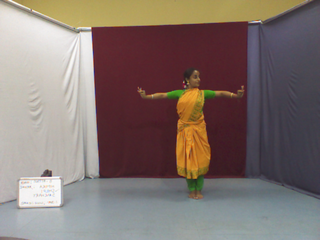}
\end{minipage} \\
(C17) & (C18) & (C19) & (C20) \\ 
\begin{minipage}{.2\textwidth}
 \includegraphics[width=2.5cm]{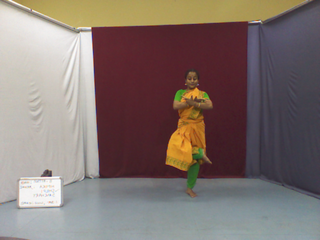}
\end{minipage}
&
\begin{minipage}{.2\textwidth}
 \includegraphics[width=2.5cm]{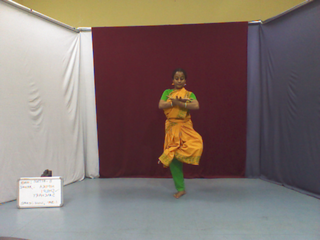}
\end{minipage}
& 
\begin{minipage}{.2\textwidth}
 \includegraphics[width=2.5cm]{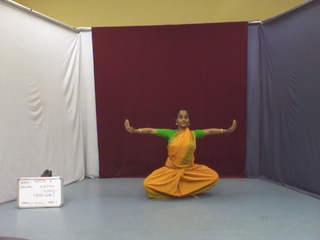}
\end{minipage}
& \\
(C21) & (C22) & (C23) & \\ 
\end{tabular}
\end{scriptsize}
\caption{23 Key Postures of {\em Natta Adavu}s with Class / Posture IDs\label{fig:natta_adavus_postures}}
\end{figure}

\renewcommand{\baselinestretch}{1}
\begin{table}[!ht]
\centering 
\caption{Data Set for Posture Recognition using 23 posture classes in Figure~\ref{fig:natta_adavus_postures} \label{tab:posture_data}}
\begin{tabular}{|l|r|r||l|r|r|} \multicolumn{6}{c}{ } \\ \hline
\multicolumn{1}{|c}{\bf Posture} & \multicolumn{1}{|c}{\bf Training} & \multicolumn{1}{|c}{\bf Test} & \multicolumn{1}{||c}{\bf Posture} & \multicolumn{1}{|c}{\bf Training} & \multicolumn{1}{|c|}{\bf Test} \\ 
\multicolumn{1}{|c}{\bf ID} & \multicolumn{1}{|c}{\bf data} & \multicolumn{1}{|c}{\bf data} & \multicolumn{1}{||c}{\bf ID} & \multicolumn{1}{|c}{\bf data} & \multicolumn{1}{|c|}{\bf data} \\ \hline \hline
 C01 & 6154 & 1457 & C13 & 235 & 80\\
 C02 & 3337 & 873	& C14 & 393 & 117\\
 C03 & 3279 & 561 & C15 & 404 & 121\\
 C04 & 1214 & 219 & C16 & 150 & 48\\
 C05 & 1192 & 268 & C17 & 161 & 51\\
 C06 & 1419 & 541 & C18 & 323 & 81\\
 C07 & 1250 & 475 & C19 & 175 & 46\\
 C08 & 284 & 112 & C20 & 168 & 43\\
 C09 & 306 & 133 & C21 & 19 & 6\\
 C10 & 397 & 162 & C22 & 21 & 6\\
 C11 & 408 & 117 & C23 & 118 & 61\\
 C12 & 229 & 84 & & & \\ \hline
 \multicolumn{6}{l}{} \\
 \multicolumn{6}{p{7.5cm}}{\scriptsize Numbers indicate the number of {\em K-frame}s. Each {\em K-frame} is given by the frame number of the RGB frame in the video. Associated depth and skeleton frames are used as needed. Various position and formation information on body parts are available for every {\em K-frame} from annotation} \\
\end{tabular}
\end{table}
\renewcommand{\baselinestretch}{1.3}

\subsubsection*{Posture Recognizer} This is a machine learning based system \cite{mallick2019posture} helps to recognize a unique posture id when RGB frame of key posture is given.  We first extract the human figure, eliminate the background, and convert the RGB into grayscale image. We next compute the {\em Histograms of Oriented Gradient} (HOG)  descriptors for each posture frame. Finally, we use HOG feature to train the same SVM classifier. There are total 23 key postures in {\em Natta Adavu}s. To recognize the postures into 23 posture classes, we use {\em One vs. Rest} type of multi-class SVM. The data set shown in Table~\ref{tab:posture_data} is used for training and testing the SVM. For testing we use the trained SVM models to predict the class labels. Our accuracy of the posture recognition is 97.95\%. 

Now we use the trained classifier to recognizer the input sequence of key postures. The key posture recognizer extract the sequence of key postures in terms of their posture IDs from the video of an {\em Adavu} performance. 

\subsubsection*{Indexing Laban Descriptor by Posture ID}
Given a posture ID, we look up the {\em Posture Ontology} to get the Laban descriptor values for the different limbs in terms of a database record.

\subsubsection*{LabanXML Generator}
From the database record of Laban descriptors an equivalent LabanXML file is generated using the definition of tags as in Section~\ref{sec:laban_xml}.

\subsubsection*{Laban Visualizer}
Since Labanotation is graphical, it is important to visualize it in terms of its icons. So we implement a converter from LabanXML to Scalable Vector Graphics (SVG). SVG is an XML-based vector image format for two-dimensional graphics with support for interactivity and animation. Like XML, SVG images can also be created and edited with any text editor, as well as with drawing software. The SVG converter is written in C++ on cygwin64 using libxml xml parser. SVG images are also rendered in PNG (using Inkscape) for easy to use offline notation.

%

\begin{figure}[!ht]
\centering
\includegraphics[width=\textwidth]{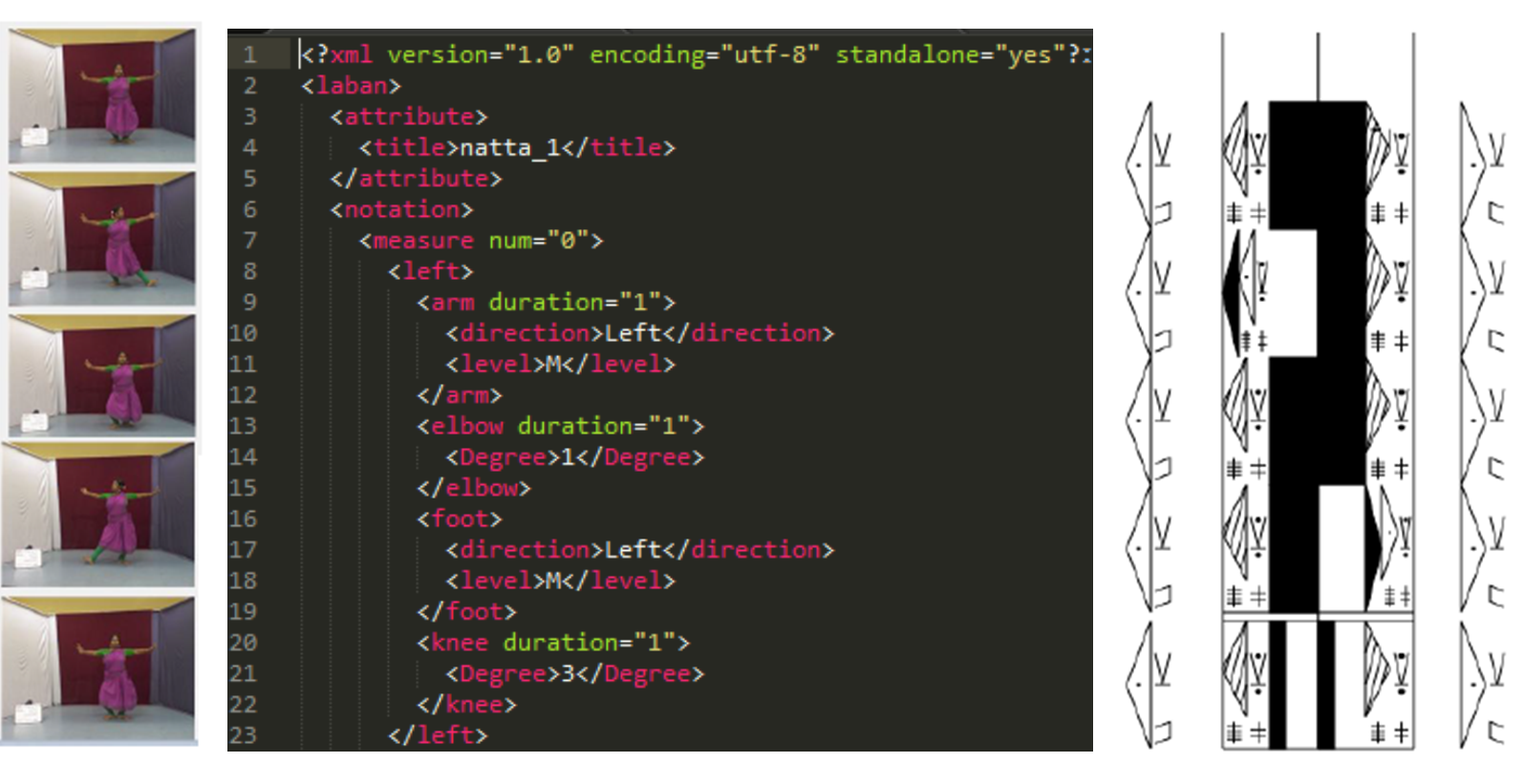}\vspace*{-0.3cm}
\caption[Transcription of {\em Natta Adavu}]{Key postures of {\em Natta Adavu Variant 1} with transcription in LabanXML (a part) and depiction in Laban Staff by our tool\label{fig:laban_result}}
\end{figure}

\subsection{Results and Discussion}
Our tool is able to generate transcription for a sequence of key frames. For given a sequence of RGB frames, Posture Recognizer generates their posture IDs. These posture IDs are mapped to corresponding cluster IDs in the laban ontology. By using posture IDs and ontology a Laban transcription for all frames is encoded in LabanXML. By using LabanXML a stack of Labans for BN Adavu key postures is generated. The Laban XML and stack of postures in Laban Staff, as generated by our tool for the sequence of key postures of {\em Natta Adavu} variation 1, are shown in Figure~\ref{fig:laban_result}. For a sequence of key frames from {\em Natta Adavu 1}, we show the transcription in Figure~\ref{fig:laban_result}. The RGB frames are shown on left and the corresponding Laban descriptors are shown on the staff on right. An initial part of the LabanXML is given in the middle.

\section{Conclusion}
In this paper, we demonstrate a system to generate parse-able representation of {\em Bharatanatyam} dance performance and document the parse-able representation using Labanotation. The system uses a unique combination of multimedia ontology and machine learning techniques. To the best of our knowledge this is the first work towards automatic documentation of dance using any notation.

In the process of developing the system, we have also presented a detailed ontology for {\em Bharatanatyam Adavu}s which is a maiden such attempt for any Indian Classical Dance. Finally, we have captured and annotated a sizable dataset for {\em Adavu}s, part of which is also available for use at:~\cite{icd_dataset}.

In future we intend to extend our work to document more fine description of each postures. We are also interested to capture movement which we have used for this study. Finally, we also want to extend our work to generate the ontology automatically guided by the grammar of the dance form.

\section*{Acknowledgment}

The authors would like to thank Tata Consultancy Services (TCS) for providing the fund and support for this work.


\bibliographystyle{acm}
\bibliography{References}



\end{document}